\documentclass[12pt]{article}
\usepackage{amsmath,amssymb,amsfonts,color,graphicx,cite,color,feynarts,soul}
\usepackage{epsfig}
\input paperdef
\graphicspath{{figs/}}

\evensidemargin \oddsidemargin
\allowdisplaybreaks

\begin{document}
\thispagestyle{empty}

\def\thefootnote{\fnsymbol{footnote}}

\begin{flushright}
IFT-UAM/CSIC-10-41\\
FTUAM-10-10
\end{flushright}

\vspace{0.5cm}

\begin{center}

{\large\sc {\bf Higgs Boson Masses in the MSSM}}

\vspace{0.4cm}

{\large\sc {\bf with Heavy Majorana Neutrinos}}

\vspace{1cm}

{\sc
S.~Heinemeyer$^{1}$%
\footnote{email: sven.heinemeyer@cern.ch}%
, M.J.~Herrero$^{2}$%
\footnote{email: maria.herrero@uam.es}%
, S.~Pe\~naranda$^{3}$%
\footnote{email: siannah@unizar.es}%

\vspace{0.4em}

~and A.M.~Rodr\'iguez-S\'anchez$^{2}$%
\footnote{email: anam.rodriguez@uam.es}%
}

\vspace*{.7cm}

{\sl
$^1$Instituto de F\'isica de Cantabria (CSIC-UC), Santander, Spain

\vspace*{0.1cm}

$^2$Departamento de F\'isica Te\'orica and Instituto de F\'isica Te\'orica,
UAM/CSIC\\
Universidad Aut\'onoma de Madrid, Cantoblanco, Madrid, Spain

\vspace*{0.1cm}

$^3$Departamento de F\'isica Te\'orica, Universidad de Zaragoza, 
Zaragoza, Spain

}

\end{center}

\vspace*{0.1cm}

\begin{abstract}
\noindent
We present a full diagrammatic computation of the one-loop corrections from the 
neutrino/sneutrino sector to the renormalized neutral $\cp$-even Higgs boson 
self-energies and the 
lightest Higgs boson mass, $\Mh$, within the context of the so-called 
MSSM-seesaw scenario. This consists of the Minimal Supersymmetric
Standard Model with the addition of massive 
right handed Majorana neutrinos and their supersymmetric partners, and  
where the seesaw mechanism is used for the
lightest neutrino mass generation. 
We explore the dependence on all the parameters involved, with
particular emphasis in the role played by the heavy 
Majorana scale. We restrict ourselves to the case of one generation of
  neutrinos/sneutrinos. For the numerical part of the study, we consider a very wide range of
values for all the parameters involved. We find sizeable corrections to
$\Mh$, which are negative in the region where the Majorana scale is
large ($10^{13}-10^{15} \gev$) 
and the lightest neutrino mass is within a range inspired by data
($0.1-1$~eV). For
some regions of the MSSM-seesaw parameter space, the corrections to $\Mh$
are substantially larger than the anticipated Large Hadron Collider  precision.
\end{abstract}

\def\thefootnote{\arabic{footnote}}
\setcounter{page}{0}
\setcounter{footnote}{0}

\newpage

%%%%%%%%%%%%%%%%%%%%%%%%%%%%%%%%%%%%%%%%%%%%%%%%%%%%%%%%%%%%%%%%%%%%%%%%%%%%%%%
%%%%%%%%%%%%%%%%%%%%%%%%%%%%%%%%%%%%%%%%%%%%%%%%%%%%%%%%%%%%%%%%%%%%%%%%%%%%%%%
\section{Introduction}
The current impressive experimental data on neutrino mass differences and 
neutrino mixing angles~\cite{pdg} indicate clearly a
signal of new physics beyond
the so far successful Standard Model of Particle Physics (SM). In order 
to incorporate the non-vanishing neutrino masses required by data
an extension of the SM with massive neutrinos is mandatory. Among the various
possibilities to extend the SM we choose here the most popular one that 
incorporates massive Majorana neutrinos and also stabilizes the
electroweak symmetry breaking scale, $v=174 \gev$, against potentially large 
radiative corrections in the presence of the new physics scale. We refer to   
the simplest version of a supersymmetric extension of the SM, the Minimal
Supersymmetric Standard Model (MSSM)~\cite{mssm}, with the addition of heavy 
right-handed Majorana neutrinos, and      
where the well known seesaw mechanism of type I~\cite{seesaw:I} is implemented 
to generate the  observed small neutrino masses. From now on we will
denote this model by ``MSSM-seesaw''. 

In this MSSM-seesaw context, 
the smallness of the light
neutrino masses, $m_{\nu} \sim \mD^2/\mM$, appears naturally due to the 
induced large suppression by the ratio of the two very distant mass scales.
Namely, the  Majorana 
neutrino mass $\mM$, that represents the new physics scale, and the Dirac
neutrino mass $\mD$, which is related to the electroweak scale via the 
neutrino Yukawa couplings $\Ynu$, by $\mD=\Ynu v \sin \beta$. The Higgs 
sector content in the MSSM-seesaw is as in the MSSM~\cite{Gunion:1986yn}, 
and $\tan \beta$ is given, as usual,  by the ratio of the two MSSM Higgs
vacuum expectation  
values (v.e.v.s). Although the present neutrino data requires two or more
neutrino generations, we shall adopt here the 
simplest case of one neutrino 
generation in order to fully understand first the role of one single Majorana
scale $\mM$, and postpone the more complex case of three generations for a
future work. In this simplified one-generation MSSM-seesaw framework, 
small neutrino masses of the order of $\mnu \sim 0.1-1$ eV can be easily 
accommodated with large Yukawa couplings, $\Ynu \sim {\cal O}(1)$, if the new
physics scale is very large, within the range $\mM \sim 10^{13}-10^{15} \gev$.
This is to be compared with the Dirac neutrino 
case where, in order to get similar small neutrino masses, extremely tiny,
hence irrelevant, Yukawa couplings of the order of 
$\Ynu \sim 10^{-12}-10^{-13}$ are required.

The hypothesis of Majorana massive neutrinos is 
very appealing for various reasons, including the interesting possibility of
generating satisfactorily baryogenesis via leptogenesis~\cite{Fukugita:1986hr}, 
and also because
they can produce
an interesting and singular phenomenology due to their potentially large Yukawa
couplings to the Higgs sector of the theory, the MSSM in the present case.  
Among the most striking phenomenological implications of these MSSM-seesaw
scenarios~\cite{Raidal:2008jk}, it is worth mentioning: 1)  the
prediction of sizeable rates for lepton 
flavor violating processes, indeed within the present experimental reach
for specific areas of the model parameters~\cite{LFV,LFV2},
2) non-negligible contributions to electric dipole moments of charged
leptons~\cite{EDM}, and 3) the occurrence of sneutrino-antisneutrino 
oscillations~\cite{Grossman:1997is} and sneutrino
flavor-oscillations~\cite{Dedes:2007ef}.

The present paper investigates another implication of heavy Majorana neutrinos 
that could be as relevant as these previously mentioned ones. More
specifically, we are interested here in the 
indirect effects of Majorana neutrinos via their radiative corrections to
the MSSM Higgs boson masses. In particular, our study will be focused on the 
radiative corrections to the lightest MSSM $\cp$-even $h$ boson mass, $\Mh$,
due to the one-loop contributions from the neutrino/sneutrino sector 
within the MSSM-seesaw framework. Previous studies in particular SUSY scenarios
and under specific assumptions on the model
parameters~\cite{Cao:2004hs,Farzan:2004cm,LFV2,Kang}  
indicate that the size of these radiative
corrections to the Higgs mass parameters in the case of extremely 
heavy Majorana neutrinos can be sizeable due to the large size of $\Ynu$.

For the estimates of the total corrections to $\Mh$ in the MSSM-seesaw,  
obviously, the one-loop 
corrections from the neutrino/sneutrino sector that we are interested here   
have to be added to the existing
MSSM corrections.
The status of radiative corrections to $\Mh$ in the
non-$\nu/\Snu$~sector, i.e.\ in the MSSM {\em without} massive
neutrinos, can be summarized as follows. Full one-loop
calculations~\cite{mhiggsf1l} have been supplemented by the leading and
subleading two-loop corrections, see~\cite{mhiggsAEC} and references
therein. Together with leading three-loop corrections~\cite{mhiggs3l}
the current precision in $\Mh$ is estimated to be 
$\sim 2 - 3 \gev$~\cite{mhiggsAEC}.

Regarding the previous estimates of neutrino/sneutrino radiative corrections
to the Higgs mass parameters the status is as follows.
In \citere{Cao:2004hs} the one-loop corrections to $\Mh$ were estimated 
within a 
split SUSY scenario where the soft-SUSY-breaking mass associated to the 
right handed neutrino, $\mR$, was chosen to be very large, of the 
order of the Majorana scale $\mM$. 
They
worked in the zero external momentum approximation and switching off the
$SU(2)\times U(1)$ gauge interactions. Besides, they 
used the mass insertion approximation for the other soft-breaking sneutrino
parameters, $A_\nu$ and $\Bnu$, associated to the trilinear coupling and
neutrino $B$-term respectively. A large and negative correction
from the neutrino/sneutrino sector of the order of a few tens of GeV was found
for $\mM= 10^{14} \gev$ and $\mR \sim {\cal O}(\mM)$. 
In \citere{Farzan:2004cm} the radiative one-loop effects 
of the neutrino $B$-term
on the Higgs mass parameters within the context of mSUGRA (with
universal scalar masses at the $\msusy$, including $\mR$) were
analyzed by means of the renormalization group equations (RGEs). They found
large effects from this $\Bnu$ term that
indeed could destabilize the electroweak symmetry breaking. By requiring a
proper breaking in this mSUGRA framework they concluded with an upper bound of 
$\Bnu \Ynu^2/(8 \pi^2) <  \msusy/\tb$. Large corrections to the Higgs soft mass
parameters within a SUSY-seesaw framework with total or partial universality
conditions have also been found by a similar
RGEs analysis in \cite{LFV2,Kang}. In \cite{LFV2} it was concluded that 
these corrections induce a considerable decrease in the physical Higgs boson 
masses which in turn enhance the rates of the Higgs-mediated LFV processes. 
In \cite{Kang} the large threshold corrections found from the heavy
neutrinos/sneutrinos were shown to affect, and even dominate at large $\Bnu$, 
the radiative breaking of the electroweak symmetry and also modify considerably 
the predictions on the neutralino dark matter abundance. 
 
In this work, we will consider instead the more general MSSM-seesaw
scenarios with no universality conditions imposed, and explore the full 
parameter space, without restricting 
ourselves just to large or small values on neither of the relevant
neutrino/sneutrino 
parameters. In principle, since the right handed Majorana neutrinos and 
their SUSY partners are $SU(2) \times U(1)$ singlets, there is no a priori 
reason why the size of their associated parameters should be related to the 
size
of the other sector parameters. 
In the numerical estimates, we will therefore explore
a wide interval for all the involved 
neutrino/sneutrino relevant input parameters. 

We will present here a full one-loop computation of the radiative
corrections to the lightest $\cp$-even Higgs boson mass from
the (one generation) 
neutrino/sneutrino sector in which we will not use any of the previous 
approximations and we will not set the external momentum to zero. The
complete set of one-loop neutrino/sneutrino contributing  
diagrams will be taken into account, with both Yukawa and gauge couplings 
switched on. We also analyze the results in several 
renormalization schemes, which will be shown to provide remarkable
differences. In addition, we present some analytical and numerical 
results in the interesting limit of very large $\mM$ as compared to all
other scales involved, which will help us in the understanding of the 
important issue of the decoupling/non-decoupling of the heavy Majorana scale. 
Our further study in the particular region of large $\mM$ and $\mR$ 
will also allow us to 
compare our results with those in \cite{Cao:2004hs}. 

Our final aim is to find out to what extent the radiative corrections computed
here enter into the measurable range. The experimental perspectives for
the Higgs mass 
measurements with precision enough to be sensitive to such sizeable radiative
corrections, as the ones found here, are indeed quite promising.  The
LHC has good prospects to discover at least one neutral Higgs boson over 
the full MSSM parameter space and a precision on the mass of a Standard
Model (SM)-like Higgs boson of $\sim 200 \mev$ are
expected~\cite{lhctdrs,atlashiggs,cmshiggs,cmsHiggs}  
(see e.g.~\cite{jakobs,schumi} for reviews). 
At the ILC a determination of the Higgs boson properties (within the
kinematic reach) will be possible, and an accuracy on the mass could
reach the $50 \mev$~level~\cite{tesla,orangebook,acfarep,Snowmass05Higgs}. 
The interplay of the LHC and the ILC in the neutral MSSM Higgs sector will
improve certainly these measurements~\cite{lhcilc,eili}.

The paper is organized as follows. In section 2, we summarize the most important
ingredients of the MSSM-seesaw scenario that are needed for the present 
computation of the Higgs mass loop corrections. These include, the setting of
the model parameters and the complete list of the Lagrangian relevant terms. A
complete set of the corresponding relevant Feynman rules in the physical basis 
is also provided here. They are collected in the Appendix A and, to
our knowledge, they  are
not available in the previous literature. We
also comment shortly in section 2 on the comparison between the Dirac and 
the Majorana cases.
In section 3 we present the renormalization procedure and emphasize the
differences between the selected renormalization schemes, specifically,
the on-shell and the ${\DRbar}$ schemes. Section~4 is devoted to the
results. First we present the 
analytical results for the renormalized Higgs boson self-energies (the main
formulas are collected in Appendix B). Then we
present the numerical results in terms of all the relevant
neutrino/sneutrino parameters that we explore exhaustively in the full plausible
range. We also include in this section a study of the behavior of the
renormalized Higgs self-energies in the large $\mM$ limit. The final
part of this section summarizes the main numerical results for the
lightest Higgs boson mass 
corrections. Finally, section 5 contains the conclusions.

%%%%%%%%%%%%%%%%%%%%%%%%%%%%%%%%%%%%%%%%%%%%%%%%%%%%%%%%%%%%%%%%%%%%%%%%%%%%%%
%%%%%%%%%%%%%%%%%%%%%%%%%%%%%%%%%%%%%%%%%%%%%%%%%%%%%%%%%%%%%%%%%%%%%%%%%%%%%%

\section{The MSSM-seesaw model}
\label{sec:nN}

\noindent
The model we are interested in here is the MSSM extended by right
handed neutrinos and their SUSY partners, and where a seesaw mechanism of 
type~I~\cite{seesaw:I} is implemented to generate the neutrino masses and
mixing angles.  
This is called usually the MSSM-seesaw model. For simplicity, as already
announced in the
introduction, 
we will 
restrict here to the one generation neutrinos/sneutrinos case although
the full compatibility with
present neutrino data for mass differences and mixing angles, requires
additional neutrino generations. Since the main idea is to analyze 
the radiative corrections from the neutrino-sneutrino sector to the
lightest Higgs mass, we restrict ourselves to the case of one
generation of neutrinos/sneutrinos. We illustrate first this
simpler case and postpone the more complex case of three
generations for a future work.

%%%%%%%%%%%%%%%%%%%%%%%%%%%%%%%%%%%%%%%%%%%%%%%%%%%%%%%%%%%%%%%%%%%%%%%%%%%%%%

\subsection{The neutrino/sneutrino sector}
\label{sec:nusnu}

The MSSM-seesaw model with one neutrino/sneutrino generation is described
in terms of the well known MSSM superpotential plus the new
relevant terms contained in:
\begin{equation}
\label{W:Hl:def}
W\,=\,\epsilon_{ij}\left[\Ynu \hat H_2^i\, \hat L^j \hat N \,-\, 
Y_l \hat H_1^i\,\hat L^j\, \hat R  \right]\,+\,
\edz\,\hat N \,\mM\,\hat N \,,
\end{equation}
where $\mM$ is the Majorana mass and  $\hat N = (\Snu_R^*, (\nu_R)^c)$ is 
the additional superfield that contains the  right-handed 
neutrino $\nu_{R}$ and its scalar partner $\Snu_{R}$. Here and in the
following $f^c$ 
denotes the particle-antiparticle conjugate (c-conjugate in short) of a fermion $f$ 
($f^c=C {\bar f}^T$) 
and $\tilde f^*$ denotes the complex conjugate of sfermion $\tilde f$.
The lepton Yukawa couplings are $Y_{l,\nu}$, and we use the convention 
$\epsilon_{12}=-1$ . The other superfields, $\hat L$ containing the  
lepton ($\nu_L, e_L$) and slepton ($\Snu_L, \tilde e_L$) 
$SU(2)$ doublets, $\hat R$ containing the lepton $(e_R)^c$ and
slepton $\tilde e_R^*$ $SU(2)$ singlets, and $\hat H _{1,2}$ containing
the Higgs boson $SU(2)$ doublets and their SUSY partners, are as in the
MSSM. We follow here the notation of ~\cite{Gunion:1986yn}.   

There are also new relevant terms in the soft SUSY breaking potential due to the
additional sneutrinos $\Snu_R$~\cite{Grossman:1997is}:
\begin{equation}
V^{\Snu}_{\rm soft}= m^2_{\tilde L} \Snu_L^* \Snu_L +
  m^2_{\tilde R} \Snu_R^* \Snu_R + (\Ynu \Anu H^2_2 
  \Snu_L \Snu_R^* + \mM \Bnu \Snu_R \Snu_R + {\rm h.c.})~.
\end{equation}

After electro-weak (EW) symmetry breaking, the charged lepton and 
Dirac neutrino masses
can be written as
\begin{equation}
m_l\,=\,Y_l\,\,v_1\,, \quad \quad
\mD\,=\,\Ynu\,v_2\,,
\end{equation}
where $v_i$ are the vacuum expectation values (VEVs) of the neutral Higgs
scalars, with $v_{1(2)}= \,v\,\cos (\sin) \be$ and $v=174 \gev$.

The $ 2 \times 2$ neutrino mass matrix is given in terms of $\mD$ and
$\mM$ by: 
\begin{equation}
\label{seesaw:def}
M^\nu\,=\,\left(
\begin{array}{cc}
0 & \mD \\
\mD & \mM
\end{array} \right)\,. 
\end{equation}
Diagonalization of $M^\nu$ leads to two mass
eigenstates, $n_i \,(i=1,2)$, which are Majorana fermions:

\begin{align}
\label{nuigenstates}
n_1 &\equiv  \nu 
     = \cos \theta (\nu_L +(\nu_L)^c)- \sin \theta (\nu_R +(\nu_R)^c) ~,
     \nonumber \\
n_2 &\equiv N 
    = \sin \theta (\nu_L+(\nu_L)^c)+ \cos \theta (\nu_R +( \nu_R)^c)
\end{align}
with the respective  mass eigenvalues given by:
\begin{equation}
\label{nuigenvalues}
m_{\nu,\, N}  = \edz \KL \mM \mp \sqrt{\mM^2+4 \mD^2} \KR~. 
\end{equation}  
It should be noticed that we have introduced an alternative notation that
makes it easier to identify the specific neutrino by its mass:
$\nu$ is the lighter one and $N$ is the heavier one.
It should also be kept in mind that with this convention $\mnu <0$ and 
$m_N > 0$, but the physical Majorana neutrino states have the proper positive
masses. 
These physical neutrinos can be reached by an additional rotation, 
$\nu \to \nu ' =e^{i\ga_5 \pi/2} \nu =-i \ga_5 \nu$, leading to 
$m_{\nu '}=|\mnu|$. 
However, we prefer to work instead with the mass eigenstates in
\refeq{nuigenstates} to avoid extra $i$ and $\ga_5$ factors in the
computation. Of course the 
final results in this work for the Higgs mass corrections are not sensitive to this choice.

The mixing angle that defines the mass eigenstates is given by,
\begin{equation}
\label{nuangle}
\tan \theta = -\frac{\mnu}{\mD}= \frac{\mD}{m_N}~.
\end{equation}
Other useful relations between the model parameters $\mD$, $\mM$  and
the physical neutrino parameters, $\mnu$, $m_N$ and $\theta$ are the
following: 
\begin{align}
\label{seesawrelations}
\sin^2 \theta &= \frac {-\mnu}{m_N-\mnu} 
               =\edz \KL 1-\frac{\mM}{\sqrt{\mM^2+4\mD^2}} \KR~, \\
\cos^2 \theta &=  \frac {m_N}{m_N-\mnu}
               =\edz \KL 1+\frac{\mM}{\sqrt{\mM^2+4\mD^2}} \KR~, \\
\mD &= \edz\sqrt{(m_N-\mnu)^2-(m_N+\mnu)^2}~,\\
\label{mDmN}
\mD^2 &= -\mnu m_N~,\\
\label{mMmN}
\mM &= \mnu+ m_N~.
\end{align}
   
Regarding the sneutrino sector, the sneutrino mass matrices for the
$\cp$-even, ${\tilde M}_{+}$,  and the $\cp$-odd, ${\tilde M}_{-}$,
subsectors are given respectively by~\cite{Grossman:1997is}: 
\begin{equation}
{\tilde M}_{\pm}^2=
\left( 
\begin{array}{cc} m_{\tilde{L}}^2 + \mD^2 + \edz \MZ^2 \cos 2 \be & 
\mD (A_{\nu}- \mu \CTb \pm \mM) \\  
\mD (A_{\nu}- \mu \CTb \pm \mM) &
m_{\tilde{R}}^2+\mD^2+\mM^2 \pm 2 \Bnu \mM \end{array} 
\right)~.
\end{equation}
The diagonalization of these two matrices, ${\tilde M}_{\pm}^2$, 
leads to four sneutrino mass eigenstates, ${\tilde n}_i \,(i=1,2,3,4)$ with 
respective $\cp$ parities 
$\cp(\tilde n_{1,2})=+1$ and $\cp({\tilde n}_{3,4})=-1$:
\begin{align} 
\label{snueigenstates}
{\tilde n}_1 &\equiv \Snu_+ = 
\sqrt{2}(\cos \theta_+  \re \Snu_L 
        -\sin \theta_+ \re \Snu_R)~, \non \\
{\tilde n}_2 &\equiv \SNu_+ =
\sqrt{2}(\sin \theta_+  \re \Snu_L 
        +\cos \theta_+ \re \Snu_R)~, \non \\
{\tilde n}_3 &\equiv \Snu_- =
\sqrt{2}(\cos \theta_-  \im \Snu_L -\sin \theta_- \im \Snu_R)~, \non \\
{\tilde n}_4 &\equiv \SNu_- =
\sqrt{2}(\sin \theta_- \im  \Snu_L +\cos \theta_- \im \Snu_R)~.
\end{align}
It should again be noted that we have introduced an alternative notation that
makes it easier to 
identify the specific sneutrino by its parity and mass: 
$\Snu_+$, $\SNu_+$ are respectively the lighter and the heavier ones
with $\cp=+1$, and $\Snu_-$, $\SNu_-$ are the lighter and the 
heavier ones with $\cp=-1$.  
The corresponding mass eigenvalues are:
\begin{align}
\label{snueigenvalues} 
  m_{{\Snu_+},{\SNu_+}}^2 &=  
\edz(\mM^2+\mL^2+\mR^2+2\mD^2+ \edz \MZ^2 \CZb + 2\Bnu \mM ) \\
&\; \mp \edz 
\sqrt{4\mD^2(\Anu-\mu \CTb + \mM)^2+(\mM^2+\mR^2-\mL^2
-\edz \MZ^2\CZb  +2\Bnu \mM)^2}~,  \non \\
 m_{{\Snu_-},{\SNu_-}}^2&= 
\edz(\mM^2+\mL^2+\mR^2+2\mD^2+ \edz \MZ^2 \CZb - 2\Bnu \mM) \\
&\; \mp \edz 
\sqrt{4\mD^2(\Anu-\mu \CTb - \mM)^2+(\mM^2+\mR^2-\mL^2
-\edz \MZ^2\CZb -2\Bnu \mM)^2}~.  \non 
\end{align}
The mixing angles in the two subsectors are given respectively by:
\begin{equation}
\label{sneuangle}
\sin 2\theta_{\pm}= \frac{2\mD(\Anu-\mu \CTb \pm \mM)}
{\sqrt{4\mD^2(\Anu-\mu \CTb \pm \mM)^2+
(\mM^2+\mR^2-\mL^2- \edz \MZ^2\CZb \pm 2\Bnu \mM )^2}}~.
\end{equation}

%%%%%%%%%%%%%%%%%%%%%%%%%%%%%%%%%%%%%%%%%%%%%%%%%%%%%%%%%%%%%%%%%%%%%%%%%%%%%%%

\subsection{The Higgs boson sector at tree-level}
\label{sec:tree}

In this subsection we summarize the Higgs-boson sector of our model at
tree-level. Contrary to the SM, in the MSSM two Higgs doublets
are required.
The  Higgs potential~\cite{hhg}
\BEA
V &=& m_{1}^2 |\cHe|^2 + m_{2}^2 |\cHz|^2 
      - m_{12}^2 (\epsilon_{ab} \cHe^a\cHz^b + \hc)  \non \\
  & & + \frac{1}{8}(g^2+g'^2) \left[ |\cHe|^2 - |\cHz|^2 \right]^2
        + \edz g^2|\cHe^{\dag} \cHz|^2~,
\label{higgspot}
\EEA
contains $m_1, m_2, m_{12}$ as soft SUSY breaking parameters;
$g, g'$ are the $SU(2)$ and $U(1)$ gauge couplings, and 
$\epsilon_{12} = -1$.

The doublet fields $H_1$ and $H_2$ are decomposed  in the following way:
\BEA
\cHe &=& \VL \cHe^0 \\[0.5ex] \cHe^- \VR \; = \; \VL v_1 
        + \ed{\wz}(\phi_1^0 - i\chi_1^0) \\[0.5ex] -\phi_1^- \VR~,  
        \non \\
\cHz &=& \VL \cHz^+ \\[0.5ex] \cHz^0 \VR \; = \; \VL \phi_2^+ \\[0.5ex] 
        v_2 + \ed{\wz}(\phi_2^0 + i\chi_2^0) \VR~.
\label{higgsfeldunrot}
\EEA
The potential \refeq{higgspot} can be described with the help of two  
independent parameters (besides $g$ and $g'$): 
$\Tb = v_2/v_1$ and $\MA^2 = -m_{12}^2(\Tb+\CTb)$,
where $\MA$ is the mass of the $\cp$-odd Higgs boson~$A$.

The diagonalization of the bilinear part of the Higgs potential,
i.e.\ of the Higgs mass matrices, is performed via the orthogonal
transformations 
\BEA
\label{hHdiag}
\VL H \\[0.5ex] h \VR &=& \ML \Ca & \Sa \\[0.5ex] -\Sa & \Ca \MR 
\VL \phi_1^0 \\[0.5ex] \phi_2^0 \VR ~, \\
\label{AGdiag}
\VL G \\[0.5ex] A \VR &=& \ML \Cb & \Sbe \\[0.5ex] -\Sbe & \Cb \MR 
\VL \chi_1^0 \\[0.5ex] \chi_2^0 \VR~,  \\
\label{Hpmdiag}
\VL G^{\pm} \\[0.5ex] H^{\pm} \VR &=& \ML \Cb & \Sbe \\[0.5ex] -\Sbe & 
\Cb \MR \VL \phi_1^{\pm} \\[0.5ex] \phi_2^{\pm} \VR~.
\EEA
The mixing angle $\al$ is determined through
\BE
\al = {\rm arctan}\KKL 
  \frac{-(\MA^2 + \MZ^2) \Sbe \Cb}
       {\MZ^2 \CQb + \MA^2 \SQb - m^2_{h}} \KKR~, ~~
 -\frac{\pi}{2} < \al < 0~.
\label{alphaborn}
\end{equation}

One gets the following Higgs spectrum:
\BEA
\mbox{2 neutral bosons},\, {\cal CP} = +1 &:& h, H \non \\
\mbox{1 neutral boson},\, {\cal CP} = -1  &:& A \non \\
\mbox{2 charged bosons}                   &:& H^+, H^- \non \\
\mbox{3 unphysical Goldstone bosons}      &:& G, G^+, G^- .
\EEA

At tree level the mass matrix of the neutral $\cp$-even Higgs bosons
is given in the $\Pe$-$\Pz$-basis 
in terms of $\MZ$, $\MA$, and $\Tb$ by
\BEA
M_{\rm Higgs}^{2} &=& \ML \mpe^2 & \mpez^2 \\ 
                           \mpez^2 & \mpz^2 \MR \non\\
&=& \ML \MA^2 \SQb + \MZ^2 \CQb & -(\MA^2 + \MZ^2) \Sbe \Cb \\
    -(\MA^2 + \MZ^2) \Sbe \Cb & \MA^2 \CQb + \MZ^2 \SQb \MR,
\label{higgsmassmatrixtree}
\EEA
which by diagonalization according to \refeq{hHdiag} yields the
tree-level Higgs boson masses

\BE
\label{rMSSM:mtree}
m_{H,h}^2=
\edz\left[\MA^2+\MZ^2 \pm\sqrt{(\MA^2+\MZ^2)^2-
4\MZ^2\MA^2\cos^2 2\be}\right] ~.
 \end{equation}
The charged Higgs boson mass is given by
\BE
\label{rMSSM:mHp}
m_{H^\pm}^2 = \MA^2 + \MW^2~.
\end{equation}
The masses of the gauge bosons are given in analogy to the SM:
\BE
\MW^2 = \edz g^2 (v_1^2+v_2^2) ;\qquad
\MZ^2 = \edz(g^2+g'^2)(v_1^2+v_2^2) ;\qquad M_\ga=0.
\end{equation}

%%%%%%%%%%%%%%%%%%%%%%%%%%%%%%%%%%%%%%%%%%%%%%%%%%%%%%%%%%%%%%%%%%%%%%%%%%%%%%%

\subsection{The interaction Lagrangian}
 
Finally the interaction Lagrangian that is relevant for the present work, 
expressed in the ($\nu_L, \nu_R$), ($\Snu_L,\Snu_R$) 
electroweak interaction basis, is given by:
\begin{equation}
\label{Lint} 
\cL_{\rm int}=\cL_{\nu \, H}+ \cL_{\nu \, Z} + \cL_{\Snu\,H} + 
\cL_{\Snu\,Z}~.
\end{equation} 
Here $\cL_{\nu \, H}$ and $\cL_{\Snu\,H}$ contain the interactions of the
neutrinos and sneutrinos with the Higgs bosons respectively; and  
$\cL_{\nu \, Z}$ and $\cL_{\Snu\,Z}$ those of the neutrinos and sneutrinos with
the $Z$~boson respectively.
 
For the various terms in \refeq{Lint} we find the following expressions:
\begin{align}
\label{LnuH}
\cL_{\nu \, H}&= -\frac{g \mD}{2\MW \sin \be}
((\overline{ \nu_L} \nu_R+\overline {\nu_R} \nu_L)
(H \sin \al + h \cos \al)
%\non \\
%&\quad
-i(\overline {\nu_L} \nu_R - \overline{ \nu_R} \nu_L)
A\cos \be )~, \\ 
\label{LnuZ}
\cL_{\nu \, Z}&= \frac{g}{2\cos \theta_W} 
\left[(\overline{\nu_L} \ga^\mu \nu_L)Z_\mu \right]~, \\
\label{LsnuH}
\cL_{\Snu\,H}&= -\frac{g \mD}{2 \MW \sin \be} \mu
\left[(\Snu_L^* \Snu_R+ \Snu_L \Snu_R^*)
(-H \cos \al  + h  \sin \al ) \right]  
\non \\
&\quad -\frac{g \mD^2}{\MW \sin \be} 
\left[(\Snu_R^* \Snu_R+ \Snu_L^* \Snu_L)
(H \sin \al + h \cos \al)\right] 
\non \\
&\quad + \frac{i g \mD}{2\MW} \mu
\left[(\Snu_L^* \Snu_R- \Snu_L \Snu_R^*) A  \right]
\non \\
&\quad -\frac{g \MZ}{2 cos \theta_W} 
\left[ (\Snu_L^* \Snu_L)
(H \cos (\al + \be) - h \sin (\al + \be))\right]
\non \\
&\quad - \frac{g \mD}{2 \MW \sin \be} \Anu
\left[(\Snu_L^* \Snu_R+ \Snu_L \Snu_R^*)
(H \sin \al + h \cos \al) \right]
\non \\
&\quad +\frac{ig \mD}{2 \MW \sin \be} \Anu 
\left[(\Snu_L^* \Snu_R- \Snu_L \Snu_R^*) 
A \cos \be \right]
\non \\ 
&\quad
-\frac{g \mD \mM}{2 \MW \sin \be}
\left[(\Snu_L \Snu_R+ \Snu_L^* \Snu_R^*)
(H \sin \al + h \cos \al) \right]
\non \\
&\quad
-i\frac{g \mD \mM}{2 \MW \sin \be} 
\left[(\Snu_L \Snu_R- \Snu_L^* \Snu_R^*)
A \cos \be \right]
\non \\
&\quad
-\frac{g^2 \mD^2}{4 \MW^2 \sin^2\be}
\left[(\Snu_L^* \Snu_L)
(H^2 \sin^2\al+ h^2 \cos^2\al+A^2 \cos^2\be+hH \sin 2\al)\right] 
\non \\  
&\quad
-\frac{g^2}{8 \cos^2 \theta_W}
\left[(\Snu_L^* \Snu_L)
(H^2 \cos 2\al - h^2 \cos 2\al - A^2 \cos 2\be -2 hH \sin 2\al)
\right]
\non \\  
&\quad
-\frac{g^2\mD^2}{4\MW^2 \sin^2 \be}
\left[ (\Snu_R^* \Snu_R)
(H^2 \sin^2\al + h^2 \cos^2\al + A^2 \cos^2\be + hH \sin 2\al)
\right]~,
\\
\label{LsnuZ}
\cL_{\Snu\,Z}&=
-\frac{ig}{2 \cos \theta_W} 
\left[ (\Snu_L^* \overleftrightarrow{\partial}^{\mu}\, 
\Snu_L) Z_\mu \right]
+\frac{g^2}{4 \cos^2 \theta_W}
\left[(\Snu_L^* \Snu_L)(Z_\mu Z^\mu) \right]~.
\end{align}  
The corresponding Feynman rules, expressed in the mass eigenstate basis, 
are collected in the Appendix A. Notice that this complete set of Feynman rules 
is, to our
knowledge, not available in the literature so far.

Some comments are in order. In the previous interaction Lagrangian, and
consequently in the 
Feynman rules, there are terms already
present in the MSSM. These are the pure gauge interactions between the
left-handed neutrinos and the $Z$~boson, given in \refeq{LnuZ}, 
those between the 'left-handed' sneutrinos and the Higgs bosons, given in 
 \refeq{LsnuH}, and those 
between the 'left-handed' sneutrinos and the $Z$~bosons, given in
\refeq{LsnuZ}. 
In addition, in this MSSM-seesaw scenario, there are interactions 
driven by the neutrino Yukawa couplings (or equivalently $\mD$ since 
$\Ynu=(g\mD)/(\wz \MW \sin\be)$), and new interactions due to the
Majorana nature  
driven by $\mM$. These genuine Majorana terms are  those in the seventh and 
eight lines of \refeq{LsnuH} and are not present in the case of Dirac
fermions. 

%%%%%%%%%%%%%%%%%%%%%%%%%%%%%%%%%%%%%%%%%%%%%%%%%%%%%%%%%%%%%%%%%%%%%%%%%%%%%%%

\subsection{Parameters and limits}
\label{sec:limits}
  
Regarding the size of the new  parameters that have been introduced 
in this model, in addition to those of the MSSM, i.e.,
$\mM$, $\mD$, $\mR$, $\Anu$ and $\Bnu$, 
there are no significant constraints. In the literature it is 
often assumed that $\mM$ has a very large value, 
$\mM \sim \cO (10^{14-15}) \gev$, in order to
get small physical neutrino masses  $|\mnu| \sim$ 0.1 - 1 eV with 
large Yukawa couplings $ \Ynu \sim \cO(1)$. This is an interesting
possibility since it can lead to important phenomenological implications
due to the large size of the radiative corrections driven by these large
Yukawa couplings.  
In this paper we will explore, however, not only these extreme values but 
the full range for $\mM$ from the
electroweak scale $\sim 10^2 \gev$ up to $\sim 10^{15} \gev$. 

On the other hand, the new soft SUSY-breaking parameters introduced in the
sneutrino sector could be unrelated to those of the MSSM, or could be
related, for instance, in the case one imposes (by hand) some kind of
universality conditions. Whereas 
the non-singlet soft mass parameter $\mL$, being common to the charged
'left handed' slepton,  is constrained by the solution
to the hierarchy problem to lie below a few TeV, the singlet soft mass 
$\mR$ is not, because it is not connected to the electroweak symmetry
breaking at tree level. The other sneutrino soft mass parameters, $\Bnu$ and 
$\Anu$ are not connected either. However, they can generate a mass-splitting
between sneutrinos and antisneutrinos which in turn and via loop corrections 
can generate neutrino mass splittings~\cite{Dedes:2007ef} that are
experimentally constrained.  
Then, if $\msusy$ represents a generic low SUSY breaking scale, with 
$\msusy \lsim \cO(10^3) \gev$ 
one expects that $|\Anu|, |\Bnu| \lsim \msusy$~\cite{Farzan:2004cm}. According
to these constraints, we will explore in this work values of 
these soft parameters ranging from the electroweak scale up to a few TeV.
Besides, and 
due to the peculiarity of the behavior with  $\mR$ and 
$\Bnu$, as will be shown later, we will explore in addition the less 
conservative but interesting possibility where 
$\mR$ or $\Bnu$ are close to $\mM$.

For illustrative purposes and a clear understanding of our full one-loop
results,  three interesting limiting cases  
will also be considered in this work.
\begin{itemize}

\item[(1)] The seesaw limit:\\
This assumes 
a large separation between the two neutrino mass scales involved, the Majorana mass and 
the Dirac mass,  $\mM\gg\mD$. Notice that both masses are different from zero,
$\mM \neq 0$ and $\mD \neq 0$, in this seesaw limit and, as we have said
above, $\Ynu$ can be large.   
The predictions  are then given in power series 
of a dimensionless parameter defined as,
\begin{align}
\xi &\equiv \frac{\mD}{\mM} \ll 1~.
\end{align}  
The light and heavy neutrino  masses are given in this limit by:
\begin{align}
m_{\nu}&= -\mD \xi + \mathcal{O}(\mD \xi^3) \simeq -\frac{\mD^2}{\mM} ~,\\ \non
m_N    &=  \mM + \mathcal{O}(\mD \xi) \simeq \mM ~.
\end{align}
Furthermore, the mixing angle $\theta$ is small in this limit and, therefore, 
$\nu$ is made predominantly of $\nu_L$ and its c-conjugate, $(\nu_L)^c$, 
whereas $N$
is made predominantly of $\nu_R$ and its c-conjugate, $(\nu_R)^c$.

In the sneutrino sector several mass scales are involved. Consequently, 
one has to set as an extra input their relative size to $\mM$. 
The simplest assumption 
is to set the value of $\mM$ to be much larger than all the other mass scales 
involved, i.e., $\mM \gg \mD, \MZ, \mu, \mL, \mR, \Bnu, \Anu$.
In this limit\ the sneutrino masses are given by:    
\begin{eqnarray}
 m_{{\Snu_+},{\Snu_-}}^2
&=& m_{\tilde{L}}^2 + 
\edz \MZ^2 \CZb \mp 2 \mD (A_{\nu} -\mu \CTb-\Bnu)\xi ~, \non \\
 m_{{\tilde N_+},{\tilde N_-}}^2  &=& \mM^{2} \pm 2 \Bnu \mM + \mR^2 + 2 \mD^2 ~.
\end{eqnarray}  
The mixing angles $\theta_{\pm}$ are small in this limit and,
therefore, ${\Snu_+}$ and ${\Snu_-}$ are made predominantly of 
${\Snu_L}$ and its c-conjugate, ${\Snu_L}^*$, whereas ${\tilde N_+}$ and 
${\tilde N_-}$ are made predominantly of ${\Snu_R}$ and its c-conjugate,
 ${\Snu_R}^*$. 

\item[(2)] The Dirac limit:\\
In this limit one sets $\mM=0$ (and $\mD \neq 0$) and one
recovers the neutrinos  
as any other fermion of the MSSM, i.e., as Dirac fermions. In the basis
that we have used in~\refeq{nuigenstates} this is manifested by the fact that 
when
$\mM=0$, the two Majorana neutrinos $\nu$ and $N$ are degenerate with
$\mnu= -\mD$ and $m_N=+\mD$,  and they combine maximally, i.e.\ with 
$\theta=\pi/4$, to form a four component Dirac neutrino with mass $\mD$. 
On the other hand, the sneutrino sector in this Dirac limit simplifies as well.
When  $\mM=0$, the real scalar fields get degenerate in pairs, 
\begin{align}
\label{snumassdiraclimit} 
  m_{{\Snu_+}}^2 &=m_{{\Snu_-}}^2=  
\edz(\mL^2+\mR^2+2\mD^2+ \edz \MZ^2 \CZb ) \\
&\quad - \edz 
\sqrt{4\mD^2(\Anu-\mu \CTb)^2+(\mR^2-\mL^2
-\edz \MZ^2\CZb  )^2} ~,  \non \\
 m_{{\SNu_+}}^2&= m_{{\SNu_-}}^2=
\edz(\mL^2+\mR^2+2\mD^2+ \edz \MZ^2 \CZb ) \\
&\quad +\edz 
\sqrt{4\mD^2(\Anu-\mu \CTb)^2+(\mR^2-\mL^2
-\edz \MZ^2\CZb )^2} ~,  \non 
\end{align}
and they combine to form two complex scalar fields,
\begin{align}
\label{snustatesdiraclimit} 
{\Snu}_1 & = \frac{1}{\sqrt{2}}({\Snu_+}+i{\Snu_-})= 
\cos {\tilde \theta}\,\, {\Snu_L} - \sin {\tilde \theta}\,\, {\Snu_R}~, \\
{\Snu}_2 & = \frac{1}{\sqrt{2}}({\SNu_+}+i{\SNu_-})= 
\sin {\tilde \theta}\,\, {\Snu_L} + \cos {\tilde \theta}\,\, {\Snu_R} 
 \end{align}
 with $m_{{\Snu}_1}=m_{{\Snu_{\pm}}}$, 
 $m_{{\Snu}_2}=m_{{\SNu_{\pm}}}$, ${\tilde \theta}=\theta_+=\theta_-$, and
\begin{equation}
\label{sneuanglediraclimit}
\sin 2{\tilde \theta}= \frac{2\mD(\Anu-\mu \CTb )}
{\sqrt{4\mD^2(\Anu-\mu \CTb )^2+
(\mR^2-\mL^2- \edz \MZ^2\CZb)^2}}~.
\end{equation} 
Notice that these two sneutrino states, ${\Snu}_{1,2}$, are equivalent to 
the usual sfermion mass eigenstates within the MSSM.  
 
In this Dirac limit it is interesting to study
the similarities in the analytical behavior of the neutrino/sneutrino
radiative corrections and the other MSSM  fermion/sfermion radiative
corrections.  In particular we are interested in the comparison with the
top/stop radiative corrections. As for the phenomenological
implications, this limit is not expected to lead to relevant numerical
results, since to get compatibility with the experimentally  tested
small neutrino masses, $|\mnu| \sim 0.1-1$ eV one needs Yukawa
couplings extremely small, $\Ynu \sim 10^{-12}-10^{-13}$.   
 
\item[(3)] The MSSM limit:\\
This limit is reached when one sets $\mD=0$ (the value of $\mM$ is not
relevant since once the Yukawa couplings are set to zero the predictions are 
absolutely independent of this mass scale)
 and one is left with a
neutrino/sneutrino  sector with just pure gauge couplings. Concretely,
there are just interactions of 
the left-handed neutrinos and the 'left-handed' sneutrinos to the $Z$~boson, 
exactly  as in the MSSM. We are interested in this limit, because we want to
compare the radiative corrections from the neutrino/sneutrino sector
within the MSSM-seesaw with those within the MSSM and to find the
interesting regions in the new parameters of the MSSM-seesaw where the
deviation from the MSSM result could be sizeable.
\end{itemize}

%%%%%%%%%%%%%%%%%%%%%%%%%%%%%%%%%%%%%%%%%%%%%%%%%%%%%%%%%%%%%%%%%%%%%%%%%%%%%%
%%%%%%%%%%%%%%%%%%%%%%%%%%%%%%%%%%%%%%%%%%%%%%%%%%%%%%%%%%%%%%%%%%%%%%%%%%%%%%%

\section{Higher-order corrections to \boldmath{$m_h$}}
\label{sec:mhiggs}

\subsection{The concept of higher order corrections in the \\
  Feynman-diagrammatic approach} 
\label{sec:FDconcept}

In the Feynman diagrammatic (FD) approach the higher-order corrected 
$\cp$-even Higgs boson masses in the MSSM, denoted here as $\Mh$ and $\MH$
 (the corresponding masses in the MSSM-seesaw model are denoted as
$\Mh^{\nu/\Snu}$ and $\MH^{\nu/\Snu}$),
are derived by finding the 
poles of the $(h,H)$-propagator 
matrix. The inverse of this matrix is given by
\BE
\left(\De_{\rm Higgs}\right)^{-1}
= - i \ML p^2 -  m_H^2 + \hSi_{HH}(p^2) &  \hSi_{hH}(p^2) \\
     \hSi_{hH}(p^2) & p^2 - m_h^2 + \hSi_{hh}(p^2) \MR~.
\label{higgsmassmatrixnondiag}
\end{equation}
Determining the poles of the matrix $\De_{\rm Higgs}$ in
\refeq{higgsmassmatrixnondiag} is equivalent to solving
the equation
\begin{equation}
\left[p^2 - m_{h}^2 + \hSi_{hh}(p^2) \right]
\left[p^2 - m_{H}^2 + \hSi_{HH}(p^2) \right] -
\left[\hSi_{hH}(p^2)\right]^2 = 0\,.
\label{eq:proppole}
\end{equation}
In perturbation theory, a (renormalized) self-energy is expanded 
as follows
\begin{align}
\hSi(p^2) &= \hSi^{(1)}(p^2) + \hSi^{(2)}(p^2) + \ldots~, \non \\
\Si(p^2)  &= \Si^{(1)}(p^2)  + \Si^{(2)}(p^2)  + \ldots~,
\end{align}
in terms of the $i$th-order contributions
$\hSi^{(i)}, \Si^{(i)}$.
In the following sections we concentrate on the one-loop corrections and
drop the order index, i.e.\ $\hSi \equiv \hSi^{(1)}$ in the following.

%%%%%%%%%%%%%%%%%%%%%%%%%%%%%%%%%%%%%%%%%%%%%%%%%%%%%%%%%%%%%%%%%%%%%%%%%%%%%%%

\subsection{One-loop renormalization}
\label{sec:renrMSSM}

In order to calculate one-loop corrections to the Higgs boson
masses, the renormalized Higgs boson
self-energies are needed. Here we follow the procedure used in
\cite{mhiggsf1l,mhcMSSMlong} (and references therein) and review it for
completeness. The parameters appearing in the Higgs
potential, (\ref{higgspot}), are renormalized as follows:
\begin{align}
\label{rMSSM:PhysParamRenorm}
  \MZ^2 &\to \MZ^2 + \dMZsq,  & \tadh &\to \tadh +
  \dtadh, \\ 
  \MW^2 &\to \MW^2 + \dMWsq,  & \tadH &\to \tadH +
  \dtadH, \notag \\ 
  M_{\rm Higgs}^2 &\to M_{\rm Higgs}^2 + \de M_{\rm Higgs}^2, & 
  \tanb &\to \tanb (1+\dtanb). \notag 
\end{align}
$M_{\rm Higgs}^2$ denotes the tree-level Higgs boson mass matrix given
in \refeq{higgsmassmatrixtree}. $\tadh$ and $\tadH$ are the tree-level
tadpoles, i.e.\ the terms linear in $h$ and $H$ in the Higgs potential.

The field renormalization matrices of both Higgs multiplets
can be set up symmetrically, 
\begin{align}
\label{rMSSM:higgsfeldren}
  \begin{pmatrix} h \\[.5em] H \end{pmatrix} \to
  \begin{pmatrix}
    1+\tfrac{1}{2} \dZ{hh} & \tfrac{1}{2} \dZ{hH} \\[.5em]
    \tfrac{1}{2} \dZ{hH} & 1+\tfrac{1}{2} \dZ{HH} 
  \end{pmatrix} \cdot
  \begin{pmatrix} h \\[.5em] H \end{pmatrix}~.
\end{align}

\noindent
For the mass counter term matrices we use the definitions
\begin{align}
  \de M_{\rm Higgs}^2 =
  \begin{pmatrix}
    \dmhsq  & \dmhHsq \\[.5em]
    \dmhHsq & \dmHsq  
  \end{pmatrix}~.
\end{align}
The renormalized self-energies, $\hSi(p^2)$, can now be expressed
through the unrenormalized self-energies, $\Si(p^2)$, the field
renormalization constants and the mass counter terms.
This reads for the $\cp$-even part,
\begin{subequations}
\label{rMSSM:renses_higgssector}
\begin{align}
\ser{hh}(p^2)  &= \se{hh}(p^2) + \dZ{hh} (p^2-m_h^2) - \dmhsq, \\
\ser{hH}(p^2)  &= \se{hH}(p^2) + \dZ{hH}
(p^2-\tfrac{1}{2}(m_h^2+m_H^2)) - \dmhHsq, \\ 
\ser{HH}(p^2)  &= \se{HH}(p^2) + \dZ{HH} (p^2-m_H^2) - \dmHsq~.
\end{align}
\end{subequations}

Inserting the renormalization transformation into the Higgs mass terms
leads to expressions for their counter terms which consequently depend
on the other counter terms introduced in~\refeq{rMSSM:PhysParamRenorm}. 

For the $\cp$-even part of the Higgs sectors, these counter terms are:
\begin{subequations}
\label{rMSSM:HiggsMassenCTs}
\begin{align}
\dmhsq &= \de\MA^2 \cos^2(\al-\be) + \de \MZ^2 \sin^2(\al+\be) \\
&\quad + \tfrac{e}{2 \MZ \sw \cw} (\dtadH \cos(\al-\be)
\sin^2(\al-\be) + \dtadh \sin(\al-\be)
(1+\cos^2(\al-\be))) \notag \\ 
&\quad + \dtanb \sinb \cosb (\MA^2 \sin 2 (\al-\be) + \MZ^2 \sin
2 (\al+\be)), \notag \\ 
\dmhHsq &= \tfrac{1}{2} (\de\MA^2 \sin 2(\al-\be) - \dMZsq \sin
2(\al+\be)) \\ 
&\quad + \tfrac{e}{2 \MZ \sw \cw} (\dtadH \sin^3(\al-\be) -
\dtadh \cos^3(\al-\be)) \notag \\ 
&\quad - \dtanb \sinb \cosb (\MA^2 \cos 2 (\al-\be) + \MZ^2 \cos
2 (\al+\be)), \notag \\ 
\dmHsq &= \de\MA^2 \sin^2(\al-\be) + \dMZsq \cos^2(\al+\be) \\
&\quad - \tfrac{e}{2 \MZ \sw \cw} (\dtadH \cos(\al-\be)
(1+\sin^2(\al-\be)) + \dtadh \sin(\al-\be)
\cos^2(\al-\be)) \notag \\ 
&\quad - \dtanb \sinb \cosb (\MA^2 \sin 2 (\al-\be) + \MZ^2 \sin
2 (\al+\be))~. \notag 
\end{align}
\end{subequations}

\bigskip
For the field renormalization we choose to give each Higgs doublet one
renormalization constant,
\begin{align}
\label{rMSSM:HiggsDublettFeldren}
  \cHe \to (1 + \tfrac{1}{2} \dZ{\cHe}) \cHe, \quad
  \cHz \to (1 + \tfrac{1}{2} \dZ{\cHz}) \cHz~.
\end{align}
This leads to the following expressions for the various field
renormalization constants in \refeq{rMSSM:higgsfeldren}:
\begin{subequations}
\label{rMSSM:FeldrenI_H1H2}
\begin{align}
  \dZ{hh} &= \sinasq \dZ{\cHe} + \cosasq \dZ{\cHz}, \\[.2em]
  \dZ{hH} &= \sina \cosa (\dZ{\cHz} - \dZ{\cHe}), \\[.2em]
  \dZ{HH} &= \cosasq \dZ{\cHe} + \sinasq \dZ{\cHz}~.
\end{align}
\end{subequations}
The counter term for $\tb$ can be expressed in terms of the vacuum
expectation values as
\begin{equation}
\de\tb = \edz \KL \dZ{\cHz} - \dZ{\cHe} \KR +
\frac{\de v_2}{v_2} - \frac{\de v_1}{v_1}~,
\end{equation}
where the $\de v_i$ are the renormalization constants of the $v_i$:
\begin{equation}
v_1 \to \KL 1 + \dZ{\cHe} \KR \KL v_1 + \de v_1 \KR, \quad
v_2 \to \KL 1 + \dZ{\cHz} \KR \KL v_2 + \de v_2 \KR~.
\end{equation}
It can be shown that the divergent parts of $\de v_1/v_1$ and 
$\de v_2/v_2$ are equal~\cite{mhiggsf1l}. Consequently, one can set 
$\de v_2/v_2 - \de v_1/v_1$ to zero.

The renormalization conditions are fixed by an appropriate
renormalization scheme. For the mass counter terms on-shell conditions
are used, leading to:
\begin{align}
\label{rMSSM:mass_osdefinition}
  \dMZsq = \re \se{ZZ}(\MZ^2), \quad \dMWsq = \re \se{WW}(\MW^2),
  \quad \de\MA^2 = \re \se{AA}(\MA^2). 
\end{align}
Here $\Si_{ZZ,WW}$ denotes the transverse part of the self-energies. 
Since the tadpole coefficients are chosen to vanish in all orders,
their counter terms follow from $T_{\{h,H\}} + \de T_{\{h,H\}} = 0$: 
\begin{align}
  \dtadh = -{\tadh}, \quad \dtadH = -{\tadH}~. 
\end{align}
For the remaining renormalization constants for $\de\tb$, $\dZ{\cHe}$
and $\dZ{\cHz}$ various renormalization schemes are
possible~\cite{tbren1,tbren2,mhcMSSMlong}.

\subsubsection*{On-shell renormalization}

One possible choice is an on-shell (OS) renormalization. The
renormalization conditions for the renormalized Higgs-boson
self-energies are
\begin{align}
\hSip_{hh}(m_h^2) &= 0~, \\
\hSip_{HH}(m_H^2) &= 0~.
\end{align}
This yields
\begin{align}
\dZ{hh}^{\os} &= - \re \Sip_{hh}(m_h^2)~,\\
\dZ{HH}^{\os} &= - \re \Sip_{HH}(m_H^2)~,
\end{align}
equivalently to
\begin{align}
\dZ{\cHe}^{\os} &= \ed{\CZa} \KL  \sinasq \re \Sip_{hh}(m_h^2)
                               - \cosasq \re \Sip_{HH}(m_H^2) \KR~, \\
\dZ{\cHz}^{\os} &= \ed{\CZa} \KL -\cosasq \re \Sip_{hh}(m_h^2)
                               + \sinasq \re \Sip_{HH}(m_H^2) \KR~.
\end{align}
For $\de\tb^{\os}$ a convenient choice is
\begin{align}
\de\tb^{\os} &= \edz \KL \dZ{\cHz}^{\os} - \dZ{\cHe}^{\os} \KR \non \\
            &= \frac{-1}{2\CZa} \KL \re \Sip_{hh}(m_h^2) 
                           - \re \Sip_{HH}(m_H^2) \KR~.
\end{align}
It should be kept in mind that this scheme can lead to 
large corrections  
to $m_h$ in the MSSM~\cite{tbren1,markusPhD}, hence worsening the 
convergence of the perturbative expansion.  Furthermore, 
it is known to provide gauge dependent corrections at the one-loop 
level~\cite{tbren2}.

\subsubsection*{\boldmath{\DRbar} renormalization}
A convenient choice which avoids the previously commented large corrections to
$m_h$ in the MSSM and is (linear) gauge independent at the one-loop level
is a \drbar\ renormalization of $\de\tb$, $\dZ{\cHe}$
and $\dZ{\cHz}$, 
\begin{subequations}
\label{rMSSM:deltaZHiggsTB}
\begin{align}
  \dZ{\cHe}^{\DRbar} &= 
       - \KKL \re \Sip_{HH \; |\al = 0} \KKR^{\rm div}, \\[.5em]
  \dZ{\cHz}^{\DRbar} &= 
       - \KKL \re \Sip_{hh \; |\al = 0} \KKR^{\rm div}, \\[.5em]
  \dtanb^{\DRbar} &= \edz \KL \dZ{\cHz}^{\DRbar} - \dZ{\cHe}^{\DRbar} \KR~.
\end{align}
\end{subequations}
The $\left[ \; \right]^{\rm div}$ terms are the ones proportional to 
$\De = 2/\eps - \ga_{\rm E} + \log(4 \pi)$, when using dimensional
regularization/reduction in $d = 4 - \eps$ dimensions; $\ga_{\rm E}$ is
the Euler constant.
The corresponding renormalization scale, $\mudim$, has to be fixed to
a certain mass scale that will be discussed below.

\subsubsection*{Modified \boldmath{\DRbar} renormalization 
                \boldmath{(m\DRbar)}}

The $\mudim$ dependence introduced in the \DRbar\ scheme can lead in the
present context  to large logarithmic corrections
$\propto \log(\mM^2/\mudim^2)$ for large values of the Majorana mass $\mM$
(as will be discussed below).
These large corrections could again worsen the convergence of the
perturbative expansion. 
One possible way out is to replace $\left[ \; \right]^{\rm div}$
by $\left[ \; \right]^{\rm mdiv}$, where the latter means 
to select not only the terms 
$\propto \De$ as in \refeqs{rMSSM:deltaZHiggsTB}, but the terms 
$\propto \De_m \equiv \De - \log(\mM^2/\mudim^2)$. This prescription for the
counterterms defines the modified ${\DRbar}$ renormalization scheme, which will
be named in this work in short as m${\DRbar}$, 
\begin{subequations}
\label{rMSSM:deltaZHiggsTBmDR}
\begin{align}
  \dZ{\cHe}^{m\DRbar} &= 
       - \KKL \re \Sip_{HH \; |\al = 0} \KKR^{\rm mdiv}, \\[.5em]
  \dZ{\cHz}^{m\DRbar} &= 
       - \KKL \re \Sip_{hh \; |\al = 0} \KKR^{\rm mdiv}, \\[.5em]
  \dtanb^{m\DRbar} &= \edz \KL \dZ{\cHz}^{m\DRbar} - \dZ{\cHe}^{m\DRbar} \KR~.
\end{align}
\end{subequations}
As will be shown below, effectively this corresponds to the particular
choice of $\mudim = \mM$. In this way the potentially large
logarithms vanish, what makes it a convenient choice. Usually this
choice is referred to in
the literature as 'decoupling the large mass scale by hand' 
(see e.g.\ \cite{decoup1,decoup2} and references therein). 

It should be kept in mind that in the \mDRbar\ scheme the parameter
$\tb = \tb^{\mDRbar}$ has a different meaning than the ``conventional''
parameter $\tb = \tb^{\DRbar}$. However, we have checked that this shift
is numerically insignificant.

%%%%%%%%%%%%%%%%%%%%%%%%%%%%%%%%%%%%%%%%%%%%%%%%%%%%%%%%%%%%%%%%%%%%%%%%%%%%%%
%%%%%%%%%%%%%%%%%%%%%%%%%%%%%%%%%%%%%%%%%%%%%%%%%%%%%%%%%%%%%%%%%%%%%%%%%%%%%%

\section{Results}
\label{sec:results}

In this section we first  
present the results of the one-loop corrections from
neutrino/sneutrino contributions to the neutral Higgs boson
renormalized self-energies within the MSSM-seesaw and then we discuss the 
derived results for the Higgs mass corrections.  

%%%%%%%%%%%%%%%%%%%%%%%%%%%%%%%%%%%%%%%%%%%%%%%%%%%%%%%%%%%%%%%%%%%%%%%%%%%%%%

\subsection{One-loop calculation of the renormalized self-energies} 

The full one-loop neutrino/sneutrino corrections to the self-energies,
$\hSi_{hh}^{\nu/\Snu}$, $\hSi_{HH}^{\nu/\Snu}$ and $\hSi_{hH}^{\nu/\Snu}$,
 entering
\refeq{eq:proppole} have been evaluated with the help of 
\fa~\cite{feynarts}\footnote{The program and the user's guide are available via {\tt{www.feynarts.de}}.} and  \fc~\cite{formcalc}. For shortness, 
in this and the next subsection these self-energies will be named simply 
as $\hSi_{hh}$, $\hSi_{HH}$, and $\hSi_{hH}$, respectively.
The new Feynman rules for the neutrino/sneutrino sector, derived in this work 
and collected in the Appendix A, have been inserted into a new model 
file\footnote{This model file is available upon request.}. 
 As regularization scheme we have used
dimensional reduction~\cite{dred},
thus preserving SUSY~\cite{dredDS,dredDS2}.

\begin{figure}[h!]
\begin{center}
\begin{tabular}{c} \hspace*{-12mm}
 \psfig{file=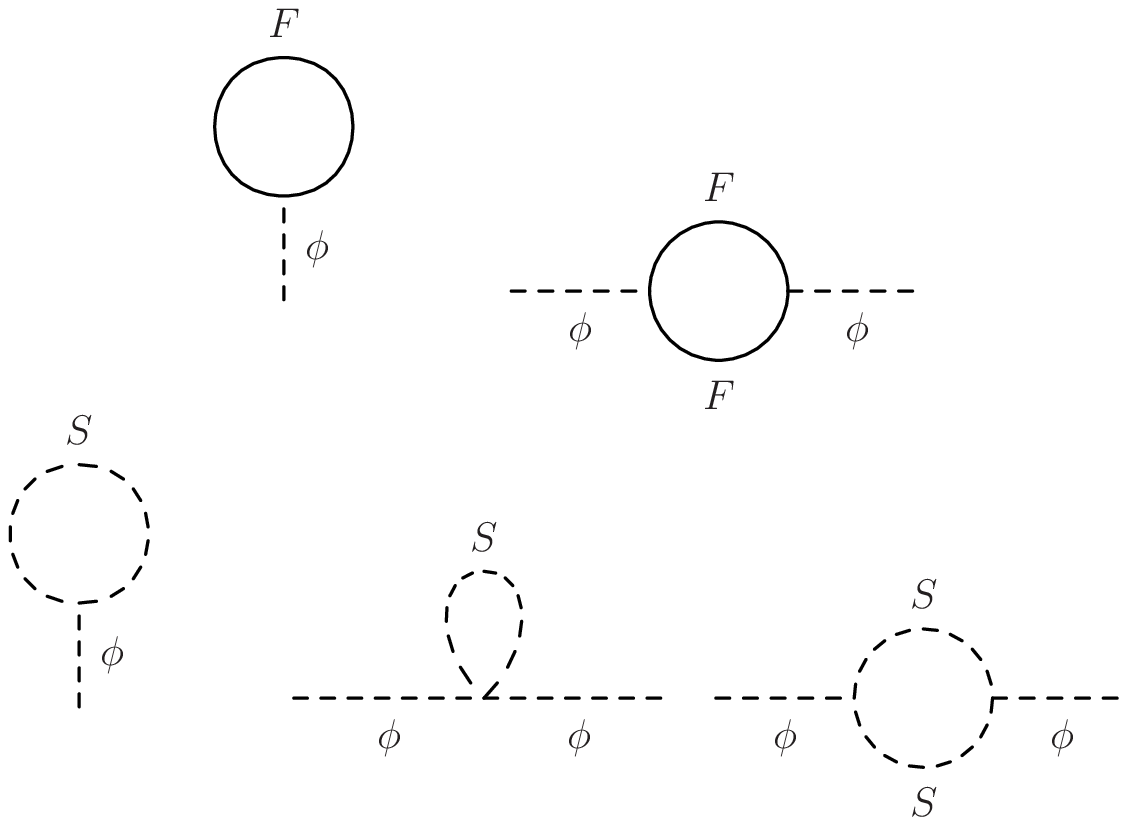,width=100mm,angle=0,clip=} \\
 \psfig{file=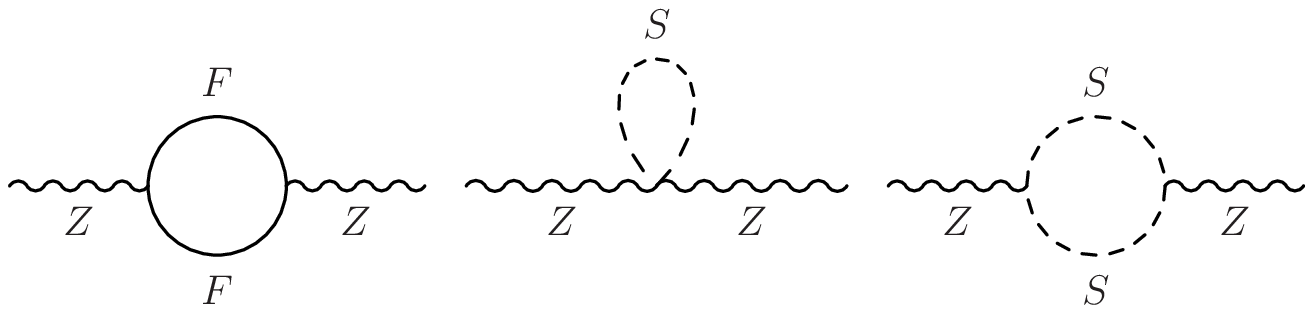,width=100mm,angle=0,clip=}
 \end{tabular}
\end{center}
\caption{Generic one-loop Feynman-diagrams contributing to the neutral Higgs bosons 
renormalized self-energies (see text)}
\label{fig:loops}
\end{figure}

The generic one-loop Feynman-diagrams contributing to the 
renormalized self-energies are depicted in \reffi{fig:loops}. They include the
two-point and one-point diagrams in the Higgs self-energies, tadpole
diagrams,  and 
the two-point and one-point diagrams in the $Z$~boson self-energy. Here the
notation is: $\phi$
refers generically to all neutral Higgs bosons, $h,H,A$;  
$F$ refers to all neutrinos $n_i$ $(i=1,2)$; $S$ refers to all
sneutrinos ${\tilde n}_i$ $(i=1,..4)$, and $Z$ refers to the $Z$~boson. 

%%%%%%%%%%%%%%%%%%%%%%%%%%% SIANNAH F I G U R E %%%%%%%%%%%%%%%%%%%%%%%%%%%%%%%%%%%%%%%
%%%%%%%%%%%%%%%%%%%%%%%%%%%%%%%%%%%%%%%%%%%%%%%%%%%%%%%%%%%%%%%%%%%%%%%%%%%%%%%% 

The analytical results for the unrenormalized self-energies and tadpoles are
collected in the Appendix B. The final analytical results for the renormalized  
self-energies are easily obtained by inserting these results into 
\refeq{rMSSM:renses_higgssector}.

We have checked that all the divergences involved in the computation 
cancel and the renormalized self-energies, $\ser{hh}(p^2)$, 
$\ser{HH}(p^2)$ and $\ser{hH}(p^2)$ in the three schemes OS, $\DRbar$, and
 m$\DRbar$ are all finite, as expected. 
We have also checked that the renormalized self-energies in the OS scheme,
are independent of the regularization scale $\mudim$, as they must be. The renormalized self-energies 
in the $\DRbar$ are  $\mudim$ dependent whereas the ones in the m$\DRbar$
scheme are $\mudim$ independent by construction. Analytically they are
related by 
$\ser{}^{{\rm m}\DRbar}(p^2) = \ser{}^{\DRbar}(p^2)|_{\mudim\,=\,\mM}$.

%%%%%%%%%%%%%%%%%%%%%%%%%%%%%%%%%%%%%%%%%%%%%%%%%%%%%%%%%%%%%%%%%%%%%%%%%%%

\subsection{Analysis of the renormalized self-energies}
\label{sec:analysis}

In the following we discuss the numerical results for the renormalized
self-energies. They are collected in Figs.~\ref{fig:renSEversusmM} 
through~\ref{fig:mDRversusp}. 
First we compare the predictions of the one-loop renormalized
self-energies in the three schemes for the full interval 
$10^3\,\,{\rm GeV}\lsim \mM \lsim 10^{15}\,\,{\rm GeV}$, and next we analyze these
exact results at large $\mM$ with the help of the simple analytical formulas 
that are obtained 
in the seesaw
limit. Then we choose the 
m$\DRbar$ scheme and show the exact numerical results of the renormalized 
self-energies as functions of all the neutrino/sneutrino parameters 
involved. Finally we conclude on the subset of most relevant parameters 
(specifically, $m_M$, $m_{\tilde R}$, $B_\nu$ and $m_\nu$) which will be
the selected ones 
to study the corrections to $\Mh$ in the next subsection.
For the final estimate of these corrections, and to
localize the regions of the
parameter space where
they can reach sizeable values, we will vary these relevant parameters within 
some selected plausible intervals. 
For the parameters which do not exhibit a relevant   
numerical effect on $\Mh$ 
(specifically, $\tb$, $M_A$, $\mu$, $m_{\tilde L}$ and $A_\nu$) we choose 
representative values. 
For
completeness, we will also comment shortly at the end of this subsection on the 
Dirac case.   
    
In order to compare systematically our predictions of the neutrino/sneutrino
sector in the MSSM-seesaw 
with those in the MSSM, we have split the full one-loop neutrino/sneutrino
result into two parts:
\begin{equation}
\label{split}
\hSi(p^2)|_{\rm full}=\hSi(p^2)|_{\rm gauge}+\hSi(p^2)|_{\rm Yukawa}~,
\end{equation}  
where $\hSi(p^2)|_{\rm gauge}$ means the contributions from pure gauge 
interactions and they are obtained by switching off the Yukawa interactions,
i.e.\ by setting $\Ynu= 0$ 
(or equivalently $\mD=0$). The remaining part is named here 
$\hSi(p^2)|_{\rm Yukawa}$ and refers to the contributions that are only 
present if $\Ynu \neq 0$. In other words, this separation splits the full
result into the common part with the MSSM, given by $\hSi(p^2)|_{\rm gauge}$, 
and the new contributions due to the presence of Majorana neutrinos with 
non vanishing Yukawa interactions, given by 
$\hSi(p^2)|_{\rm Yukawa}$. Thus, by comparing the size of these two 
parts, within the allowed parameter space region, we will localize the 
areas where $\hSi(p^2)|_{\rm Yukawa} \gg \hSi(p^2)|_{\rm gauge}$, which  
will therefore indicate a significant departure from the MSSM result.   

\subsubsection*{Dependence on \boldmath{$\mM$}}

\begin{figure}[ht!]
   \begin{center} 
     \begin{tabular}{cc} \hspace*{-12mm}
  	 \psfig{file=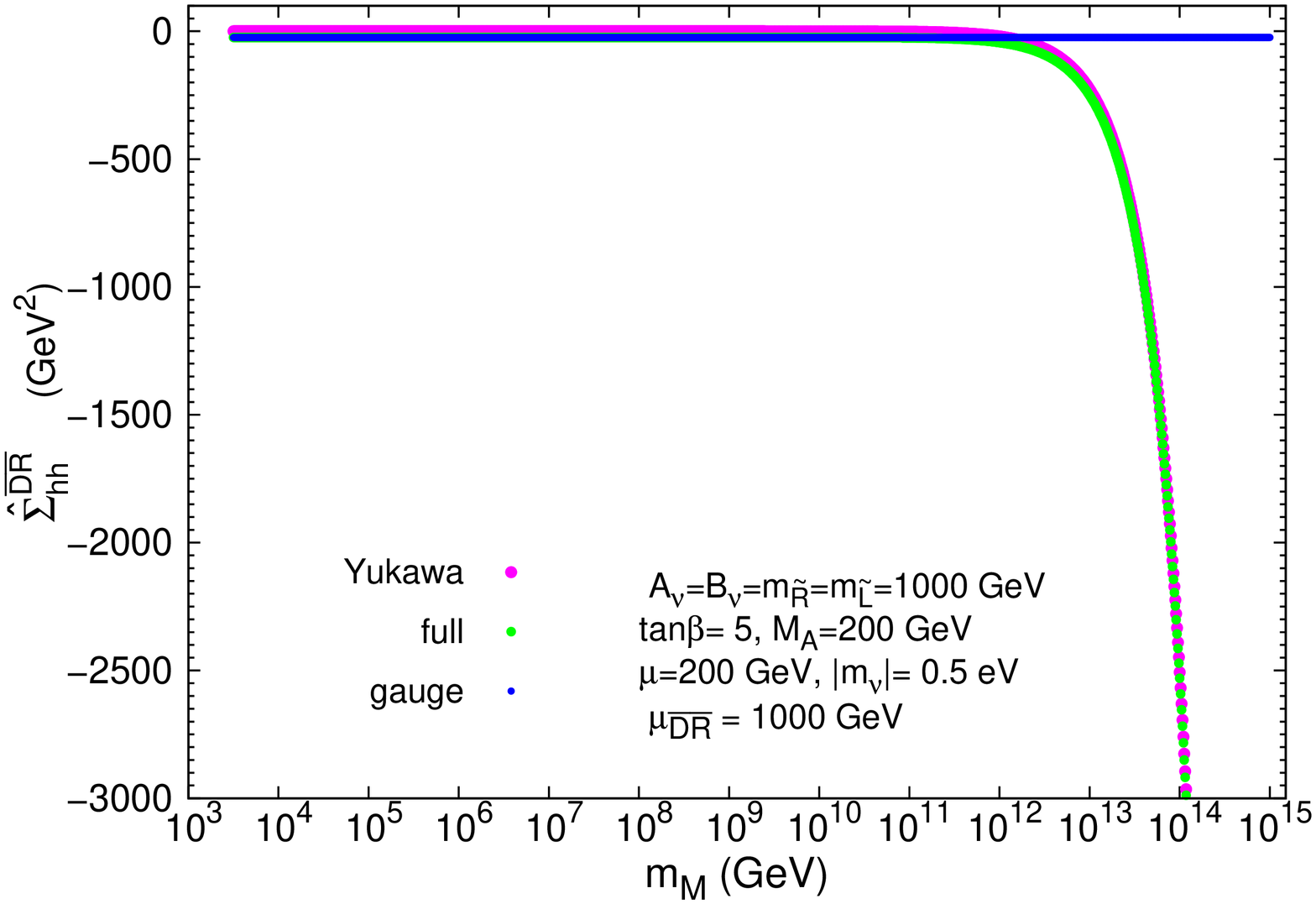,width=60mm,angle=270,clip=}
  &
  	 \psfig{file=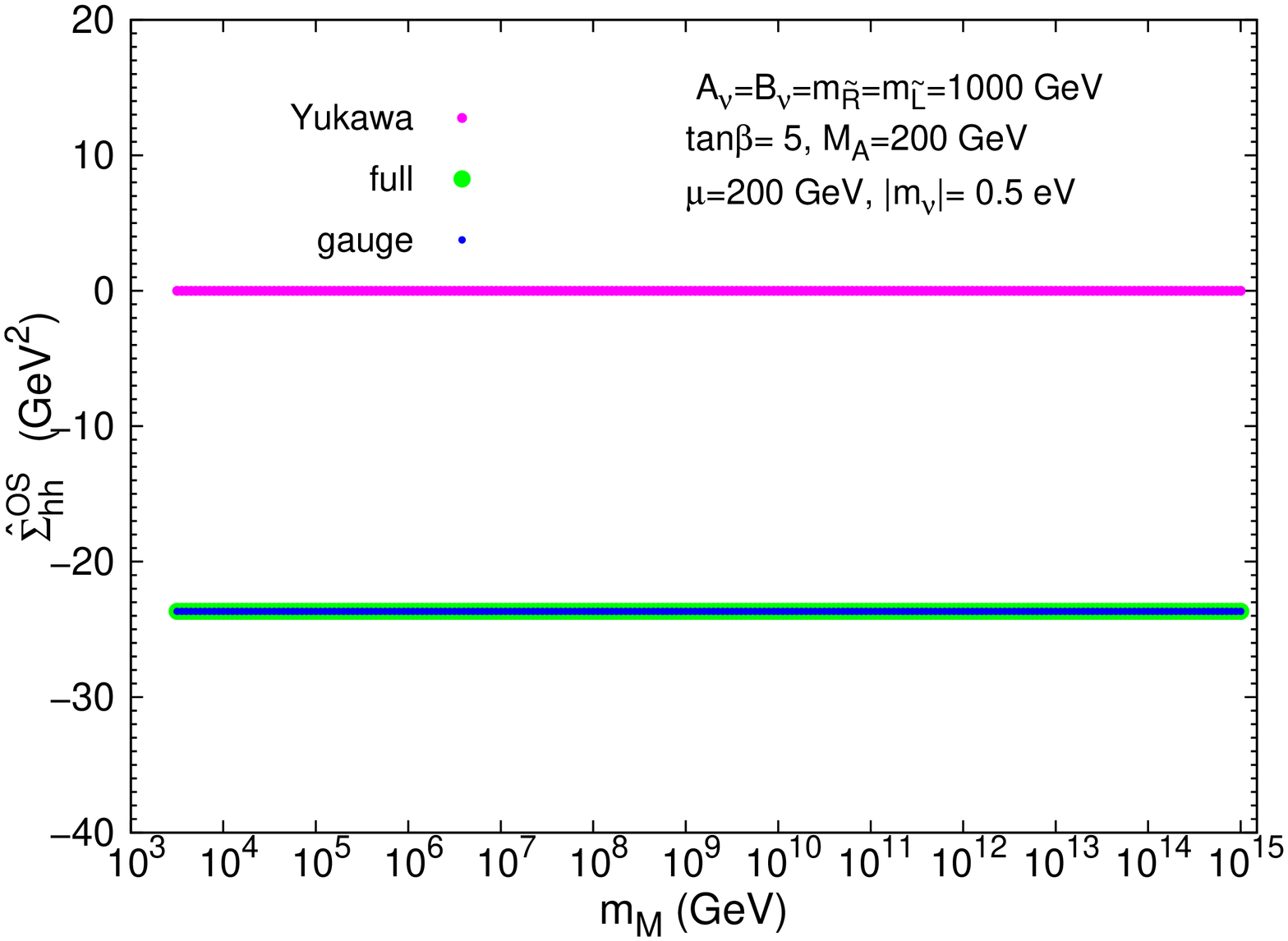,width=60mm,angle=270,clip=}
	 \\
	\hspace{-1.3cm}   \psfig{file=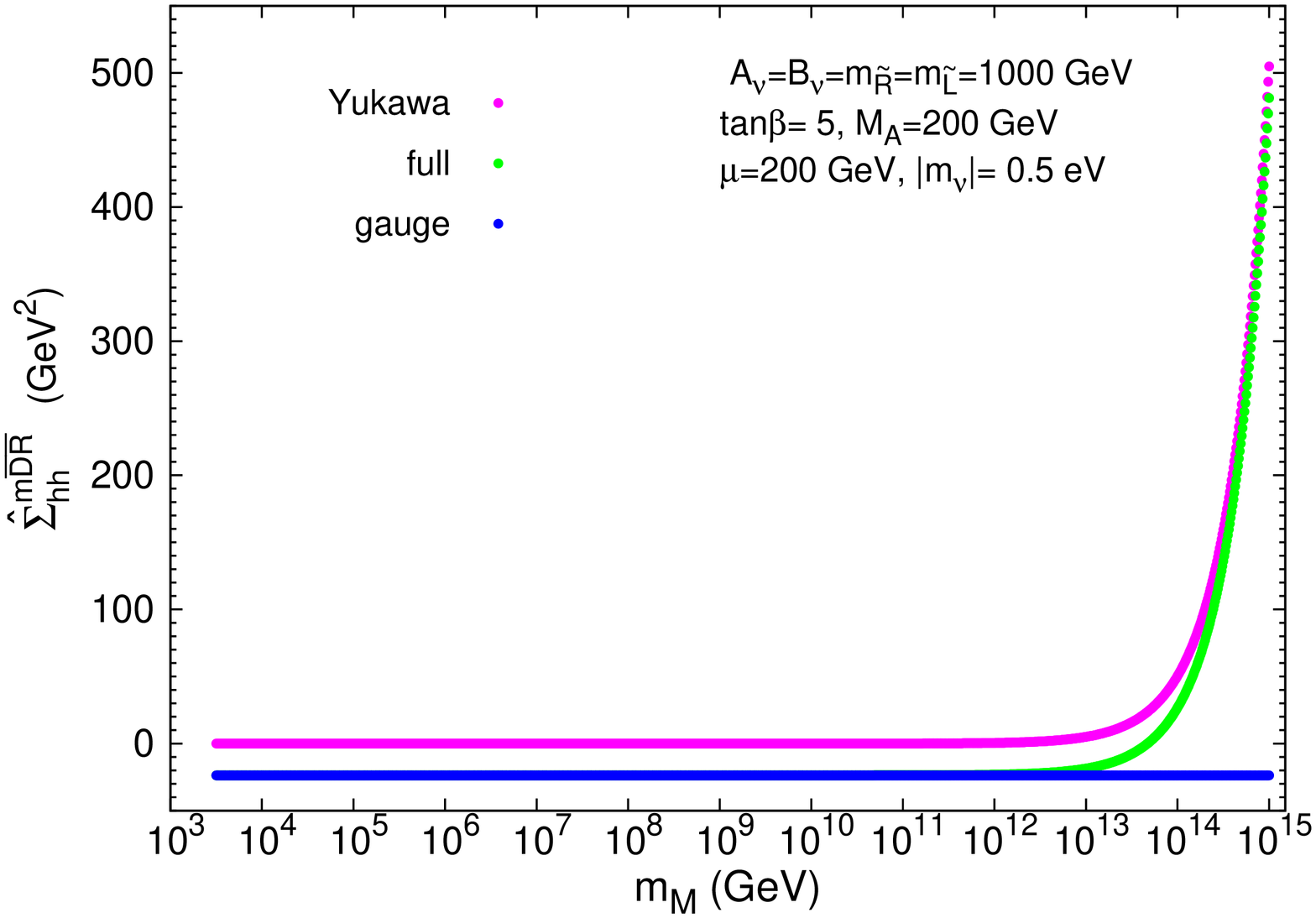,width=60mm,angle=270,clip=}
  &
  	 \psfig{file=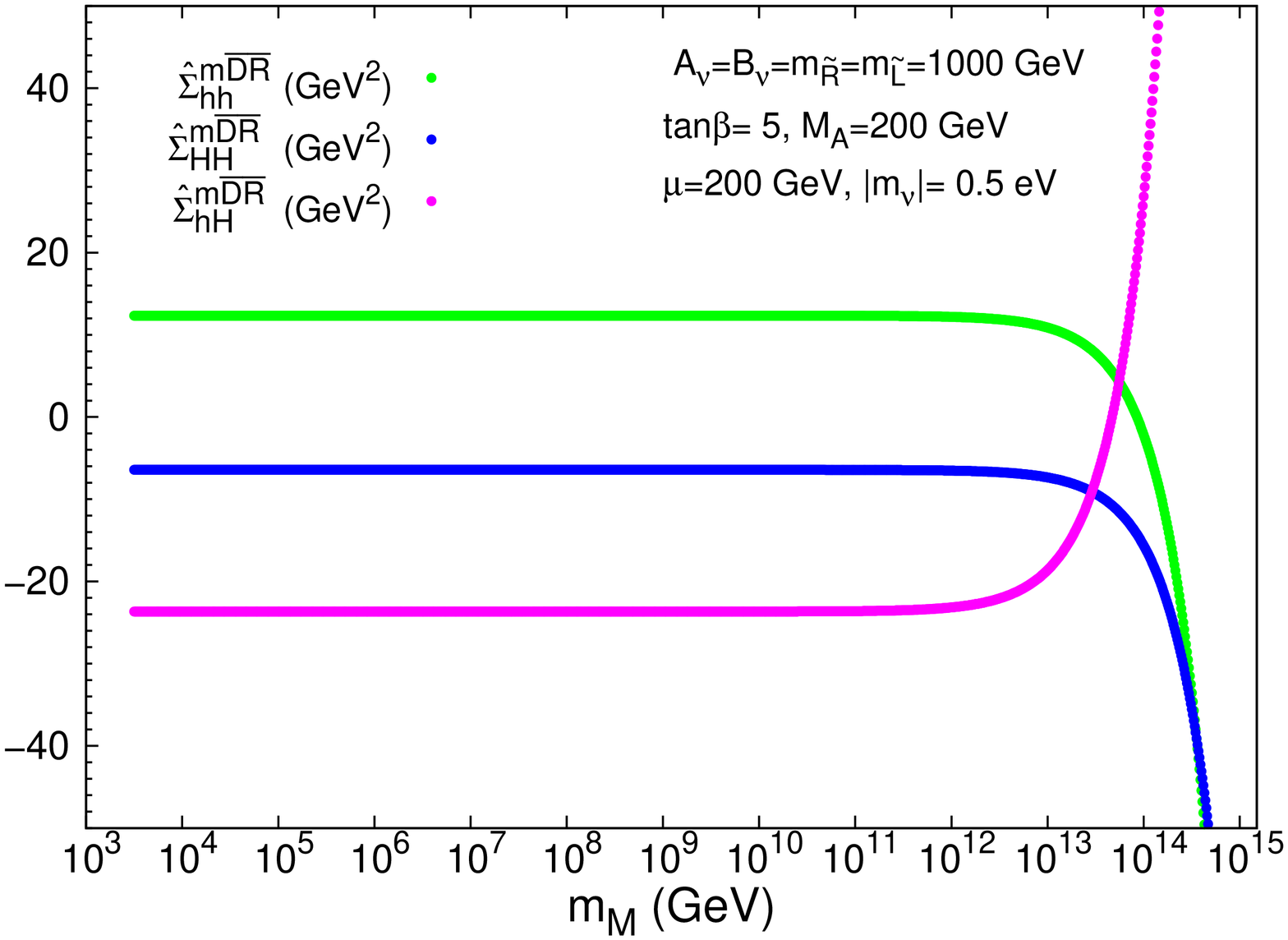,width=60mm,angle=270,clip=}
	
     \end{tabular}
\caption{Renormalized Higgs boson self-energies as a function of  $\mM$ and  
     comparison between the three considered schemes. Upper left panel:
     $\hSi_{hh}^{\DRbar}(p^2)$. Upper right panel: 
     $\hSi_{hh}^{\rm OS}(p^2)$. Lower left panel: $\hSi_{hh}^{{\rm
     m}\DRbar}(p^2)$. Lower right panel: 
     $\hSi_{hh}^{{\rm m}\DRbar}(p^2)$, $\hSi_{HH}^{{\rm m}\DRbar}(p^2)$ 
     and $\hSi_{hH}^{{\rm m}\DRbar}(p^2)$. All self-energies are evaluated at
     $p^2= (116 \gev)^2.$
      }
\label{fig:renSEversusmM}
   \end{center}
 \end{figure}

We show in \reffi{fig:renSEversusmM} the predictions for
$\hSi_{hh}(p^2)$ as a function of 
$\mM$ in the three schemes: $\DRbar$ (upper left plot), OS (upper right plot),
and m$\DRbar$ (lower left plot). 
In these plots we have considered an extremely wide 
range for the $\mM$ values, from $10^3 \gev$ up to $10^{15} \gev$, and fixed the
physical light neutrino mass to $|\mnu| = 0.5 \ev$. Consequently, $\mD$
is derived from $\mM$ and $\mnu$ by using \refeq{mDmN} and \refeq{mMmN}. The other
parameters are fixed as indicated in the figure. In this and in the following 
figures we have fixed $p^2$ in the self-energies to a particular value, 
corresponding to an approximation of the higher-order corrected value of $\Mh$ 
for the input 
MSSM parameters set in each figure, 
see below.
The numerical values used here and in the following for the SUSY
parameters are representative
values (as will also be shown below). Therefore, despite choosing only a
few values for the parameters, the results obtained can be considered as
more general.

%%%%%%%%%%%%%%%%%%%%%%%%%% F I G U R E %%%%%%%%%%%%%%%%%%%%%%%%%%%%%%%%%%%%%%%

%%%%%%%%%%%%%%%%%%%%%%%%%%%%%%%%%%%%%%%%%%%%%%%%%%%%%%%%%%%%%%%%%

In the three mentioned plots in \reffi{fig:renSEversusmM} one can see that the
numerical value of the full result is nearly
constant with $\mM$ in the three schemes from $\mM= 10^3 \gev$ up to 
$\mM \sim 10^{12} \gev$. Furthermore, 
 this constant value is approximately the same in the three schemes 
 (the differences are below $\sim 10^{-2}$ ${\rm GeV}^2$), and is totally dominated 
by the 'pure gauge contributions'. Thus, for 
$10^{3} \gev \lsim  \mM \lsim 10^{12}\gev$ the result in the MSSM-seesaw
nearly coincides with the result in the MSSM, irrespectively of the scheme. 
For the choice  of input parameters in this plot, we get
$\hSi_{hh}|_{\rm full} \simeq \hSi_{hh}|_{\rm gauge}  \simeq  -23.67 \gev^2$.  

For larger values 
of $\mM$ in the range $10^{12} \gev\,<\, \mM\, <\,10^{15}\gev$, 
there are, however, remarkable differences between the three considered 
schemes, and the main differences come clearly from the 'Yukawa
contributions'.  Whereas  $\hSi_{hh}^{\rm OS}|_{\rm full}$  
is apparently constant with $\mM$, also for 
$\mM> 10^{12} \gev$, $|\hSi_{hh}^{\DRbar}|_{\rm full}|$ and 
$|\hSi_{hh}^{m\DRbar}|_{\rm full}|$ 
grow noticeably with $\mM$ at these large $\mM$ values.  The numerical
value of $\hSi_{hh}^{\DRbar}|_{\rm full}$ is negative 
for $\mM> 10^{12} \gev$ and gets large values in this range, where they
are totally dominated by the 'Yukawa contributions'. 
For instance, for $\mM =10^{13} \gev$, we get
$\hSi_{hh}^{\DRbar}|_{\rm full}\simeq \hSi_{hh}^{\DRbar}|_{\rm Yukawa} \simeq 
-250 \gev^2$, and for $\mM =10^{14} \gev$, we get 
$\hSi_{hh}^{\DRbar}|_{\rm full} \simeq \hSi_{hh}^{\DRbar}|_{\rm Yukawa} \simeq 
-3000 \gev^2$. In the m$\DRbar$ scheme, the result is negative up to 
$5 \times 10^{13} \gev$ and then becomes positive and large for 
$\mM>5 \times 10^{13} \gev$. Notice that, the absolute value in the
m$\DRbar$ scheme at large $\mM$ is always  
smaller than in the $\DRbar$ scheme, due to the commented cancellation 
of the large logarithms $\log(\mM/\mudim)$ corresponding to the choice
$\mudim=\mM$. Notice also that, in spite of this cancellation, the 
size of the corrections in m$\DRbar$,
are still large for large enough $\mM$ values. For instance, for 
$\mM= 10^{15} \gev$, we get dominance of the 'Yukawa contributions'  
$\hSi_{hh}^{m\DRbar}|_{\rm full} \simeq \hSi_{hh}^{m\DRbar}|_{\rm Yukawa} 
\simeq 500 \gev^2$. In contrast, for $\mM= 10^{14} \gev$, the 'Yukawa
contributions' and the 'pure gauge contributions', compete since  
 $\hSi_{hh}^{m\DRbar}|_{\rm Yukawa} \simeq 60 \gev^2$ and 
 $\hSi_{hh}^{m\DRbar}|_{\rm gauge} \simeq -24 \gev^2$ leading to 
 $\hSi_{hh}^{m\DRbar}|_{\rm full} \simeq 36\gev^2$.  
 
In the lower right plot of \reffi{fig:renSEversusmM} we compare 
$\hSi_{hh}^{m\DRbar}|_{\rm full}$ to the other two renormalized
self-energies, $\hSi_{HH}^{m\DRbar}|_{\rm full}$ and 
$\hSi_{hH}^{m\DRbar}|_{\rm full}$.
One can observe that the three self-energies behave
qualitatively very similarly with $\mM$, being approximately constant for 
$\mM<10^{12} \gev$ and growing (in modulus) with $\mM$ for  
$10^{12} \gev\,<\, \mM\, <\,10^{15}\gev$. For the choice of
parameters in this plot, $|\hSi_{hh}^{m\DRbar}|_{\rm full}|$ is larger
than the others in the full explored $\mM$ range. This will be relevant
for the forthcoming estimate of the one-loop radiative corrections to~$\Mh$.
 
The previously commented growing behavior of the renormalized
self-energies with $\mM$ is a consequence of the corresponding growing
behavior of the neutrino Yukawa interactions with $\mM$, see
\refeq{mDmN} and \refeq{mMmN}. This is a well 
known feature of the seesaw models that,
in order to get the light neutrino masses $\mnu$ in agreement with data, one 
must impose for each input $\mM$ value the proper 
$\Ynu$ (and therefore $\mD$) to precisely match the experimentally inspired
input $\mnu$. $\Ynu$ is therefore not an input but an output in this
approach, and according to \refeq{mDmN} and \refeq{mMmN} $\Ynu$ grows with 
$\mM$ as $\Ynu \propto \sqrt{\mM}$. The
behavior of the 
renormalized self-energies with $\mM$ is, consequently, the result of the two
competing facts, the increase of $\Ynu$ with $\mM$ and the decreasing
with $\mM$ 
from the neutrino and sneutrino propagators in the loops.

%%%%%%%%%%%
\subsubsection*{Dependence on \boldmath{$\mM$} in the seesaw limit}

In order to illustrate more clearly the behavior with $\mM$, we have analyzed in 
more detail the renormalized self-energies in the seesaw limit, as defined in
section \ref{sec:nN}. As the increase with $\mM$ starts at very large 
$\mM>10^{12} \gev$ values (i.e.\ much larger than the other scales, 
$\mM \gg \mD, \MZ, \MA, \mu, \mL, \mR, \Bnu, \Anu$), 
one expects that this limit should approximate pretty well the full result 
and show its same main features.   

For the computation of the renormalized self-energies in this seesaw limit, we
have performed a systematic expansion of the exact result in
powers of the seesaw parameter $\xi=\mD/\mM$. In order to reduce the number of
parameters, and for a clearer interpretation of the results, we have set in 
this expansion, $\Anu=\mu=\Bnu=0$ (which is justified, see below) and
we have assumed universal soft SUSY breaking masses, i.e., 
$\mL= \mR=\msusy$. 

The analytical expressions 
for these
expanded renormalized self-energies are of the generic form:
\begin{equation}
\hSi(p^2)=\left(\hSi(p^2)\right)_{\mD^0}+\left(\hSi(p^2)\right)_{\mD^2}+
\left(\hSi(p^2)\right)_{\mD^4} + \ldots ~,
\label{seesawser}
\end{equation}
where, $\left(\hSi(p^2)\right)_{\mD^0}$ is the first term in the
expansion, i.e.\ 
$\cO(\xi^0)$, $\left(\hSi(p^2)\right)_{\mD^2}$ is the next term,
i.e.$\cO(\xi^2)$, $\left(\hSi(p^2)\right)_{\mD^4}$ is the term of
$\cO(\xi^4)$, etc. 
It should be noticed that there are no terms with odd powers of
$\xi$. The first term in this expansion is precisely the 
pure gauge contribution, 
$\KL \hSi(p^2) \KR_{\mD^0}= \hSi(p^2)|_{\rm gauge}$. Therefore, it
approximates 
the result in the MSSM and the rest approximates the Yukawa part, 
\begin{align}
\left(\hSi(p^2)\right)_{\rm MSSM} &\simeq 
\left(\hSi(p^2)\right)_{\mD^0}~,
\non \\
\left(\hSi(p^2)\right)_{\rm Yukawa} &\simeq 
\left(\hSi(p^2)\right)_{\mD^2}+\left(\hSi(p^2)\right)_{\mD^4} + \ldots~.
\end{align}
In order to get simple formulas, we have expanded in addition
each term in the series in \refeq{seesawser} in powers of the other
small dimensionless parameters, namely, $\MZ/\mM$, $\MA/\mM$, $p/\mM$ 
and $\msusy/\mM$.

The result of the previous seesaw expansion (we just show
the leading terms; terms suppressed by factors $1/\mM^2$ respect to these
leading ones are not relevant and, therefore, are not included)
for each of the three considered renormalization schemes is as follows.

\subsubsection*{\boldmath{\order{\mD^0}}}

\begin{subequations}
\label{mD0}
\begin{align}
\KL \hSi_{hh}^{\DRbar}(p^2) \KR_{\mD^0} &= 
\frac{g^2 \MZ^2 \sin^2(\al +\be)}
     {1152 \cw^2 \msusy^2 \pi^2}
 \Big[ -20 \msusy^2 + 3 p^2 
   + 12 \msusy^2 \log  \frac{\MZ^2}{\msusy^2} \Big] \\
\label{mD0b}
\KL \hSi_{hh}^{\mDRbar}(p^2) \KR_{\mD^0} &= 
\KL \hSi_{hh}^{\DRbar}(p^2) \KR_{\mD^0} \\
\KL \hSi_{hh}^{\rm OS}(p^2) \KR_{\mD^0}&=
\KL \hSi_{hh}^{\DRbar}(p^2) \KR_{\mD^0} 
+ \frac{g^2 \MZ^2}{3072 \cw^2 \msusy^2 \pi^2}
\Big[ 4\left(p^2- m_h^2\right)\left(\cos2\alpha\cos2\beta-1\right) \nonumber \\ 
&+\sec2\alpha\sin2\beta\left(M_A^2\left(\sin4\beta-\sin4\alpha\right)-M_Z^2\sin4(\alpha+\beta)\right)\Big] 
\end{align}  
\end{subequations}

\subsubsection*{\boldmath{\order{\mD^2}}}

\begin{subequations}
\label{mD2} 
\begin{align}
\KL \hSi_{hh}^{\DRbar}(p^2) \KR_{\mD^2} &=
  \frac{g^2 \mD^2}{64  \pi^2\MW^2 \sin^2\be }  
\KKL 1-\log  \frac{\mM^2}{\mu_{\DRbar}^2}  \KKR \KKL -2 M_A^2 \cos^2(\al-\be)\cos^2\be \right.\nonumber \\
&\left.+2 p^2 \cos^2\al - M_Z^2 \sin\be\sin(\al+\be)\left(2 \left(1+\cos^2\be\right)\cos\al-\sin2\beta\sin\al \right) \KKR
\label{mD2a}
 \\ 
\label{mD2b}
\KL \hSi_{hh}^{\mDRbar}(p^2) \KR_{\mD^2} &=
\KL \hSi_{hh}^{\DRbar}(p^2) \KR_{\mD^2\Big|\mu_{\DRbar} = \mM} \\
  \KL \hSi_{hh}^{\rm OS}(p^2) \KR_{\mD^2}&=
\frac{g\mD^2}{768 \pi^2 M_W^2 p^2 \mM^2 }\Big[12 \msusy^2\Big[ M_A^2 p^2 \left(2\cos^2(\al-\be)\cot^2\be-\cot\be\sin2(\al-\be)\right)\non \\
&-2m_h^2 p^2 \cos^2\al\csc^2\be-4 M_Z^2 p^2 \cos\al\csc\be\sin(\al+\be)+4 M_Z^4 \sin^2(\al+\be) \non \\
&+2 M_Z^2 p^2 \sin^2(\al+\be)  -M_Z^2 p^2 \cot\be\sin2(\al+\be)-4 M_Z^2 p^2 \sin^2(\al+\be)\log\frac{M_Z^2}{m_M^2}\non \\
&+4 M_Z^4\sin^2(\al+\be)\log\frac{p^2}{m_M^2}-\log\frac{\msusy^2}{m_M^2}\Big[2m_h^2 p^2 \cos^2\al\csc^2\be +4 M_Z^4 \sin^2(\al+\be)\non \\
&- M_Z^2 p^2 \left(2\sin^2(\al+\be)- \cot\be\sin2(\al+\be)+4 \cos\al\csc\be\sin(\al+\be)\right)\non \\
&+ M_A^2 p^2\left(\cot\be\sin2(\al-\be)-2\cot^2\be\cos^2(\al-\be)\right) \Big]\Big]\non \\
& +p^2 \Big[8 M_A^4 \cos^2(\al-\be)\cot^2\be+ 8\cos^2\al\left(3 M_Z^2\left(m_h^2-p^2\right)+p^2\csc^2\be\left(3m_h^2-p^2\right)\right)\non \\
&+24 M_Z^2 p^2\cos\al\csc\be\sin(\al+\be)+12 M_A^2 M_Z^2\cos^2\be\cos2\be\sec2\al\non \\
&+12 M_Z^4\sin^2(\al+\be)(-1+2\log\frac{M_Z^2}{p^2})+3\cot\be\Big[-2 M_A^2 M_Z^2\sin2\al\non \\
&+2\sec2\al\big[-M_A^2\sin2(\al-\be)\left(-M_A^2+2m_h^2-M_Z^2+M_A^2\cos2\al\right)\non\\
&+M_Z^2 \sin2(\al+\be)\left(M_A^2-2 m_h^2+M_Z^2-M_A^2\cos2\al-M_Z^2\cos2(\al+\be)\right)\big]\Big]\Big]\Big]
\end{align} 
\end{subequations}

\subsubsection*{\boldmath{$\cO(\mD^4)$}}

%%%%%%%%
\begin{subequations}
\label{mD4} 
\begin{align}
\KL \hSi_{hh}^{\DRbar}(p^2) \KR_{\mD^4}&= \frac{g^2 \mD^4 }{128 \pi^2 \MW^2 m_M^2 p^4}\Big[4 M_Z^2 p^2 \left(p^2-M_Z^2\right)\log\frac{\msusy^2}{m_M^2}\sin^2(\al+\be) \non \\
&+8 M_A^2 p^4 \cos^2(\alpha-\beta)\cot^2\beta \log\frac{M_A^2}{m_M^2}+4
\left(2\msusy^2-3 M_Z^2\right)p^4\sin^2(\al+\be)\log\frac{M_Z^2}{m_M^2}\non \\
&+ 8 p^4\csc^2\be\big[ M_A^2\cos^2\be\cos^2(\al-\be)-p^2 \cos^2\al\big]\non \\
&+8 M_Z^2\sin(\al+\be)p^4\big[2\cos\al\csc\be-\sin(\al+\be)\big] \non \\
&+ 4 \msusy^2\log\frac{\msusy^2}{m_M^2}\big[p^4\left(-1+\cos2(\al+\be)-4\cos^2\al\csc^2\be\right) \non \\
&+8 M_Z^2 p^2 \cos\al\csc\be\sin(\al+\be)-2 M_Z^4\sin^2(\al+\be)\big] \non \\
&- 4 \log\frac{p^2}{m_M^2}\big[
2 p^6 \cos^2\al\csc^2\be+4 M_Z^2 p^2\left(2 \msusy^2-p^2\right)\cos\al\csc\be\sin(\al+\be)\non \\
&- M_Z^4 \sin^2(\al+\be)\left(2 \msusy^2+ p^2\right)\big]- 8 \msusy^2\big[2 p^4\cos^2\al\csc^2\be \non \\
&+4 M_Z^2 p^2 \cos\al \csc\be\sin(\al+\be)+\sin^2(\al+\be)\left(M_Z^4-p^4\right)\big]\Big]
 \\ 
\KL \hSi_{hh}^{{\rm m}\DRbar}(p^2) \KR_{\mD^4} &= 
\KL \hSi_{hh}^{\DRbar}(p^2) \KR_{\mD^4} \\
%(1/(32 CW^2 mM^2 MZ^2 \[Pi]^2))g^2 mD^4 (Cot[ B] (2 + Log[MHH2/mM^2]) Sec[2 A] Sin[A]^2 (MA02 Sin[2 (A - B)] + MZ^2 Sin[2 (A + B)]) - 
 %  - Cos[A]^2 (2 + Log[Mh02/mM^2]) (2 (Mh02 - p2) Csc[B]^2 + Cot[B] Sec[2 A] (MA02 Sin[2 (A - B)] + MZ^2 Sin[2 (A + B)])))
\KL \hSi_{hh}^{\rm OS}(p^2) \KR_{\mD^4}&= \KL \hSi_{hh}^{\DRbar}(p^2) \KR_{\mD^4} +
\frac{g^2 \mD^4 }{32\pi^2 \MW^2  \mM^2}\Big[\cot\be\sec2\al\sin^2\al\big[M_A^2\sin2(\al-\be)\non \\
&+M_Z^2\sin2(\al+\be)\big]\Big[2+\log\frac{m_H^2}{m_M^2}\Big]-\cos^2\al\Big[2+\log\frac{m_h^2}{m_M^2}\Big]\big[2 (m_h^2-p^2)\csc^2\be \non \\
&+\cot\be\sec2\al\big[M_A^2\sin2(\al-\be)+M_Z^2\sin2(\al+\be)\big]\big]\Big]
\end{align}
\end{subequations}

%%%%%%

%%%%%%%%%%%%%%%%%%%%%%%%%% F I G U R E %%%%%%%%%%%%%%%%%%%%%%%%%%%%%%%%%%%%%%%
%%%%%%%%%%%%%%%%%%%%%%%%%% F I G U R E %%%%%%%%%%%%%%%%%%%%%%%%%%%%%%%%%%%%
 \begin{figure}[h!]
   \begin{center} 
     \begin{tabular}{cc} \hspace*{-12mm}
  	\psfig{file=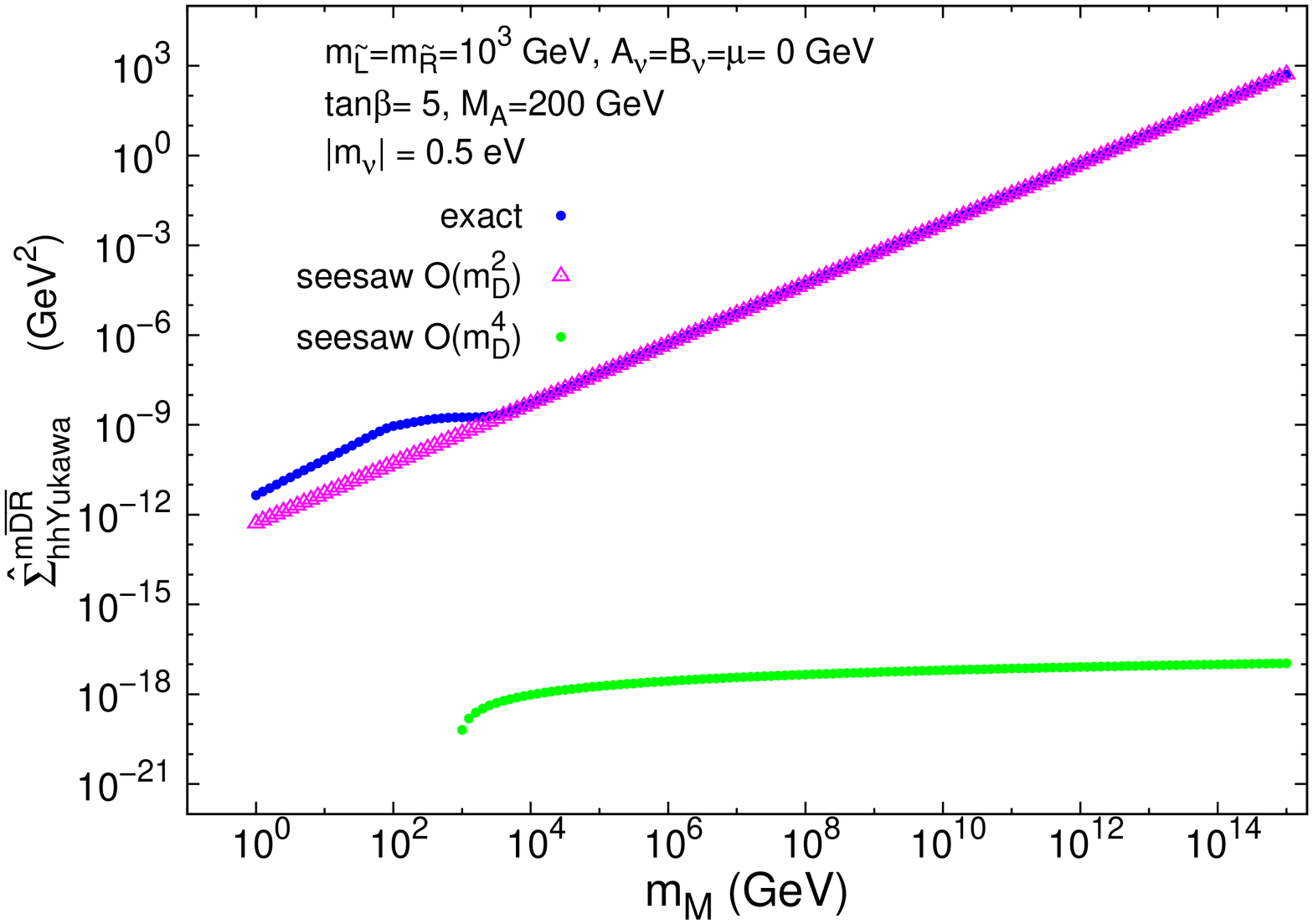,width=60mm,angle=270,clip=} 
	&
        \psfig{file=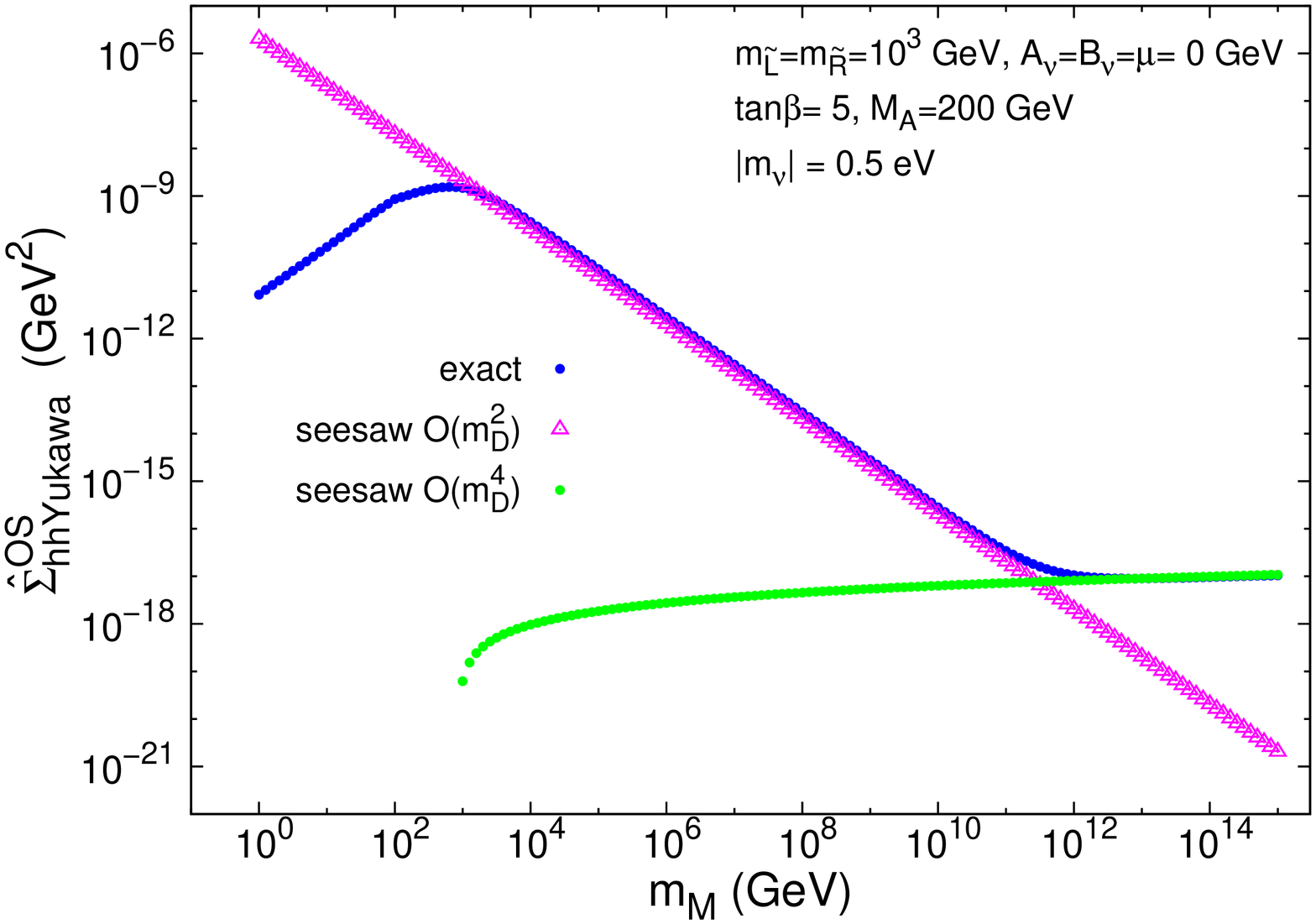,width=60mm,angle=270,clip=} 	  
       \end{tabular}
     \caption{Comparison between the predictions from the seesaw expansion and
     the exact results for the Yukawa part. Left panel: m$\DRbar$
     scheme. Right panel: OS scheme. In both panels, $p^2=(116 \gev)^2$.}
     \label{fig:seesawversusfull} 
   \end{center}
 \end{figure}
%%%%%%%%%%%%%%%%%%%%%%%%%%%%%%%%%%%%%%%%%%%%%%%%%%%

From these formulas the qualitatively different behavior of the 
renormalized Higgs-boson self-energies on the Majorana mass scale $\mM$
can be understood. 
The main difference between the OS scheme and the 
\DRbar/\mDRbar\ schemes  appears in the Yukawa part, especially in the
term of \order{\mD^2}. At the various orders the comparison of the three
schemes is given as follows.
 
At the leading order in the seesaw expansion, \order{\mD^0} in
\refeq{mD0}, the results 
in the $\DRbar$ and  m$\DRbar$ schemes coincide. This is indeed a consequence
of the fact that, at this order,  
 $\hSi_{hh}^{\DRbar}(p^2)$ turns out to be $\mudim$ independent.
The result in the OS scheme differs from these later by a term of order 
$g^2\MZ^2M_{\rm EW}^2/\msusy^2$, where $M_{\rm EW}^2$ refers generically 
to the involved masses of the order of the electroweak scale, i.e., $\MA^2$, 
$p^2$, $\MZ^2$, $m_{h\, {\rm tree}}^2$. Furthermore, this difference  turns out to be
numerically extremely  small. This explains why, for low values
of the Majorana scale, where the \order{\mD^0} term of the expansion
dominates, the predictions from the three schemes are nearly 
indistinguishable.   
 
At the next order in the seesaw expansion, \order{\mD^2} in \refeq{mD2},
the OS result differs  
substantially from the \DRbar\ and \mDRbar\ schemes. First, 
the OS result is extremely suppressed with respect to the \DRbar\ and \mDRbar\
results at large $\mM$. This is due to the fact that the leading
contribution, i.e.\ of the order of $g^2\mD^2M_{\rm EW}^2/\MZ^2$, vanishes
in the OS whereas it is present in the other schemes.  
As can be seen in \refeq{mD2}, the first non
vanishing contribution  contains an extra factor 
$\sim \msusy^2/\mM^2$ which can be extremely small for 
$\mM \gg \msusy$. This remarkable difference of the OS result has its origin
in the different values of the $\de Z_{hh}$ and $\de \tb$
counterterms. More specifically, by computing their finite parts  
in the OS scheme and in the seesaw limit, we get
\begin{align}
\label{wavefunctiontanbetaOS}
 \de^{\rm OS}Z_{hh}|_{\rm finite} &=
-\frac{g^2 \mD^2 \cosasq}
     { 32 \cw^2 \MZ^2 \pi^2 \sin^2\be}
\KKL 1-\log  \frac{\mM^2}{\mu_{\DRbar}^2}  \KKR 
+ \cO\KL \frac{M_{\rm EW}^2,\msusy^2}{\mM^2}\KR ,
 \\
\de^{\rm OS}\tb|_{\rm finite}&=
-\frac{ g^2 \mD^2}
     {64 \cw^2 \MZ^2 \pi^2 \sin^2\be}
\KKL 1-\log  \frac{\mM^2}{\mu_{\DRbar}^2}  \KKR
+\cO \KL \frac{M_{\rm EW}^2,\msusy^2}{\mM^2} \KR~.
\end{align}
These finite contributions lead to the cancellation of the 
above commented leading contributions.

In the \DRbar\ scheme, we get an explicit
logarithmic dependence on $\mM$, concretely as 
$-\log(\mM^2/\mudim^2)$. By construction this term is absent in the \mDRbar\
result. Therefore, the main difference between these two schemes
\DRbar\ and  \mDRbar\ is this logarithmic contribution that can be sizeable for very large
$\mM \gg \mudim$.  
 
The results at the next to next order in the seesaw expansion,
\order{\mD^4} in \refeq{mD4}, show that they all go (leaving apart the
logarithms) as  
$g^2 \mD^4 (M_{\rm EW}^2, \msusy^2)/(\MZ^2 \mM^2)$. Therefore the \order{\mD^4}
terms are extremely suppressed in the three schemes, and
consequently they are not relevant in the large $\mM$ regime.

All the above commented analytical features of the seesaw expansion have also
been checked numerically, 
as it is illustrated in \reffi{fig:seesawversusfull}. In this figure we
show separately the \order{\mD^2} and $\cO(\mD^4)$ contributions and the
exact Yukawa prediction in both the \mDRbar\ (left plot) and 
OS scheme (right plot).\footnote{
It should be kept in mind that due to the different renormalization of $\tb$
the meaning of this input parameter is different in OS and in the \mDRbar\
scheme. In order to perform a real numerical comparison a transition from
$\tb \equiv \tb^{\mDRbar} \to \tb^{\rm OS}$ would have to be performed.
However, here we are interested in the qualitative behavior 
and we do not consider this shift.}
One clearly observes the dominance of the \order{\mD^2}
over the $\cO(\mD^4)$ in the \mDRbar\ scheme
by many orders of magnitude in the full explored $\mM$ range. One also
sees that the \order{\mD^2} result approximates extremely well the exact
Yukawa result for $\mM \gsim 10^4 \gev$. In contrast, in the OS scheme, the
\order{\mD^2} term dominates just up to about $\mM= 10^{10} \gev$, but 
then for larger values the $\cO(\mD^4)$ dominates. In this plot it is
also manifested that the exact Yukawa result in the OS
is well approximated by the \order{\mD^2} term in the interval
$10^3\gev< \mM< 10^{11}\gev$ 
and by the $\cO(\mD^4)$ term for $\mM> 10^{12}\gev$. At this large
values, however, the size of the correction is extremely small 
(below $10^{-17}$ ${\rm GeV}^2$), hence,
irrelevant. It is also clear from this plot that the numerical results for the 
$\cO(\mD^4)$ contributions are similar in the three schemes.

%%%%%%%%%%%%%%%%%%%%%%%%%% F I G U R E %%%%%%%%%%%%%%%%%%%%%%%%%%%%%%%%%%%%
 \begin{figure}[h!]
   \begin{center} 
     \begin{tabular}{cc} \hspace*{-12mm}
  	\psfig{file=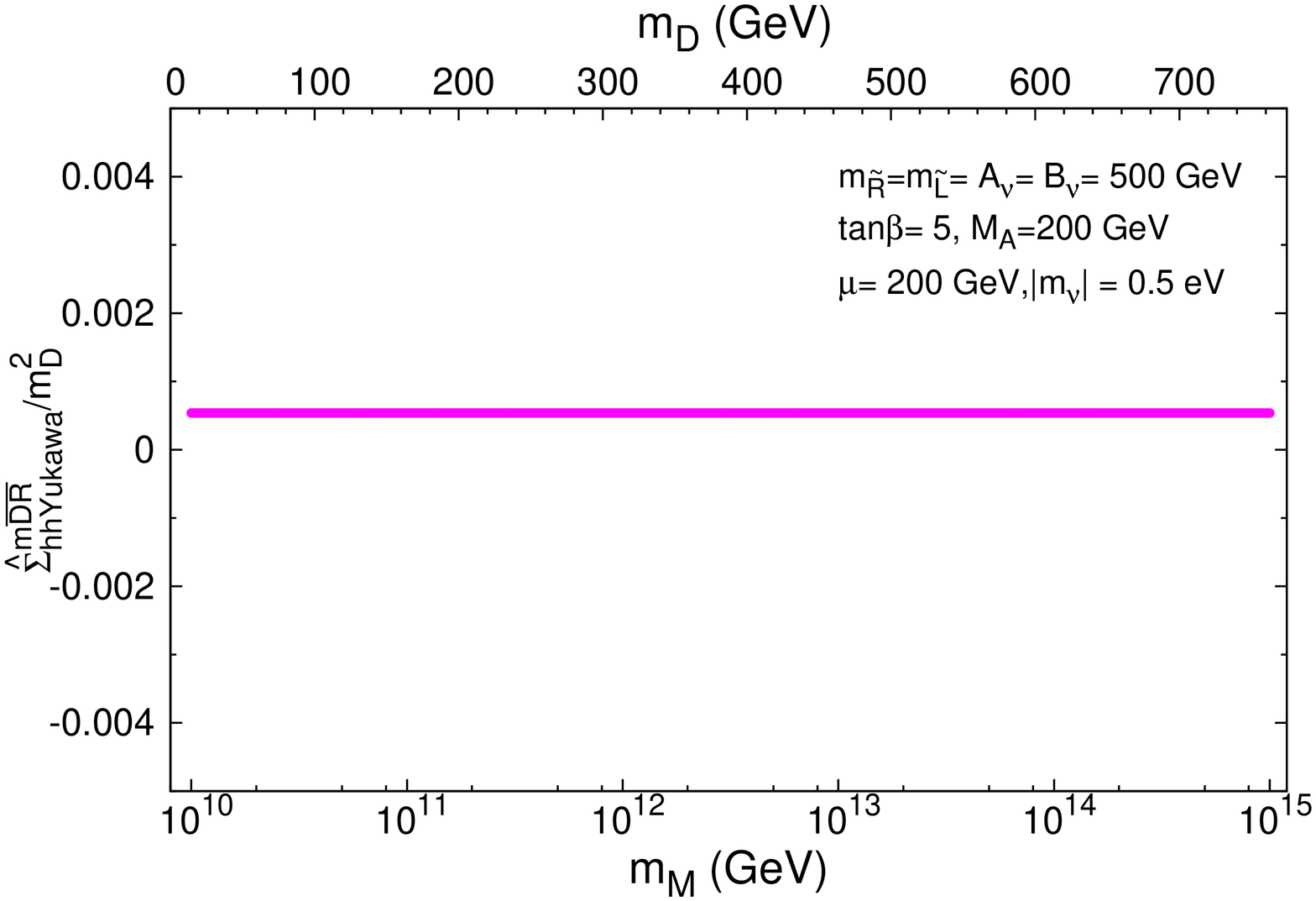,width=60mm,angle=270,clip=} 
	&
        \psfig{file=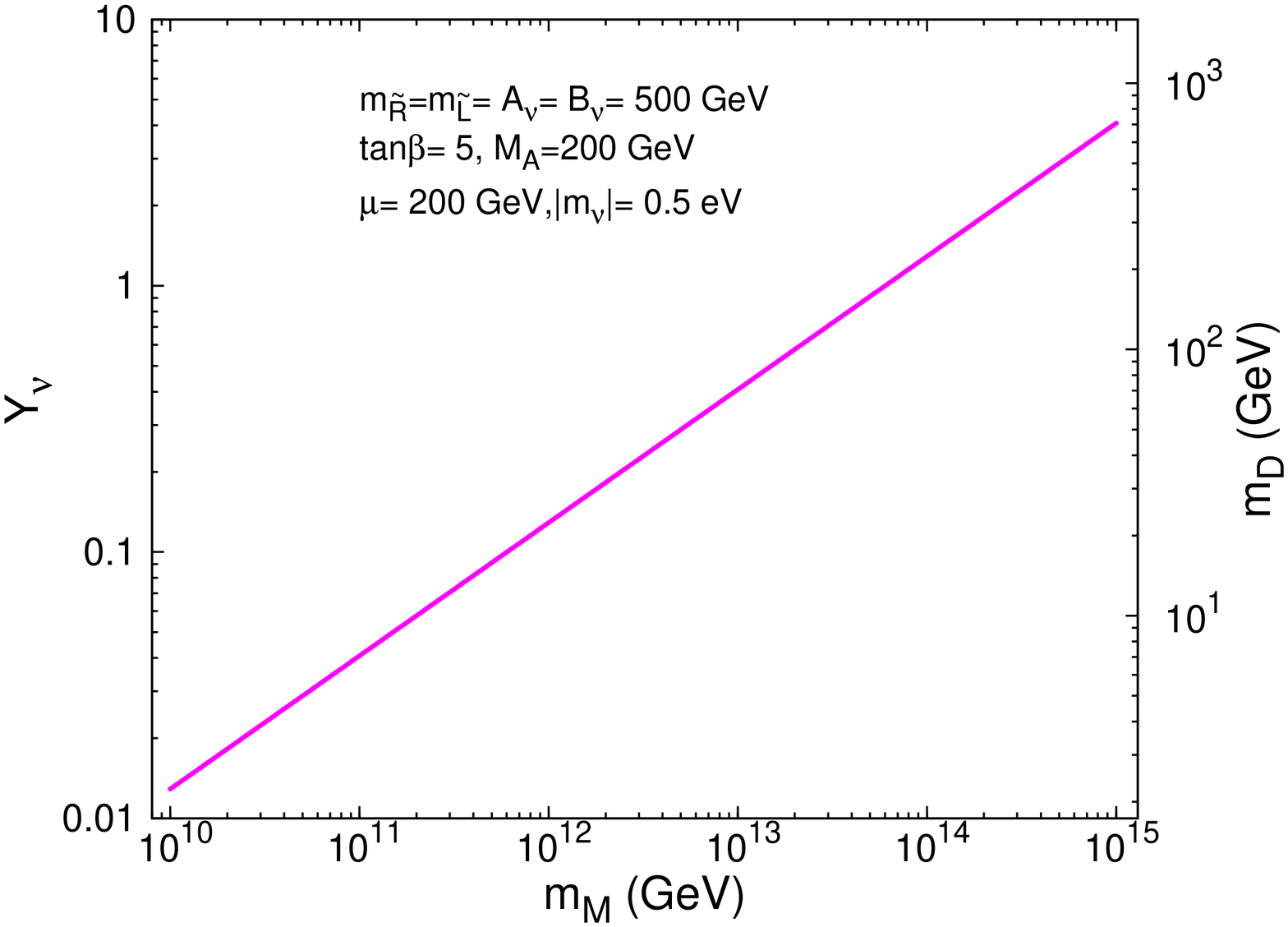,width=60mm,angle=270,clip=} 	  
       \end{tabular}
 \caption{Left panel: Decoupling/Non-decoupling behavior  of the one-loop 
 neutrino/sneutrino corrections to the
 renormalized lightest Higgs boson self-energy at large $\mM$ in the
     \mDRbar\ scheme. Right panel: Dependence of the neutrino Yukawa coupling 
     (and $\mD$) with $\mM$.} 
     \label{fig:Decoupling}
   \end{center}
 \end{figure}
%%%%%%%%%%%%%%%%%%%%%%%%%%%%%%%%%%%%%%%%%%%%%%%%%%%
 
From the definition of the three renormalization schemes, see
\refse{sec:renrMSSM}, and our analytical and numerical analysis in this
section we conclude that the \mDRbar\ scheme is best suited for higher-order
calculations in MSSM-seesaw model. The other two schemes can lead to
unphysically large corrections at the one-loop level.
We will focus in the following on this
scheme, and the numerical evaluation of $\Mh^{\nu/\Snu}$, see \refse{sec:Mh-MSSM-seesaw},
will be performed solely in this ``preferred'' scheme.

Finally, in this context, we discuss the 
decoupling or non-decoupling behavior of the neutrino/sneutrino 
one-loop radiative corrections with the Majorana scale. According to 
\reffis{fig:renSEversusmM} and \ref{fig:seesawversusfull}, the Yukawa part 
of the renormalized self-energy in the \mDRbar\ scheme 
grows with $\mM$. However, this does not constitute by itself a proof of
non-decoupling of $\mM$ in the
radiative corrections to $\hSi_{hh}^{\mDRbar}$ for asymptotically 
large $\mM$. To analyze this question, we have to investigate 
separately the behaviors of $\hSi_{hh}^{\mDRbar}$ and $\mD$ with
$\mM$, since in the way the seesaw mechanism is implemented here, as we have
mentioned before, $\mD$
(or equivalently $\Ynu$) is not an input but an output and it grows
proportional to $\sqrt{\mM}$. 
To analyze these two behaviors separately we show in the left plot of 
\reffi{fig:Decoupling} the 
ratio $(\hSi_{hh}^{\mDRbar})_{\rm Yukawa}/\mD^2$ versus $\mM$ 
(and $\mD$), and in the right plot we show the predictions of the Yukawa
coupling (and $\mD$) as a function of $\mM$. 
The latter one exhibits the (trivial) result of $\Ynu \propto \sqrt{\mM}$
as expected.
In the left plot a constant behavior of the ratio
$(\hSi_{hh}^{\mDRbar})_{\rm Yukawa}/\mD^2$ is clearly manifested,
which means that the growing of $(\hSi_{hh}^{\mDRbar})_{\rm Yukawa}$ 
with $\mM$ is exclusively due to the growing of $\Ynu$ (or $\mD$) 
with $\mM$. However, still this ratio turns
out to be non-vanishing for asymptotically 
large $\mM$, and constant with $m_D$, 
as can be seen in \reffi{fig:Decoupling}. Therefore, a non-decoupling
constant behavior  
must be concluded in the Majorana case from all this discussion. 
 This constant, on the other hand, is very well approximated by the
coefficient multiplying the 
factor $\mD^2$ in the $\hSi_{hh}^{\mDRbar}(p^2)_{\mD^2}$ result of
\refeq{mD2}. 

In order to understand this issue better, we
compare this analytical result, showing a constant behaviour of the
renormalized Higgs boson self-energy in the $m_M  \to \infty$ limit when
$Y_\nu$ is kept fixed, with the corresponding result in the 
Dirac case.  
For simplification in this analytical
comparison we focus just on the ${\cal O}(p^2m_D^2)$ terms and use
the electroweak basis for neutrinos and sneutrinos%
\footnote{The computation in this case reduces to just the evaluation 
of one type of loop diagrams, the sunset 
diagrams, 2nd and 5th in \reffi{fig:loops}.}%
%Fig.1 (counting from left to right and from top to bottom), with external h and: 1) internal $\nu_L,\nu_R$, 
%and 2) internal $\Snu_L,\Snu_R$ propagators, respectively
.~The results 
at ${\cal O}(p^2m_D^2)$ for the renormalized self-energies in the $\DRbar$
scheme for the Majorana and Dirac cases are:
\begin{eqnarray}
 \hSi_{hh}^{{\rm Majorana},\DRbar}(p^2)&=&\frac{g^2m_D^2p^2\cos^2\alpha}
 {32\pi^2M_W^2\sin^2\beta}(\frac{1}{2}-\log\frac{m_M^2}{\mu_{\DRbar}^2}) 
 \nonumber \\
 &&
+ \frac{g^2m_D^2p^2\cos^2\alpha}{64\pi^2M_W^2\sin^2\beta} 
 %\nonumber 
 \\
 \hSi_{hh}^{{\rm Dirac},\DRbar}(p^2)&=&\frac{g^2m_D^2p^2\cos^2\alpha}
 {32\pi^2M_W^2\sin^2\beta} 
 (2-\log\frac{p^2}{\mu_{\DRbar}^2})
 %\nonumber 
 \end{eqnarray} 
where the first and second lines in 
$\hSi_{hh}^{{\rm Majorana},\DRbar}(p^2)$ are the contributions
from neutrinos and sneutrinos respectively. It should be noticed 
that the ${\cal O}(p^2m_D^2)$ sneutrino contributions come exclusively from
the new couplings 
$g'_{h\Snu_L\Snu_R}=-\frac{igm_Dm_M\cos\alpha}{2M_W\sin\beta}$, which are not
present in the Dirac case. It should also be noticed that this result  
in the Majorana case translates into our ${\cal O}(p^2m_D^2)$ term in
(\ref{mD2a}). The comparison of the two formulas shows
that the result of the Majorana 
case for low momenta, $p^2 \ll m_M^2$, does not coincide with the result of
the Dirac case.

From the right plot in  \reffi{fig:Decoupling} we can also conclude on
  the range 
of $\mM$ values where the neutrino Yukawa couplings get too large and
potentialy non-perturbative. The concrete crossing line to set the
perturbativity region is not uniquely defined, 
but it should be considered around $\Ynu \sim \cO(1)$. For instance, by
setting the crossing at $ \Ynu^2/(4\pi)=1.5$ ($\Ynu= 4.34$) we get
perturbativity  
for $\mM<10^{15} \gev$, and by setting it at $\Ynu=1.5$ it is got for 
$\mM<10^{14} \gev$. In the following of this subsection we set 
$\mM=10^{14} \gev$ as our reference value.        
    
\subsubsection*{Dependence on \boldmath{$\tb$, $M_A$, $\mu$, $\mL$, $\mR$, $\Anu$, $\mnu$,
$\Bnu$ and $p$}} 

\begin{figure}[h!]
   \begin{center} 
     \begin{tabular}{cc} \hspace*{-12mm}
  	\psfig{file=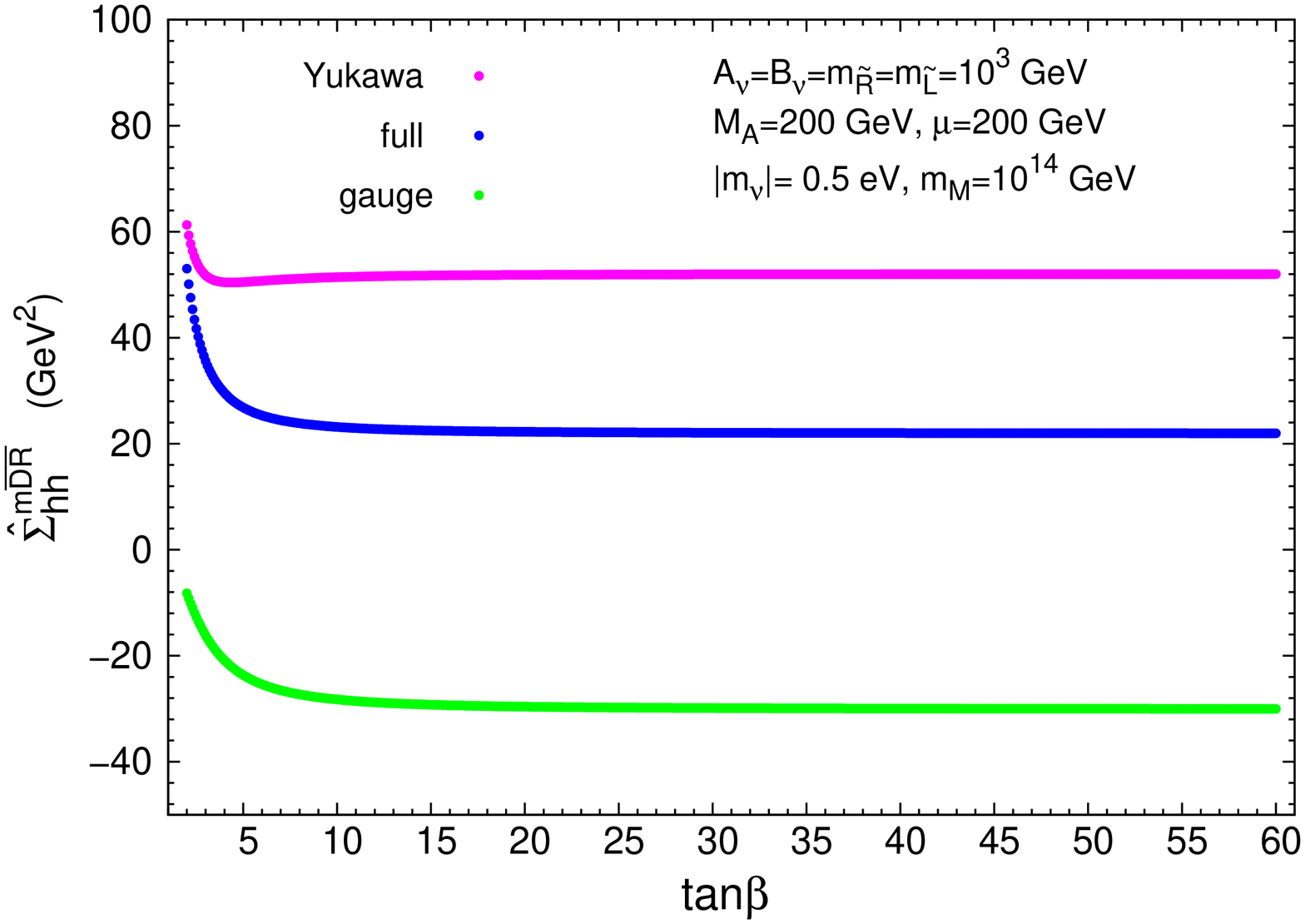,width=60mm,angle=270,clip=} 
	&
        \psfig{file=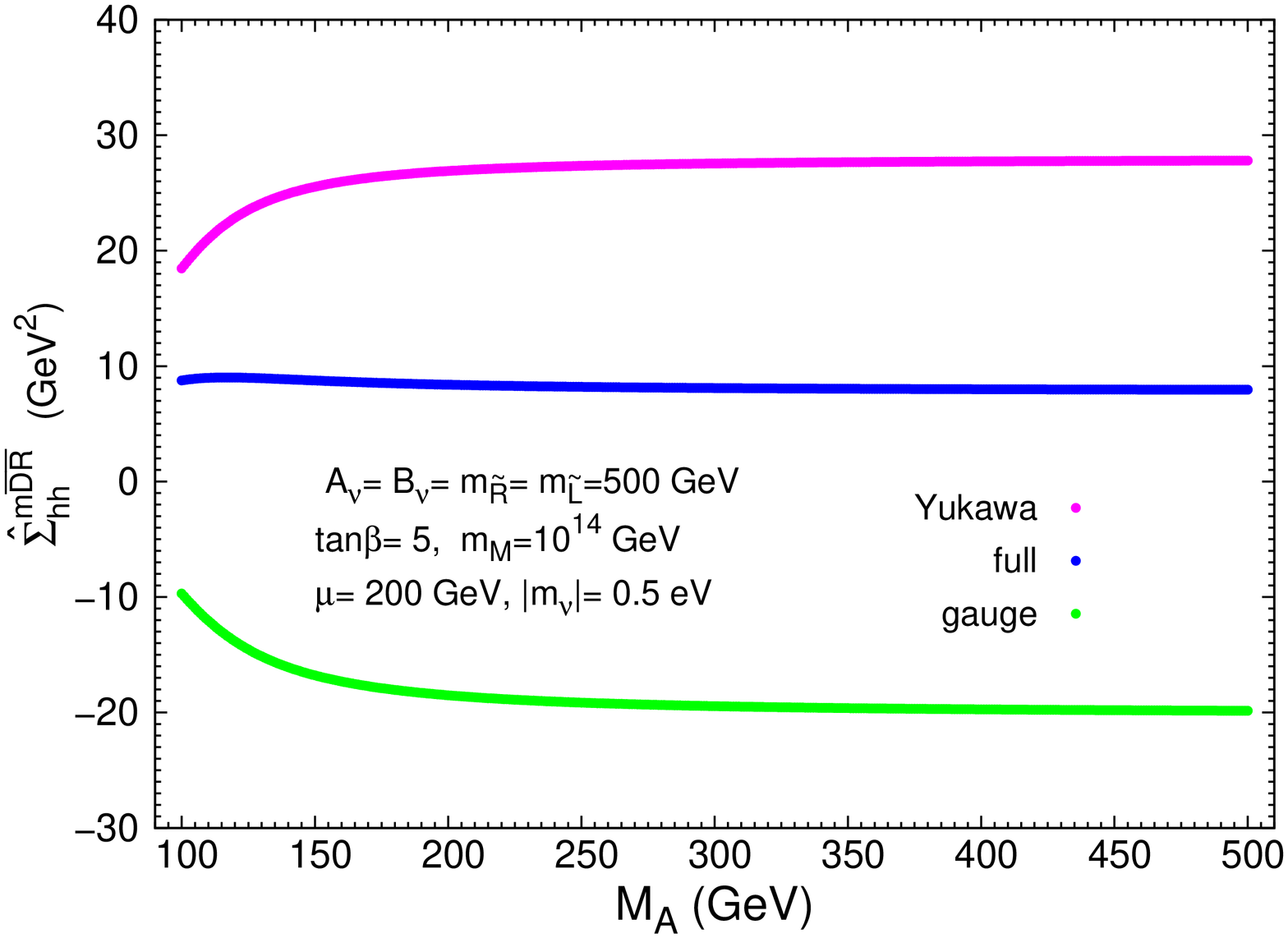,width=60mm,angle=270,clip=} 	  
       \end{tabular}
\caption{Left panel: $\hSi_{hh}^{\mDRbar}(p^2)$ as a function of $\tb$.
     Right panel: $\hSi_{hh}^{\mDRbar}(p^2)$ as a function of $M_A$. In
     the left (right) panel, $p^2=(116\,\, {\rm GeV})^2$ 
     ($p^2=(105 \gev)^2)$.}
     \label{fig:mDRversustbandmA} 
   \end{center}
 \end{figure}

\begin{figure}[h!]
   \begin{center} 
     \begin{tabular}{c} \hspace*{-12mm}
 \psfig{file=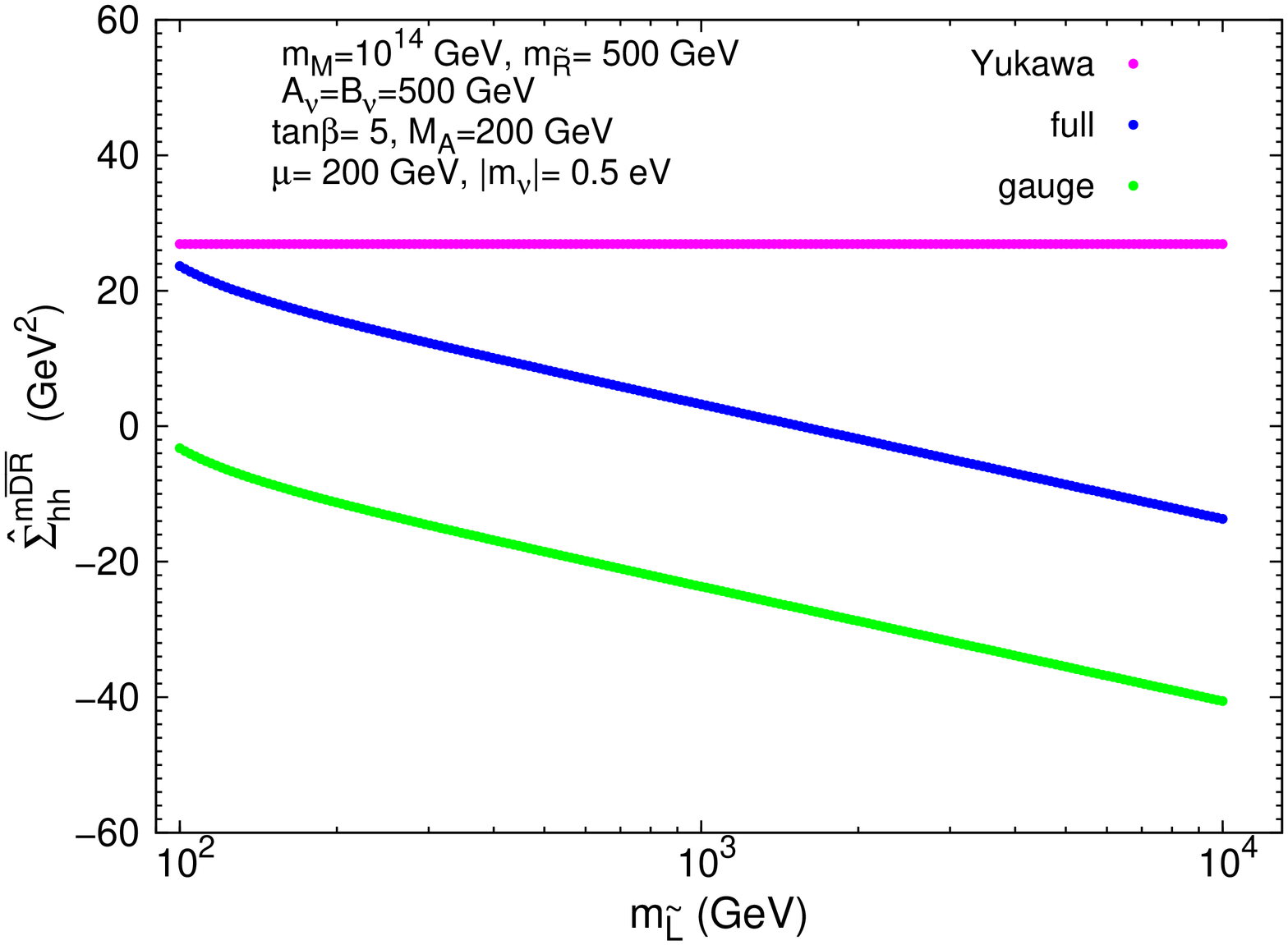,width=60mm,angle=270,clip=} 	  
       \end{tabular}
\caption{$\hSi_{hh}^{\mDRbar}(p^2)$ as a function of $\mL$; we have set
$p^2=(105 \gev)^2$.}
     \label{fig:mDRversusmL} 
   \end{center}
 \end{figure}

\begin{figure}[h!]
   \begin{center} 
     \begin{tabular}{cc} \hspace*{-12mm}
  	\psfig{file=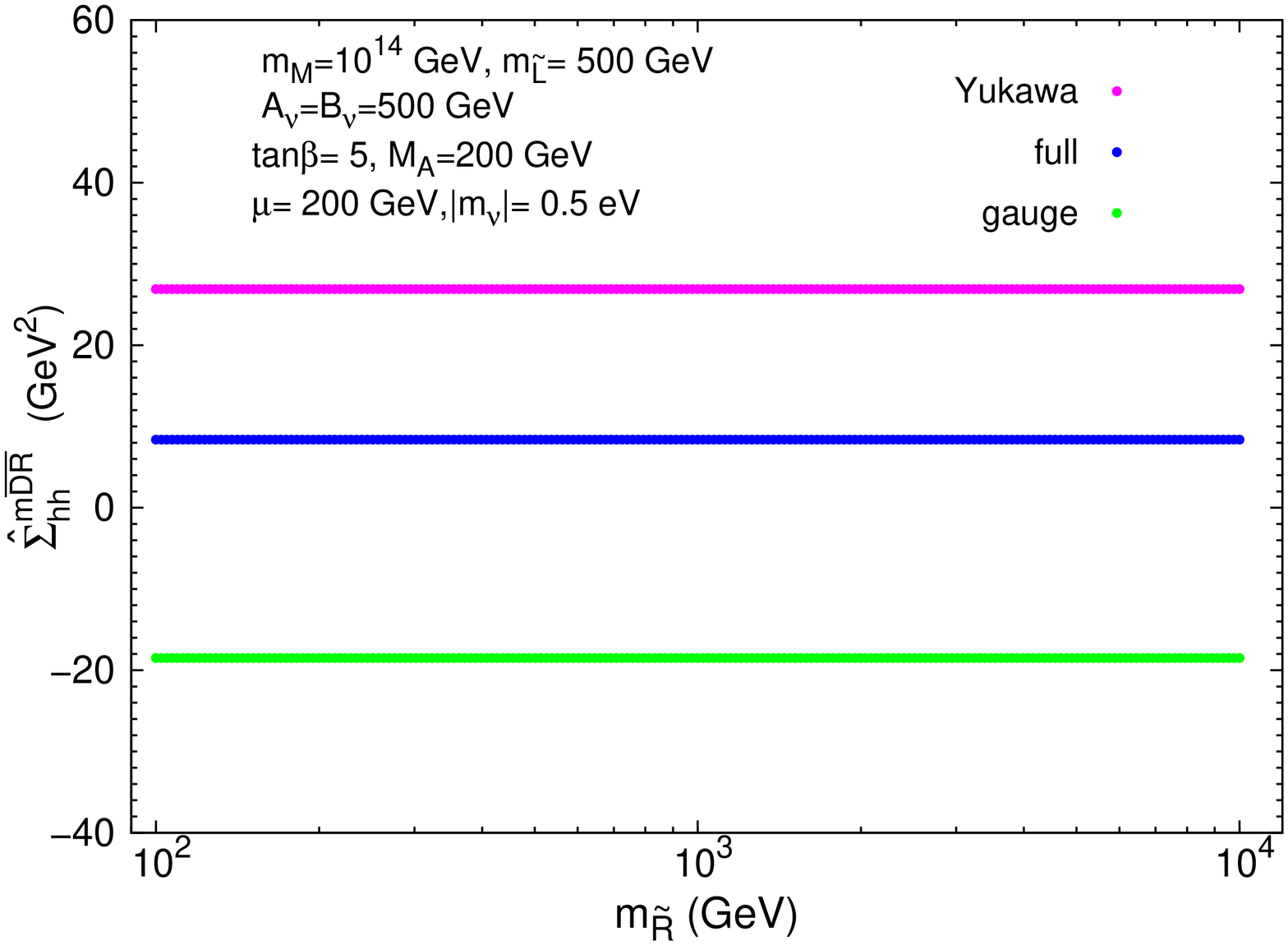,width=60mm,angle=270,clip=} 
	&
        \psfig{file=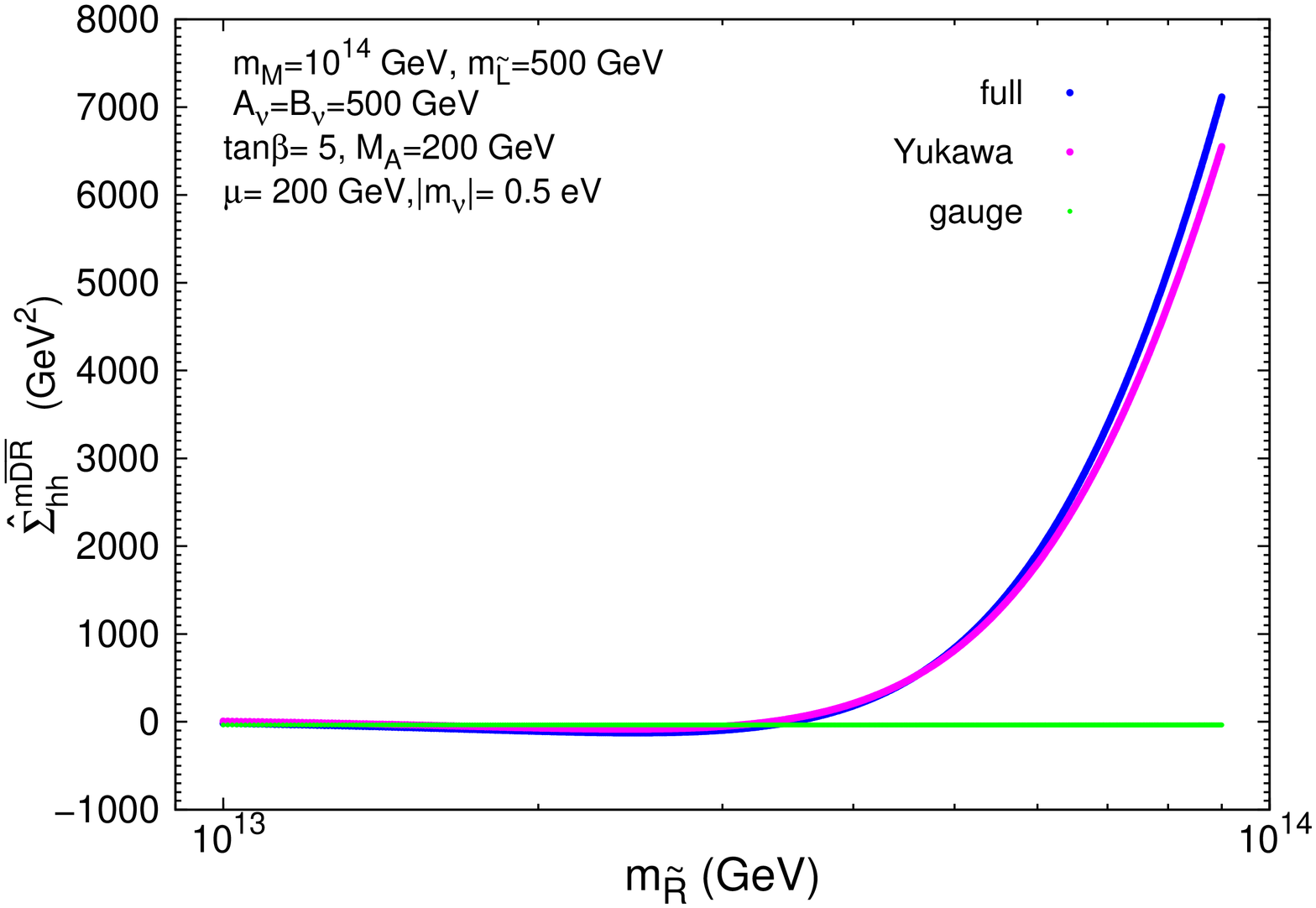,width=60mm,angle=270,clip=} 	  
       \end{tabular}
\caption{ $\hSi_{hh}^{\mDRbar}(p^2)$ as a function of $\mR$. Left
  panel: low mass values $10^2\gev <\mR  < 10^4 \gev$. Right panel:
  high mass values $10^{13}\gev < \mR< 10^{14} \gev$. In
  both panels we have set $p^2=(105 \gev)^2$.}
     \label{fig:mDRversusmR} 
   \end{center}
 \end{figure}

\begin{figure}[h!]
   \begin{center} 
     \begin{tabular}{cc} \hspace*{-12mm}
  	\psfig{file=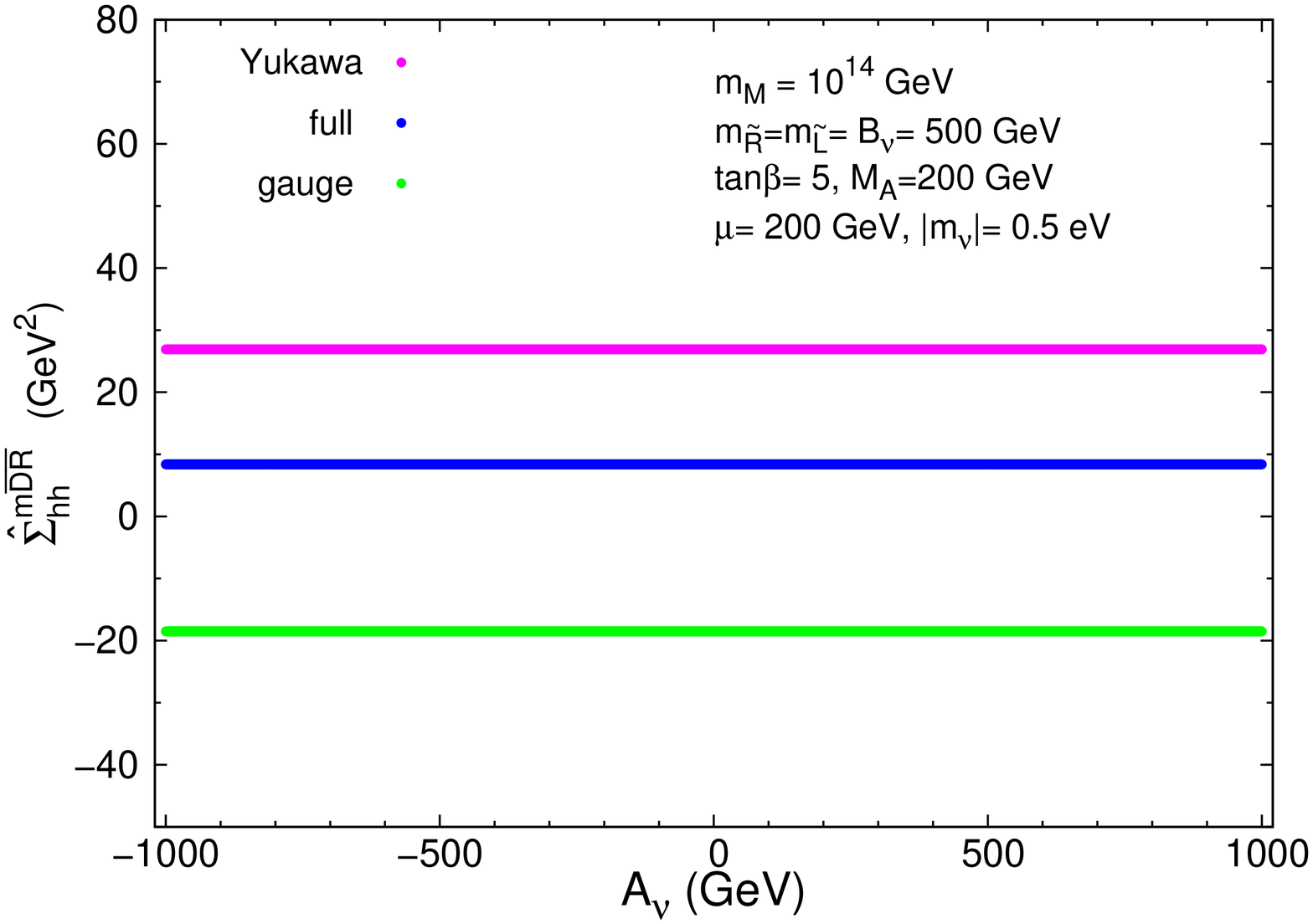,width=60mm,angle=270,clip=} 
	&
        \psfig{file=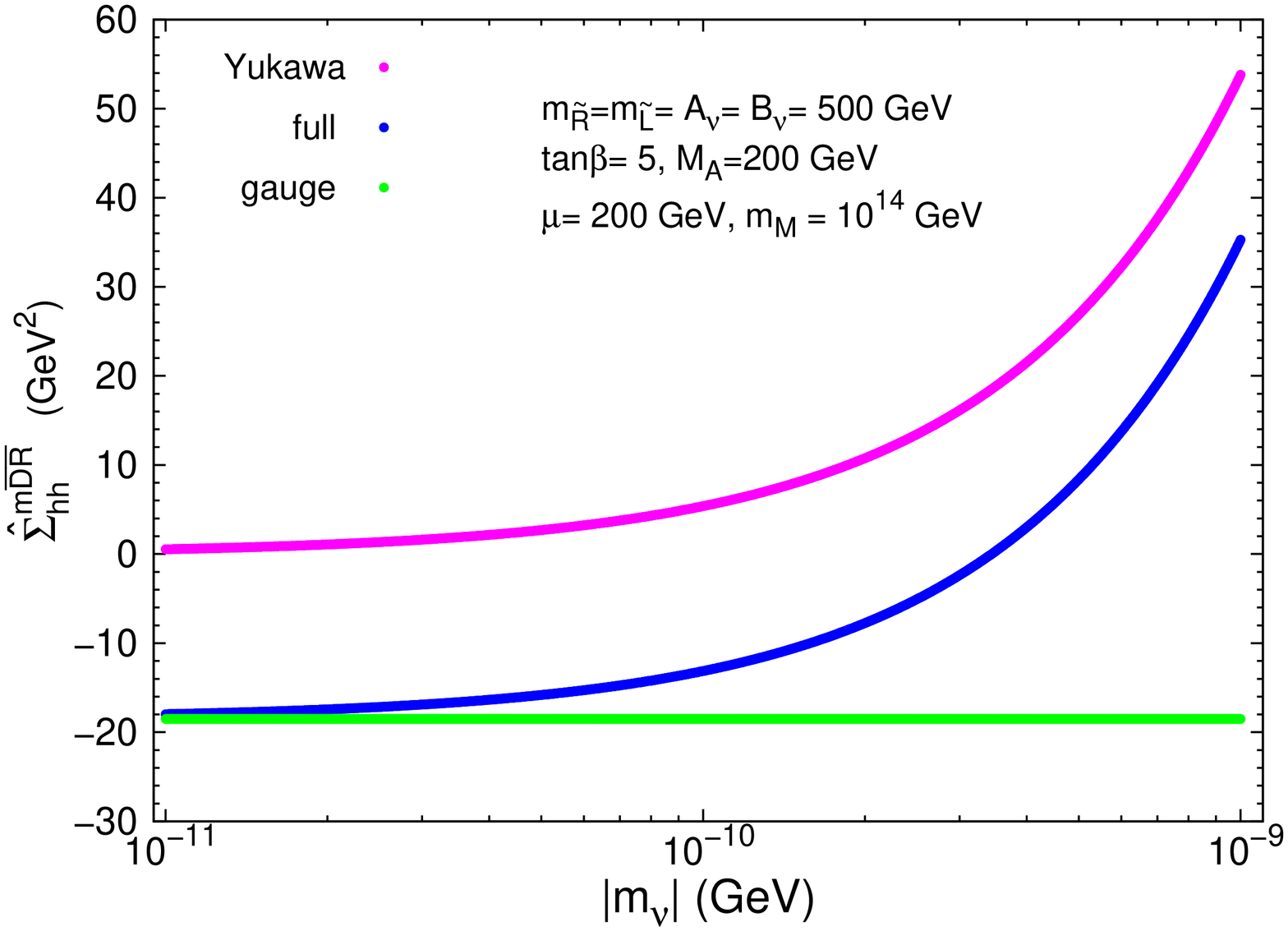,width=60mm,angle=270,clip=} 	  
       \end{tabular}
\caption{Left panel: $\hSi_{hh}^{\mDRbar}(p^2)$ as a function of $\Anu$.
     Right panel: $\hSi_{hh}^{\mDRbar}(p^2)$ as a function of $|\mnu|$. In
     both panels we have set $p^2=(105 \gev)^2$. }
     \label{fig:mDRversusAnuandmnu}  
   \end{center}
 \end{figure}

\begin{figure}[h!]
   \begin{center} 
     \begin{tabular}{cc} \hspace*{-12mm}
  	\psfig{file=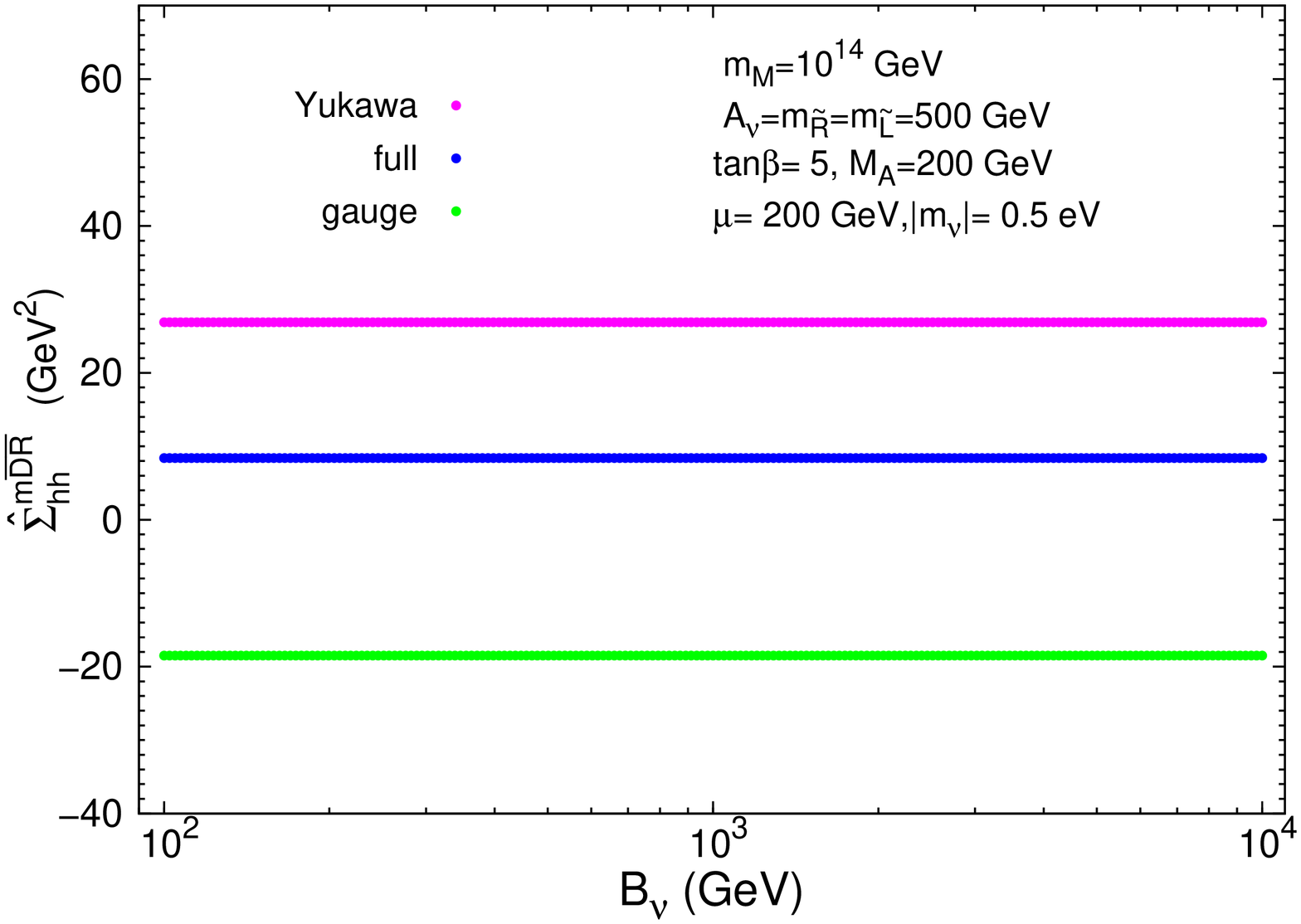,width=60mm,angle=270,clip=} 
	&
        \psfig{file=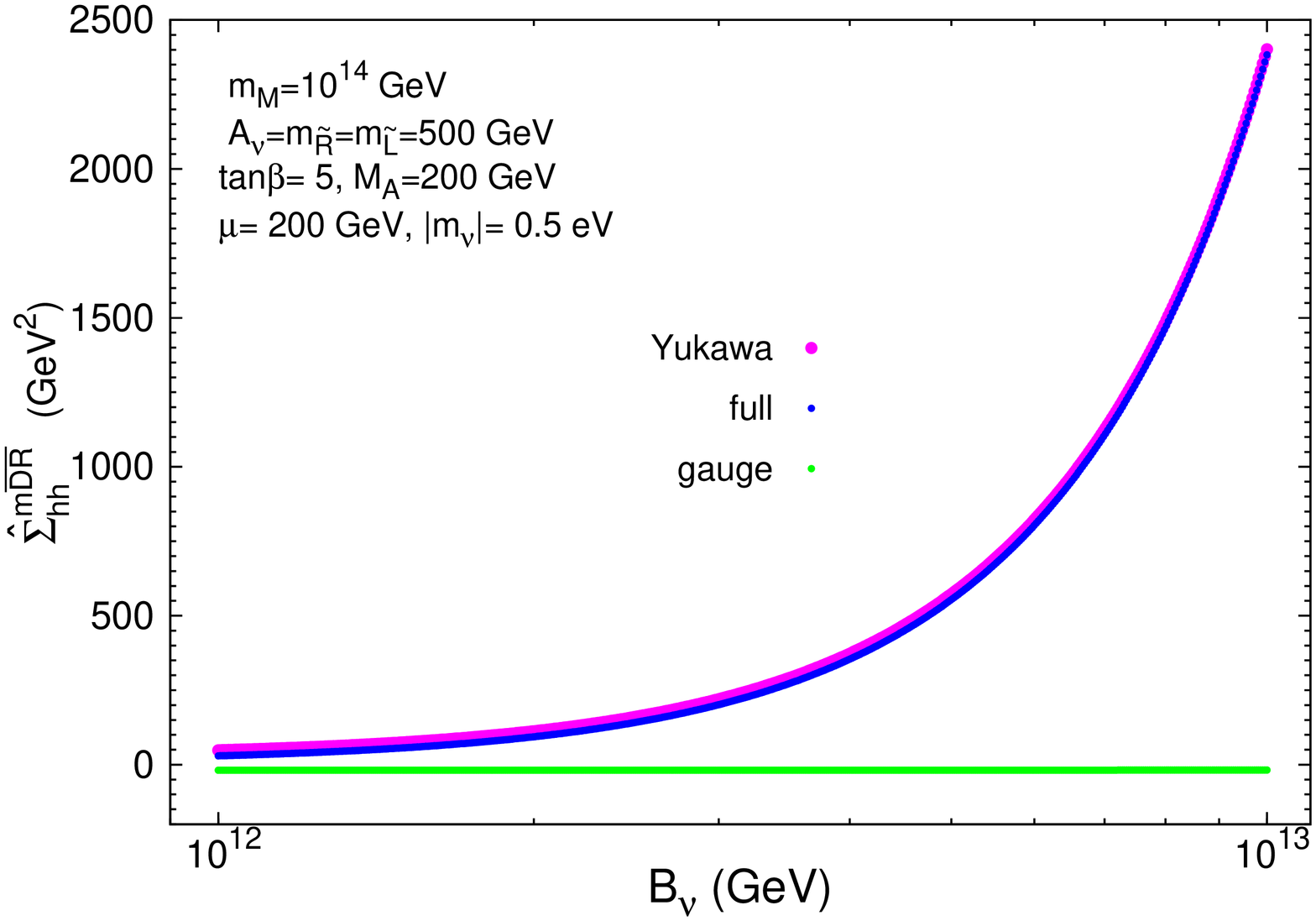,width=60mm,angle=270,clip=} 	  
       \end{tabular}
     \caption{ $\hSi_{hh}^{\mDRbar}(p^2)$ 
      as a function of $\Bnu$. Left panel: low $\Bnu$ values,
      $10^2\gev < \Bnu < 10^4 \gev$. Right panel:
      high $\Bnu$ values, $10^{12}\gev < \Bnu < 10^{13} \gev$. In
     both panels we have set $p^2=(105 \gev)^2$.}
     \label{fig:mDRversusBnu}
   \end{center}
 \end{figure}

\begin{figure}[h!]
   \begin{center} 
     \begin{tabular}{cc} \hspace*{-12mm}
  	\psfig{file=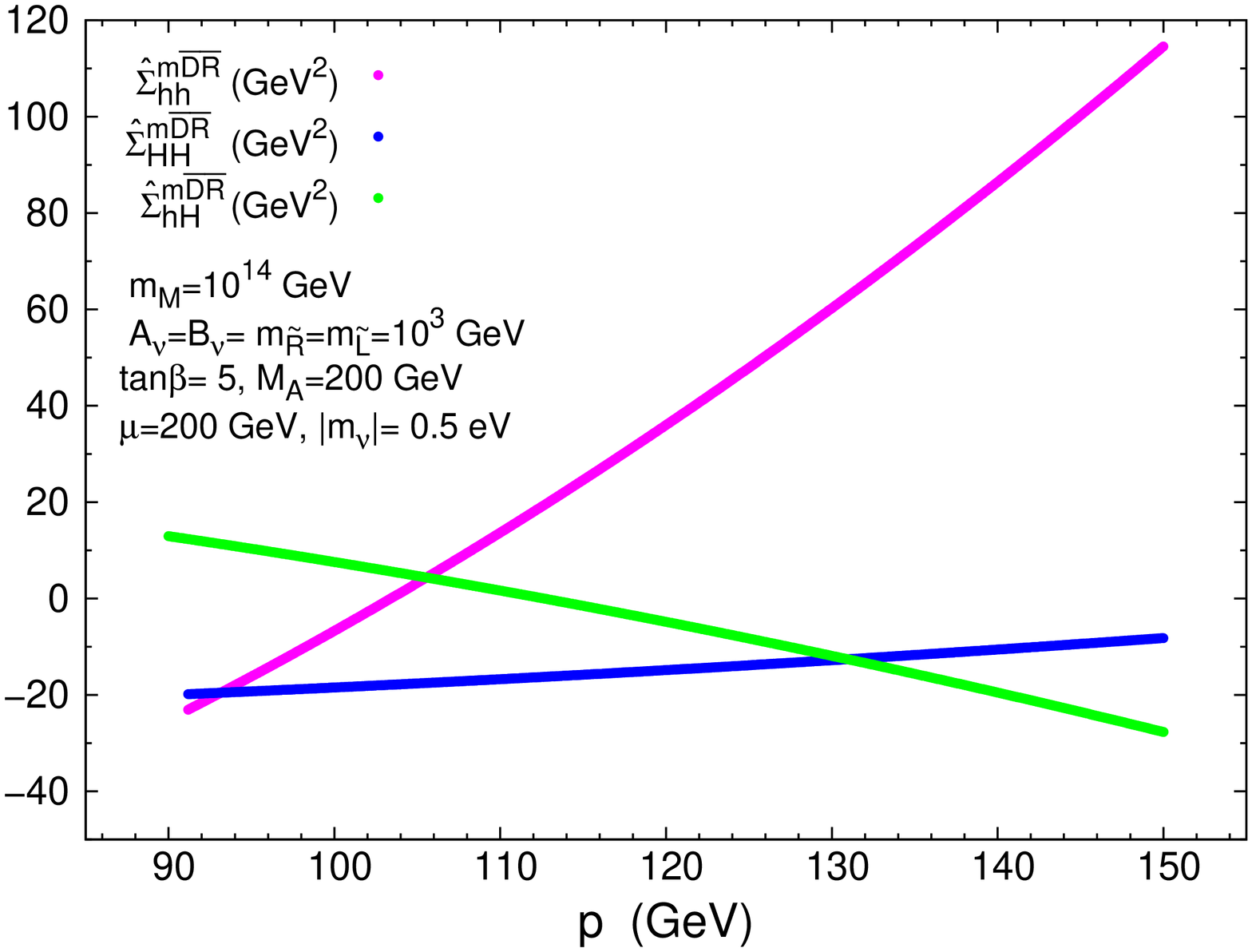,width=60mm,angle=270,clip=} 
	&
        \psfig{file=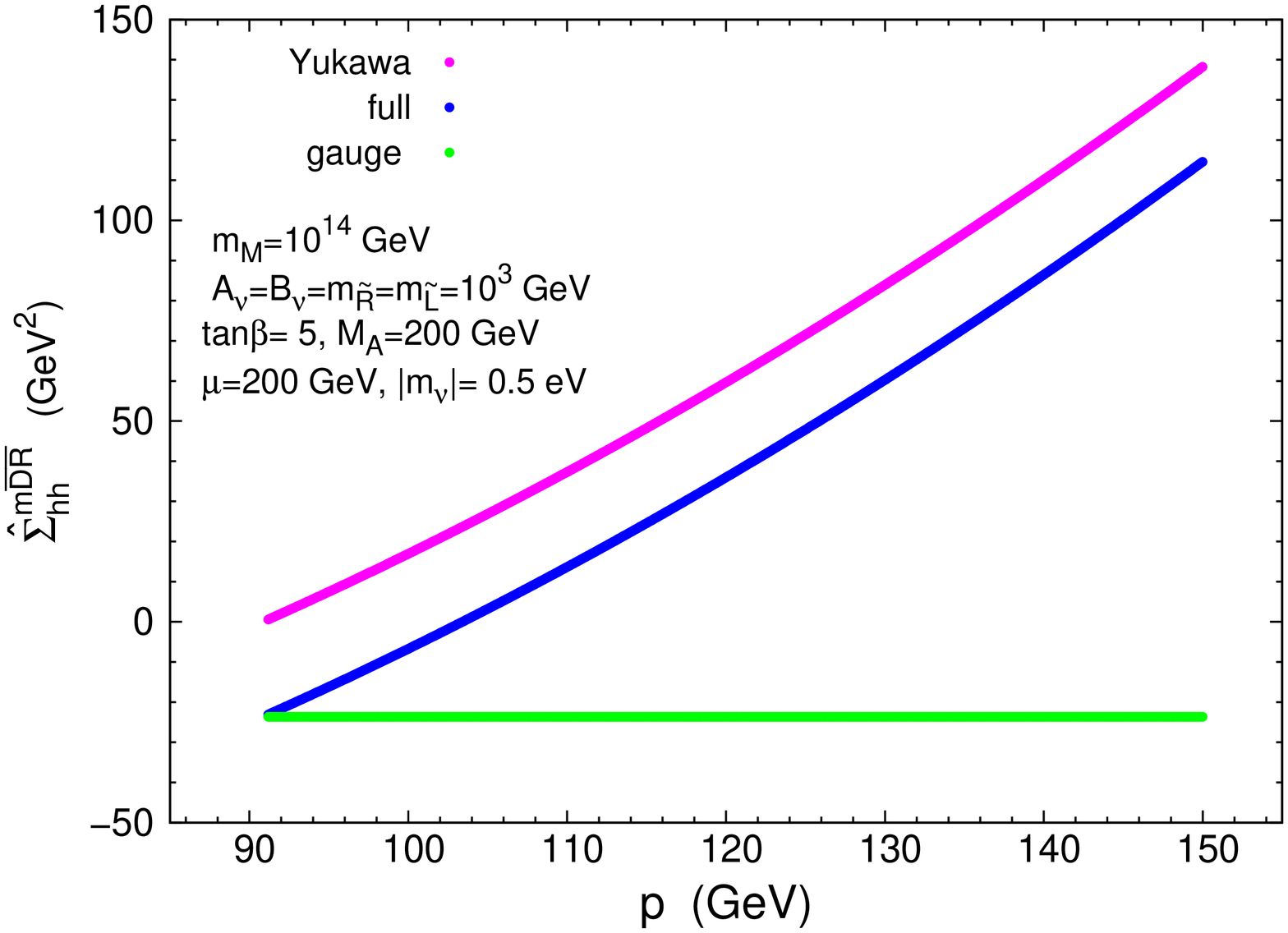,width=60mm,angle=270,clip=} 	  
       \end{tabular}
\caption{Left panel: $\hSi_{hh}^{\mDRbar}(p^2)$, $\hSi_{hh}^{\mDRbar}(p^2)$ 
     and $\hSi_{hh}^{\mDRbar}(p^2)$ as a function of  the external 
     momentum $p$. Right panel: the two contributions 
     $\hSi_{hh}^{\mDRbar}(p^2)_{\rm gauge}$
      $\hSi_{hh}^{\mDRbar}(p^2)_{\rm Yukawa}$ and the full
     result are shown separately.}
     \label{fig:mDRversusp} 
   \end{center}
 \end{figure}

The behavior of the renormalized self-energy in the m$\DRbar$ scheme with the
other parameters entering in this computation are shown in 
\reffis{fig:mDRversustbandmA} - \ref{fig:mDRversusp}. In all these
plots we have included separately the gauge, Yukawa and total results for
comparison.

First, the behavior with $\tb$ is analyzed in the left plot of
\reffi{fig:mDRversustbandmA}. It exhibits basically the expected features
that can be inferred from the loop corrections of an up-type
fermion/sfermion. The neutrino/sneutrino 
one-loop radiative corrections reach their maximum value at the lowest 
considered
value of $\tb$, $\tb =2$ in this plot. For $\tb > 5$ the
dependence is nearly flat. There are no relevant differences 
between the behaviors with $\tb$ of the Yukawa and the gauge parts. 
From now on, we will set $\tb=5$ as our reference value. 

%%%%%%%%%%%%%%%%%%%%%%%%%% F I G U R E %%%%%%%%%%%%%%%%%%%%%%%%%%%%%%%%%%%%

%%%%%%%%%%%%%%%%%%%%%%%%%%%%%%%%%%%%%%%%%%%%%%%%%%%%%

The behavior with $M_A$ is displayed in the right panel of Fig.\ref{fig:mDRversustbandmA}. Again we
see no relevant differences with respect to the well known behavior in the 
MSSM. For $M_A$ larger that 150 GeV the total contribution from the 
neutrino/sneutrino sector to the 
renormalized self-energy is nearly flat with $M_A$. In the following we will take $M_A=
200 \gev$ as our reference value. 

The dependence with the soft SUSY breaking mass of the `left handed'
$SU(2)$ doublet, 
$\mL$, is shown in \reffi{fig:mDRversusmL}. We
see that the gauge contribution is negative and increases in modulus 
with increasing $\mL$, whereas the
Yukawa contribution is positive and
nearly insensitive to changes of $\mL$ in the investigated interval, 
$10^2 \gev < \mL < 10^4\gev$. The total neutrino/sneutrino
corrections, at these selected values of the model parameters, are 
positive and decreasing with $\mL$ for $10^2 \gev < \mL < 2 \times 10^3\gev$ 
and then become negative and increasing in modulus with  $\mL$ for 
$2 \times 10^3\gev < \mL < 10^4\gev$.

%%%%%%%%%%%%%%%%%%%%%%%%%% F I G U R E %%%%%%%%%%%%%%%%%%%%%%%%%%%%%%%%%%%%

The behavior with the soft SUSY breaking parameter of the `right handed' 
sector $\mR$ is shown in \reffi{fig:mDRversusmR}. 
In the left plot a mass scale similar to the other soft SUSY-breaking
parameters is investigated, whereas in the right plot values of $\mR$
closer to $\mM$ are explored. It should be reminded that these
values are not constrained by data. 
An interesting feature can be observed at large values of $\mR$.
The contributions to the renormalized self-energy stay flat up to about 
$\mR \sim 10^{13} \gev$. Above this mass scale the Yukawa part grows
rapidly, reaching very large values at $\mR \sim 10^{14} \gev$ 
of around  $\hSi_{hh}^{\mDRbar} \sim 7000 \gev^2$.

%%%%%%%%%%%%%%%%%%%%%%%%%% F I G U R E %%%%%%%%%%%%%%%%%%%%%%%%%%%%%%%%%%%%
%%%%%%%%%%%%%%%%%%%%%%%%%%%%%%%%%%%%%%%%%%%%%%%%%%%%

The behavior with the new soft SUSY-breaking trilinear coupling $\Anu$
is shown in the left plot of \reffi{fig:mDRversusAnuandmnu}. 
The full result, the gauge, and Yukawa 
parts  are nearly independent on this parameter in the
studied  interval, $-1000 \gev < \Anu < 1000 \gev$. Although 
not shown explicitly, we have also studied the behavior with $\mu$ and got 
the same `flat' behavior for $-1000 \gev < \mu < 1000 \gev$.
This justifies our choice $\Anu = \mu = 0$ in our seesaw expansion
above.   
 
%%%%%%%%%%%%%%%%%%%%%%%%%% F I G U R E %%%%%%%%%%%%%%%%%%%%%%%%%%%%%%%%%%%%
%%%%%%%%%%%%%%%%%%%%%%%%%%%%%%%%%%%%%%%%%%%%%%%%%%

The behavior with the lightest neutrino mass, $\mnu$, is demonstrated in
the right plot of \reffi{fig:mDRversusAnuandmnu}. 
One can see that the Yukawa part is quite 
sensitive to this mass that we have varied in a plausible and 
compatible with data range. The growing of the result with $|\mnu|$, for 
fixed $\mM$, is the consequence of the growing of $\Ynu$ (or $\mD$) with 
$|\mnu|$ since in this model they are correlated, as shown in 
(\ref{mDmN}) and (\ref{mMmN}).

%%%%%%%%%%%%%%%%%%%%%%%%%% F I G U R E %%%%%%%%%%%%%%%%%%%%%%%%%%%%%%%%%%%%
%%%%%%%%%%%%%%%%%%%%%%%%%%%%%%%%%%%%%%%%%%%%%%%%%
 
The behavior with $\Bnu$ is analyzed in \reffi{fig:mDRversusBnu}. We
have found a flat result with this new soft parameter for most of the
explored range, 
except at very large values, $\Bnu> 10^{12} \gev$, as shown in the
right plot. For these large values the 
Yukawa part grows noticeably with $\Bnu$ and dominates largely the total
result, leading to large radiative corrections. For instance, for the
parameters chosen in this figure and $\Bnu = 10^{13} \gev$, we found 
$\hSi_{hh}^{\mDRbar} \sim 2400\gev^2$. The question whether such large
values of $\Bnu$ are realistic depends on the particular
models and universality conditions. However, such an analysis is beyond
the scope of our paper.
On the other hand, if we apply the bounds that are imposed in 
order to avoid destabilizing
the electroweak symmetry breaking~\cite{Farzan:2004cm}, leading to  
$\Bnu \Ynu^2/(8 \pi^2) <  \msusy/\tb$, one gets an upper limit 
on $\Bnu$. For $\Ynu \sim 1$, $\msusy \sim 1000 \gev$ and $\tb \sim 5$
one finds 
$\Bnu < 1.6 \times 10^4 \gev$. For this range the renormalized
Higgs-boson self-energy is nearly independent of $\Bnu$. From now on, we will 
choose 
$\Bnu = 500 \gev$ as our reference value.

%%%%%%%%%%%%%%%%%%%%%%%%%% F I G U R E %%%%%%%%%%%%%%%%%%%%%%%%%%%%%%%%%%%%
%%%%%%%%%%%%%%%%%%%%%%%%%%%%%%%%%%%%%%%%%%%%%
 
Finally, we show in \reffi{fig:mDRversusp} the behavior with $p^2$, 
the square of the external momentum of the Higgs boson self-energies, 
which
is a relevant issue for the discussion of the radiative corrections to
the Higgs-boson masses (see the next subsection). 
The three renormalized self-energies, $\hSi_{hh}$, $\hSi_{HH}$ and 
$\hSi_{hH}$, are clearly
dependent on $p^2$, but the most sensitive one is $\hSi_{hh}$. It is clear 
from this figure that setting $p^2=0$ in the renormalized self-energies does 
not provide a good approximation for the estimate of the
radiative corrections to the Higgs boson mass from the neutrino/sneutrino sector in the present case of
Majorana neutrinos.
One can also see that mainly the Yukawa part is responsible for this
sensitivity to $p^2$. Setting the 
proper $p^2$ in order to estimate realistically the Higgs mass corrections will 
be discussed in the next subsection.

\subsubsection*{The Dirac case} 
Finally, we perform a comparison between the case of massive Majorana
neutrinos (as analyzed so far) and the case of Dirac neutrinos. In order
to analyze the Dirac case,  
we have computed the one-loop neutrino/sneutrino contributions to the 
renormalized lightest Higgs boson self-energy for $\mM=0$. The
analytical results for this Dirac case are collected in Appendix C. We have
chosen here the \DRbar\ scheme, since due to the absence of $\mM$
  no large logarithmic corrections are expected, and a comparison to
  existing calculations can readily be performed. 
 \begin{figure}[h!]
   \begin{center} 
     \begin{tabular}{c} \hspace*{-12mm}
  	\psfig{file=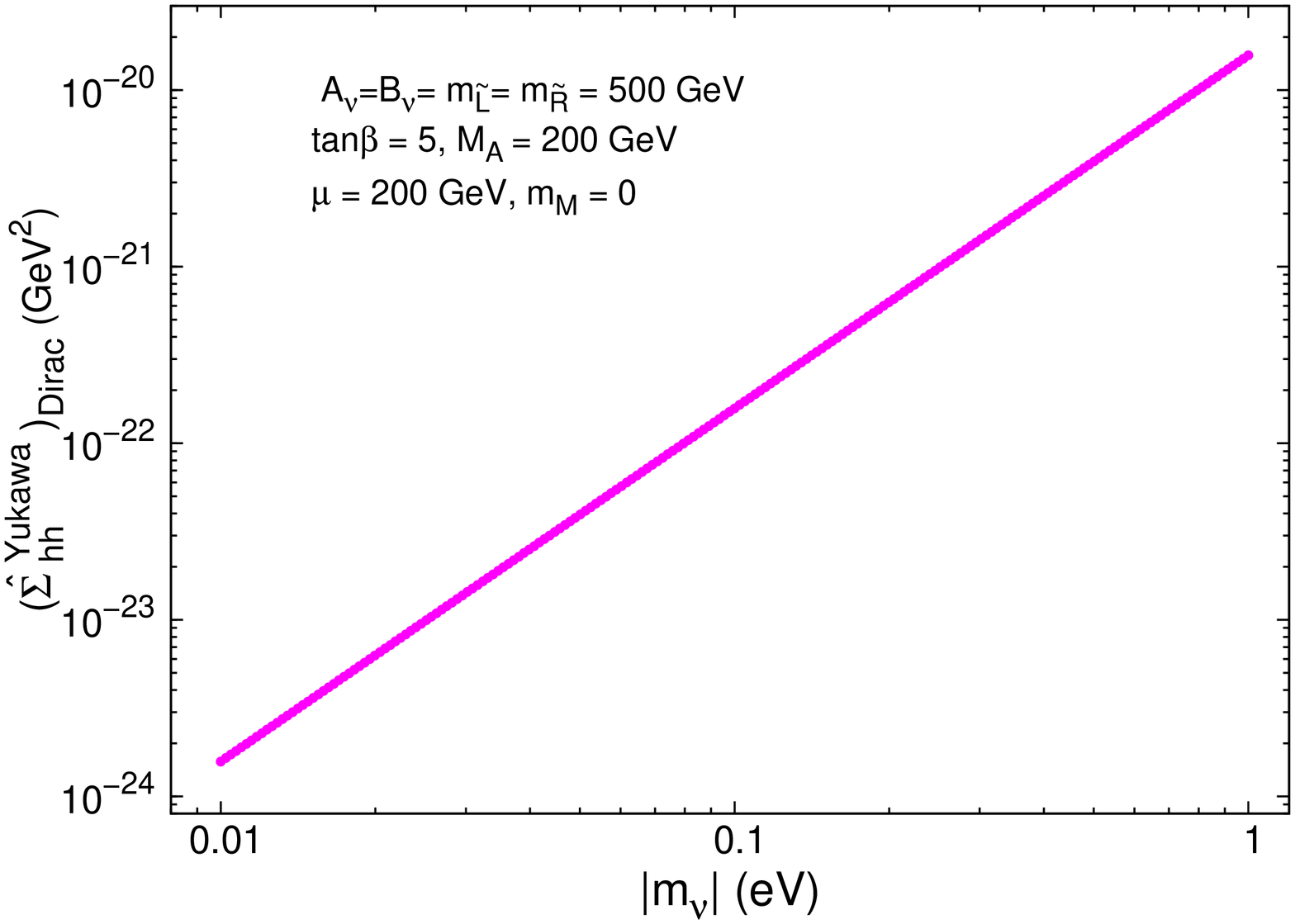,width=60mm,angle=270,clip=}
      \end{tabular}
\caption{One-loop corrections to the Yukawa part of the lightest Higgs boson 
renormalized self-energy from the neutrino/sneutrino sector in the case of 
massive Dirac neutrinos}
     \label{fig:Diraccase} 
   \end{center}
 \end{figure} 
First, we have checked the finiteness of
the result. Second, we have also checked that the obtained formulas agree 
with the well known result of the one-loop radiative corrections from other massive
fermion/sfermion sectors of the MSSM, with the obvious corresponding changes
of fermion/sfermion parameters and quantum numbers. In particular, it can be
seen that the 
formulas in Appendix C coincide 
with the one-loop corrections from the MSSM top/stop sector by replacing, 
correspondingly, the neutrino $SU(2)\times U(1)$ quantum numbers by the top
quark ones, $\mD$ by $m_t$, 
 $m_{{\Snu_{\pm}}}$ $(=m_{{\Snu}_1})$ by $m_{{\tilde t}_1}$, 
 $m_{{\SNu_{\pm}}}$ $(=m_{{\Snu}_2})$ by $m_{{\tilde t}_2}$, 
 $\theta_{\pm}$ $(={\tilde \theta})$ by ${\tilde {\theta_t}}$ and by adding the
 proper color factor, $N_C=3$.

 As for the numerical estimate, we present in Fig.\ref{fig:Diraccase} 
 the result of the 
 Yukawa contributions from the  
 one-loop neutrino/sneutrino radiative corrections to 
 the renormalized self-energy, 
 $(\ser{hh}^{\rm Yukawa})_{\rm Dirac}$,  as a function of the physical neutrino 
 mass, $|\mnu|= \mD$. The regularization scale has been fixed here to 
 $\mudim=100 \gev$ and the external momentum to $p=116 \gev$. 
As in the Majorana case, we consider an interval for the neutrino mass
inspired by experimental data, 
$0.01\,\, {\rm eV} \lsim |\mnu| \lsim 1 \,\,{\rm eV}$.
 In this plot we see clearly that, as expected, these Yukawa contributions 
 are extremely small (below $10^{-20}$ ${\rm GeV}^2$) and are fully dominated 
 by the gauge part which we have also estimated, for the chosen
 parameters in this plot, leading  
 to $(\ser{hh}^{\rm gauge})_{\rm Dirac}= -18.5 \,{\rm GeV^2}$. Notice that this gauge part is similar in both Majorana and
 Dirac cases, as can 
 be seen in the right plot of Fig.\ref{fig:mDRversusAnuandmnu}. In summary, 
 the radiative corrections from the massive neutrinos/sneutrinos in the Dirac
 case are phenomenologically irrelevant and therefore this case is totally 
 indistinguishable from the MSSM with 
 massless neutrinos.

%%%%%%%%%%%%%%%%%%%%%%%%%%%%%%%%%%%%%%%%%%%%%%%%%%%%%%%%%%%%%%%%%%%%%%%%%%%%%%

\subsection{Estimate of the one-loop corrections from neutrino/sneutrino
sector to \boldmath{$M_h$} within the MSSM-seesaw} 
\label{sec:Mh-MSSM-seesaw}

We recall that the anticipated LHC precision of the mass of an
  SM-like Higgs boson is $\sim 200 \mev$, and that at
the ILC an accuracy on the mass could
reach the $50 \mev$~level. These experimental precisions set the goal
for the theoretical accuracies.

As outlined in \refse{sec:FDconcept} the higher-order corrected light
MSSM Higgs-boson mass is obtained as a pole from \refeq{eq:proppole},
i.e.\ where $p^2 = \Mh^2$. 
A realistic evaluation requires to take into account all known higher-order
corrections to the renormalized Higgs-boson self-energies~\cite{reviews}. 
In order to simplify our analysis, but to maintain the high accuracy we
follow a slightly different strategy. 
For a given set of SUSY parameters we first calculate $\Mh$ and $\MH$ in the 
MSSM with the
help of \fh~\cite{feynhiggs,mhiggslong,mhiggsAEC,mhcMSSMlong}. In this
way all relevant known higher-order corrections are included, but no
$\nu/\Snu$ contributions are taken into account yet. This
corresponds to a `diagonalization' of the $\cp$-even Higgs sector in the
MSSM without heavy Majorana (s)neutrinos.
In a second step we search for the poles of 
\begin{equation}
\left[p^2 - \Mh^2 + \hSi_{hh}^{\nu/\Snu}(\Mh^2) \right]
\left[p^2 - \MH^2 + \hSi_{HH}^{\nu/\Snu}(\Mh^2) \right] -
\left[\hSi_{hH}^{\nu/\Snu}(\Mh^2)\right]^2 = 0 ~,
\label{eq:proppole-new}
\end{equation}
where, $\hSi_{hh,HH,hH}^{\nu/\Snu}$ denote the full corrections to the
renormalized Higgs-boson self-energies from the $\nu/\Snu$ sector,
obtained in the \mDRbar\ scheme as described in the present work. 
The pole, the light Higgs mass including the $\nu/\Snu$
corrections (i.e.\ in the MSSM-seesaw model), 
is denoted by $\Mh^{\nu/\Snu}$.
This `re-diagonalization' now effectively takes into account the full
result of the MSSM-seesaw. 
The momentum in the self-energies is fixed to the value $\Mh$ as
obtained with \fh, since it is expected that the new contributions only
give a relatively small correction to this $\Mh$. In a more elaborate
analysis the renormalized self-energies should be evaluated with
free~$p^2$. However, we expect only a very minor effect from fixing the
external momentum to this value. In the near future the results of the new
neutrino/sneutrino corrections will be implemented into the code \fh.

%%%%%%%%%%%%%%%%%%%%%%%%%% F I G U R E %%%%%%%%%%%%%%%%%%%%%%%%%%%%%%%%%%%%
\begin{figure}[h!]
   \begin{center} 
     \begin{tabular}{cc} \hspace*{-12mm}
  	\psfig{file=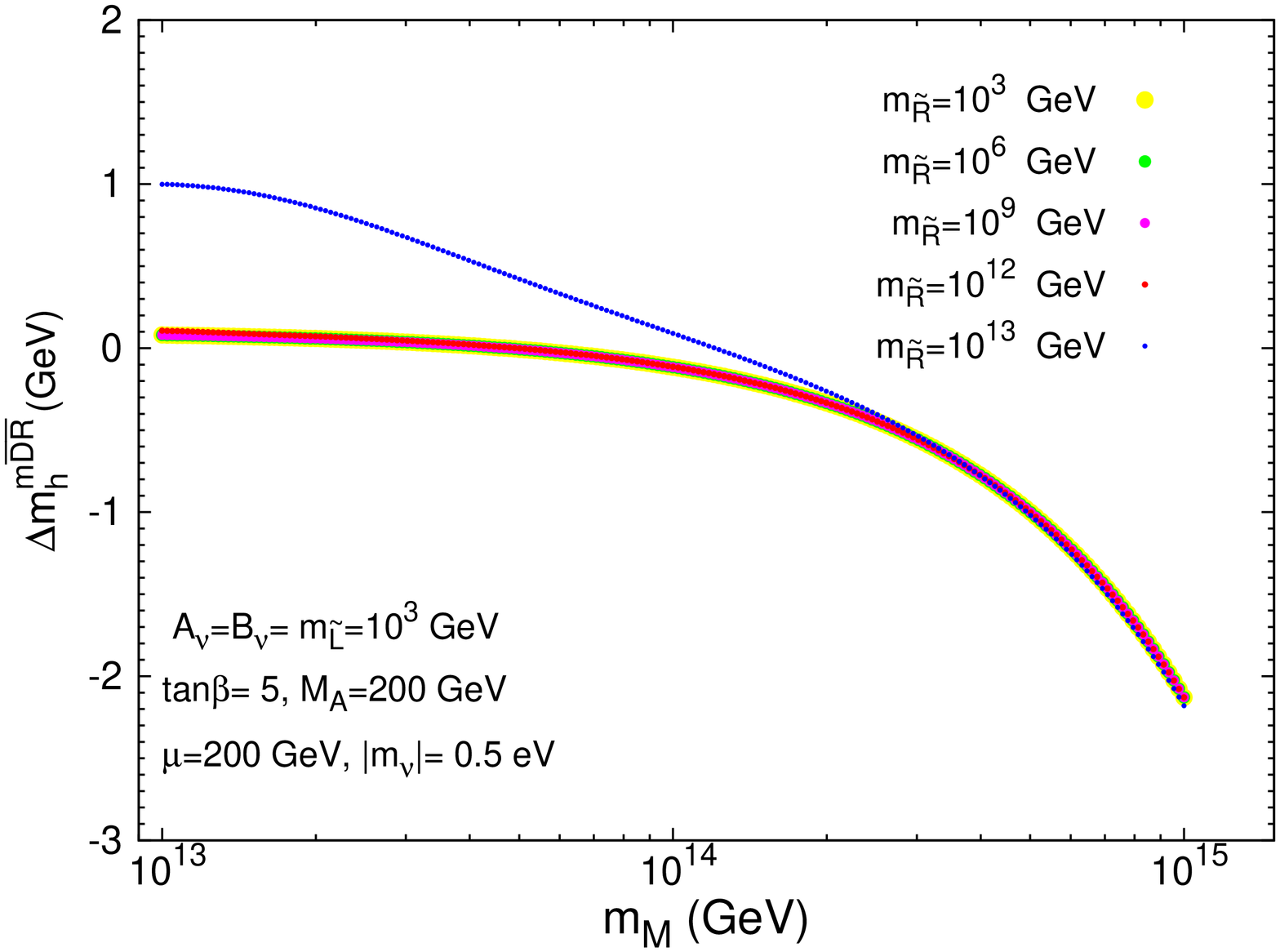,width=60mm,angle=270,clip=} 
	&
        \psfig{file=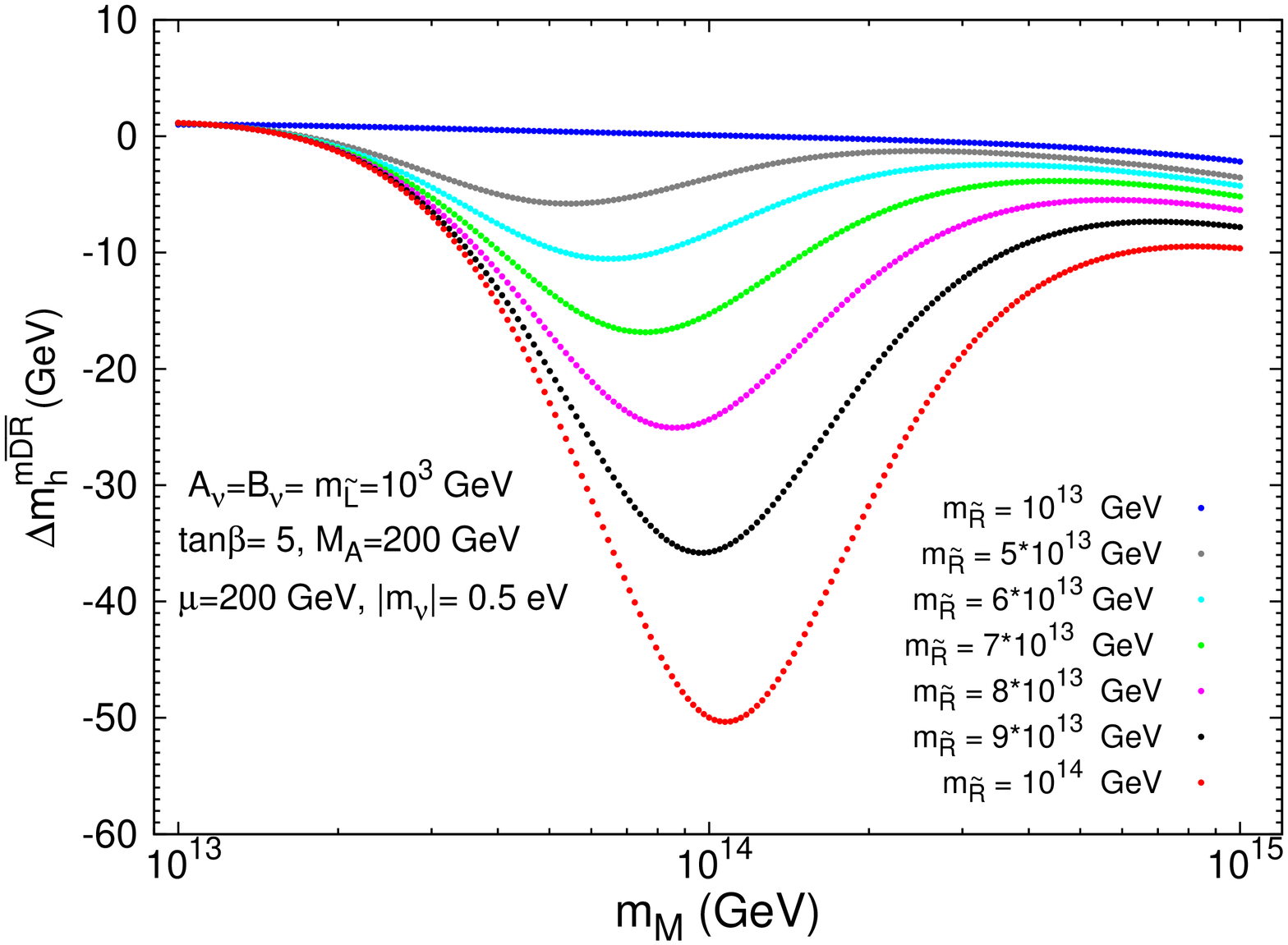,width=60mm,angle=270,clip=} 	  
       \end{tabular}
\caption{One-loop corrections to the lightest Higgs boson mass from the
neutrino/sneutrino sector as a function of the heavy Majorana mass for various 
choices
of the soft mass ${\mR}$. Left panel: ${\mR} < 10^{13} \gev$. 
Right panel:$10^{13}\,\,{\rm GeV}<{\mR}<10^{14}\,\,{\rm GeV}$.} 
     \label{fig:mhcorrectionsmR} 
   \end{center}
 \end{figure}
%%%%%%%%%%%%%%%%%%%%%%%%%%%%%%%%%%%%%%%%%%%%%%%%%%%%%%%%%%%%%%%%%%%%%%%%%%

\begin{figure}[h!]
   \begin{center} 
     \begin{tabular}{cc} \hspace*{-12mm}
  	\psfig{file=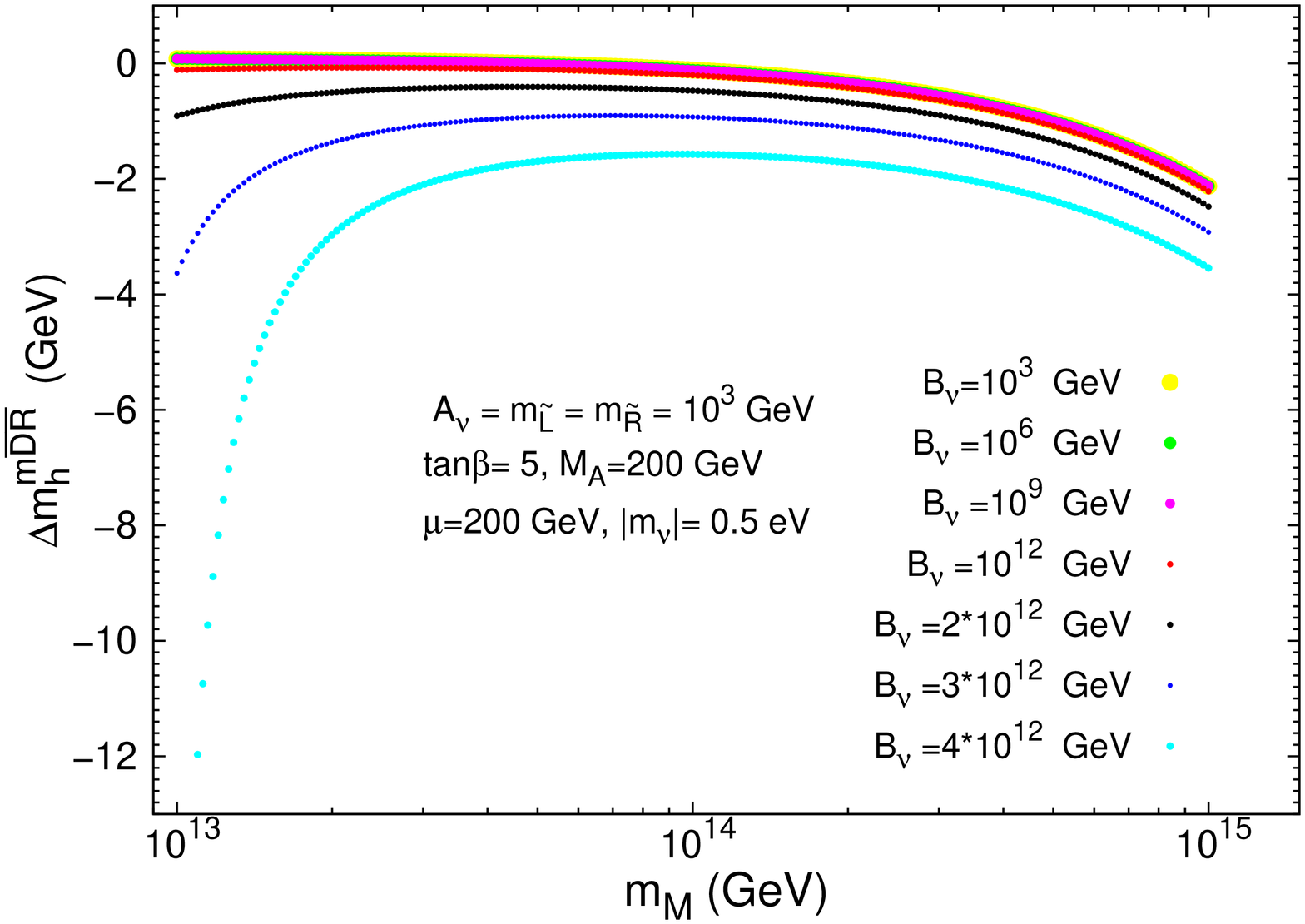,width=60mm,angle=270,clip=} 
&
        \psfig{file=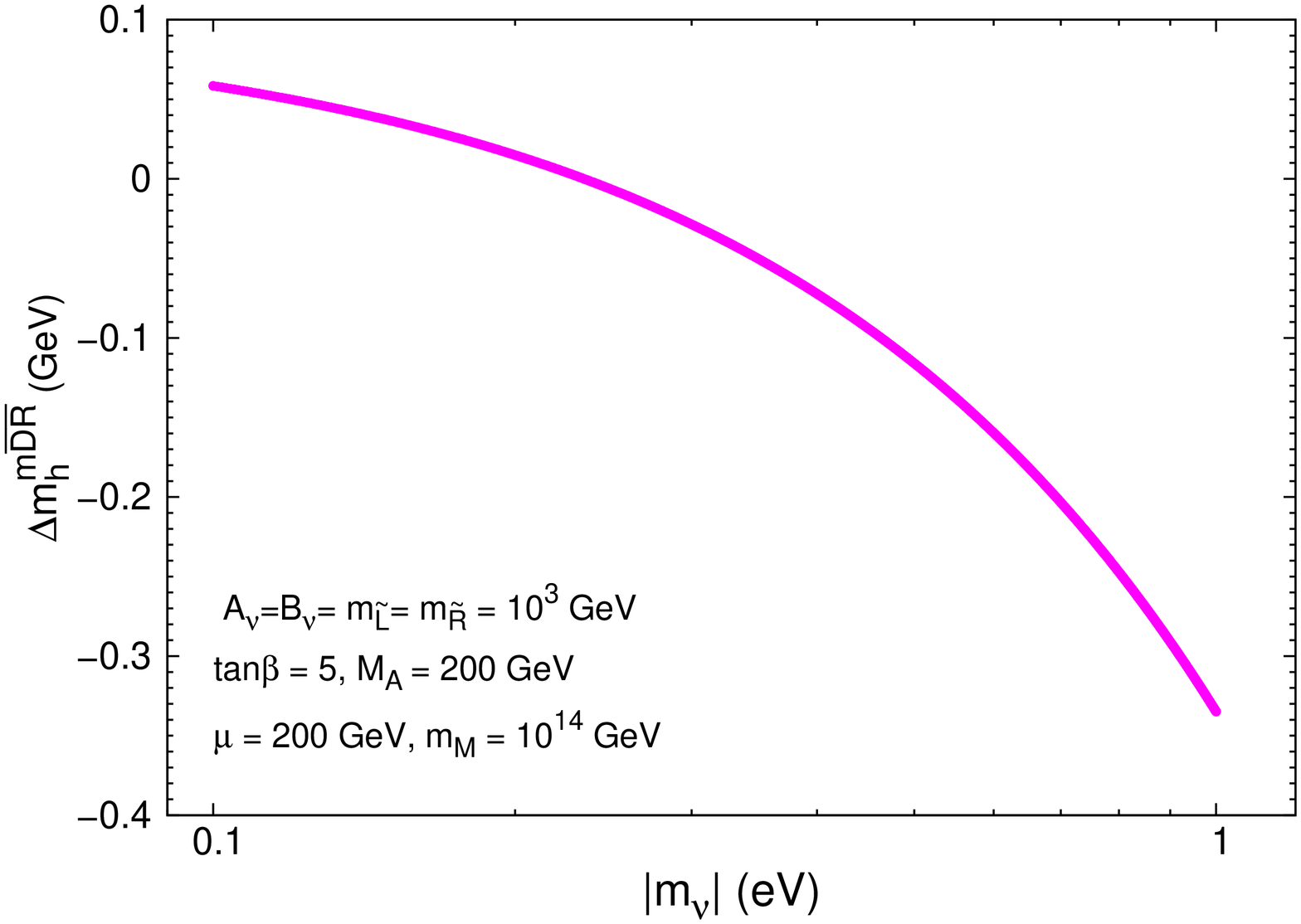,width=60mm,angle=270,clip=}
      \end{tabular}
\caption{Left panel: One-loop corrections to the lightest Higgs boson mass from the
neutrino/sneutrino sector as a function of the heavy Majorana mass, $\mM$, for various 
choices
of the soft $B$-parameter, $10^{3}\,\,{\rm GeV}<\Bnu<4 \times 10^{12}\,\,{\rm GeV}$. 
Right panel: Dependence of the Higgs mass corrections with the lightest
neutrino mass, $|\mnu|$.}
     \label{fig:mhcorrectionsBnumnu} 
   \end{center}
 \end{figure}

\begin{figure}[h!]
   \begin{center} 
     \begin{tabular}{c} \hspace*{-12mm}
  	\psfig{file=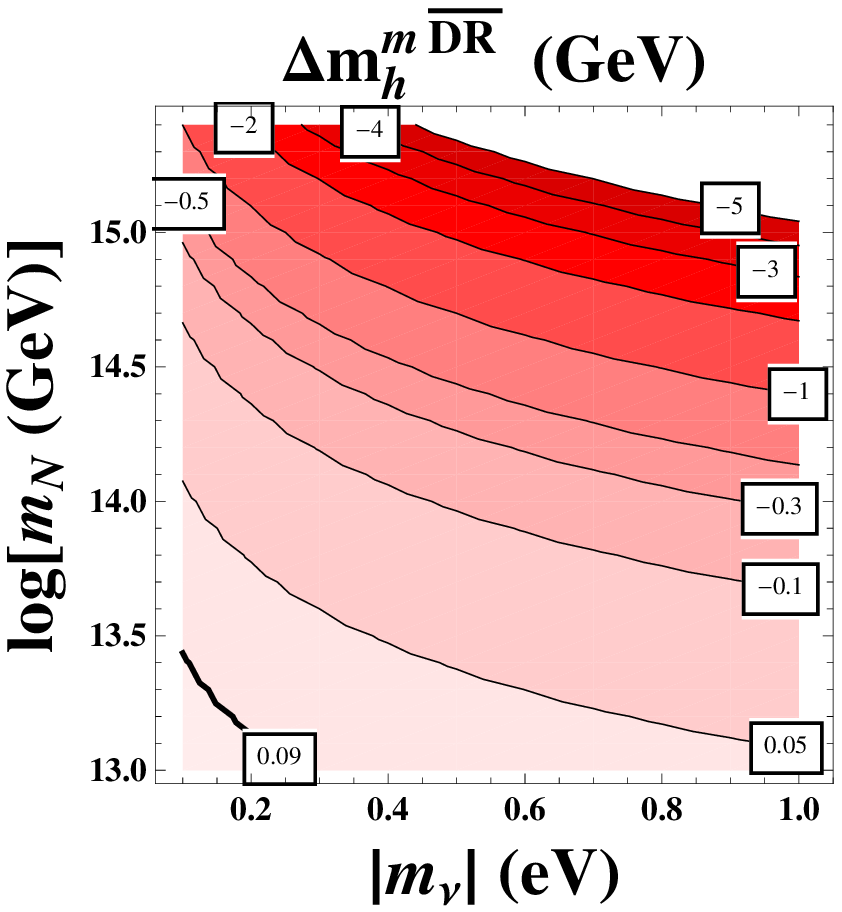,width=90mm,clip=}   
       \end{tabular}
     \caption{Contour-lines for the Higgs mass corrections from the
     neutrino/sneutrino sector as a function of the physical  
     Majorana neutrino masses, light $|\mnu|$ and heavy $m_N$. The other parameters are 
     fixed to: $A_\nu=\Bnu=m_{\tilde L}=\mR=
     10^3 \gev$, $\tb=5$, $M_A=200 \gev$ and $\mu=200 \gev$.}  
   \label{masscontours1} 
   \end{center}
 \end{figure}

\begin{figure}[h!]
   \begin{center} 
     \begin{tabular}{c} \hspace*{-12mm}
  	\psfig{file=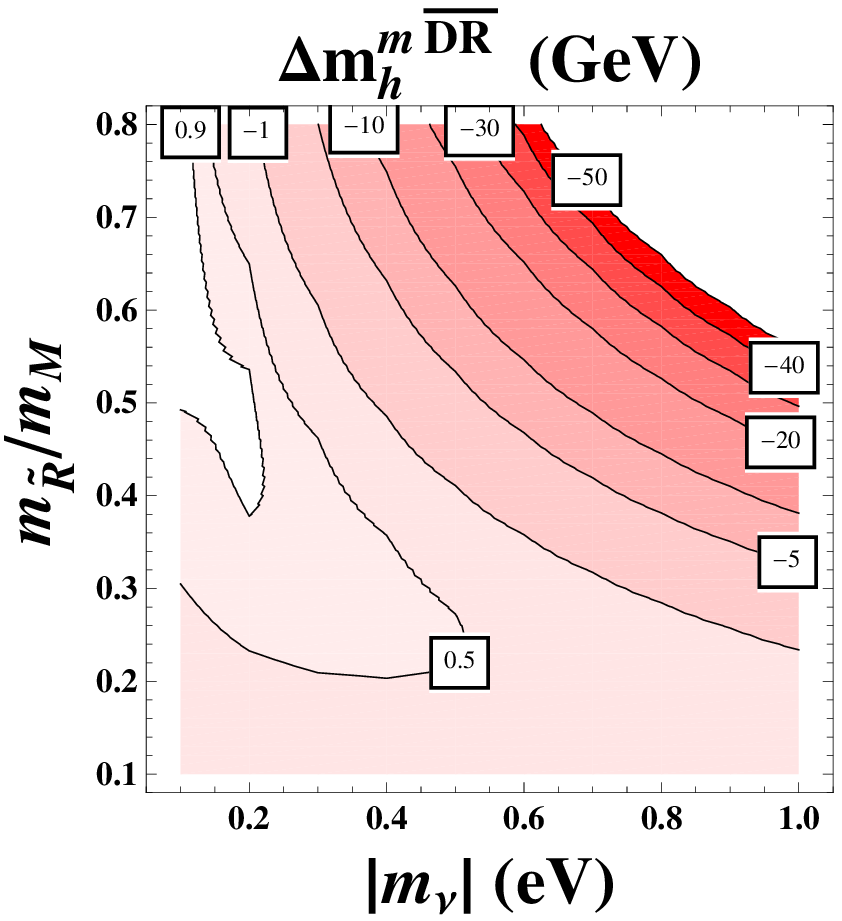,width=90mm,clip=}   
       \end{tabular}
     \caption{Contour-lines for the Higgs mass corrections from the
     neutrino/sneutrino sector as a function of the ratio 
     $\mR/\mM$ and the lightest Majorana neutrino mass $|\mnu|$. 
     The other parameters are fixed to: $\mM= 10^{14} \gev$, $A_\nu=\Bnu=m_{\tilde L}=
     10^3 \gev$, $\tb=5$, $M_A=200 \gev$ and $\mu=200 \gev$}  
   \label{masscontours_largemR} 
   \end{center}
 \end{figure}

The numerical results for $\Delta m_h^{\mDRbar} := \Mh^{\nu/\Snu} - \Mh$ 
are summarized in
Figs.~\ref{fig:mhcorrectionsmR} through~\ref{masscontours_largemR}. 
 We have chosen here to explore the Higgs mass predictions as a function 
of just the most relevant model parameters which, according to 
our previous exhaustive analysis of the renormalized Higgs-boson 
self-energies, are going to
provide the most interesting/sizeable corrections. These
are: the Majorana mass $\mM$ (or, equivalently, the heaviest physical
Majorana neutrino 
mass $m_N$), the soft SUSY breaking parameters $\mR$ and $\Bnu$ 
and the lightest physical Majorana neutrino
mass $\mnu$. As for the numerical values of these relevant parameters, we
focus here in the following intervals: 
$10^{13}\,\,{\rm GeV} \leq \mM \leq 10^{15}\,\,{\rm GeV}$, 
$0.1\,\,{\rm eV} \leq |\mnu|\leq  1 \,\, {\rm eV}$,
$10^{3}\,\,{\rm GeV} \leq \mR \leq \mM$ and
$10^{3}\,\,{\rm GeV} \leq B_{\nu} \leq 4 \times 10^{12}\,\,{\rm GeV}$.
For the remaining model parameters, $\tb$, $M_A$, $\mu$, $m_{\tilde L}$ and $A_\nu$,  
we choose here the same 
reference values as in the previous subsection. The corresponding predictions
for other choices of the parameters can be easily inferred from our previous results of the
renormalized self-energies.

%%%%%%%%%%%%%%%%%%%%%%%%%% F I G U R E %%%%%%%%%%%%%%%%%%%%%%%%%%%%%%%%%%%%

%%%%%%%%%%%%%%%%%%%%%%%%%%%%%%%%%%%%%%%%%%%%%%%%%%%%%%%%%%%%%%%%%%%%%%% 
In Fig.~\ref{fig:mhcorrectionsmR} we show the predictions for 
$\Delta m_h^{\mDRbar}$
as a function of the Majorana mass $\mM$, for several input $\mR$
values. As a general feature, the Higgs mass corrections for the reference
parameter values in the left plot are positive and below 0.1 GeV  if
$\mM \lsim 5 \times 10^{13} \gev$ and $\mR< 10^{12} \gev$. For 
larger Majorana mass values,  the corrections get
negative and grow up to a few GeV. For instance,  $\Delta m_h^{\mDRbar}= -2.15$
GeV for $\mM= 10^{15} \gev$. 
The results in the right plot show that for larger
values of the soft mass, $\mR\gsim 10^{13} \gev$ the Higgs mass
corrections are negative and can be sizeable, a few tens of GeV, reaching their 
maximum values at $\mR \simeq \mM$. For instance, for 
$\mR=\mM= 10 ^{14} \gev$ we get a very large correction, 
$\Delta m_h^{\mDRbar}=-50 \gev$. This last large negative value 
is in agreement with the prediction in \citere{Cao:2004hs} for the same
corresponding input values of the parameters in their split SUSY scenario.
It should be noticed that, in the case of such large corrections our approximation of
  \refeq{eq:proppole-new} is not accurate enough to obtain a precise
  result for $\Mh^{\nu/\Snu}$. However, our method still yields an
  indication of the size of the corrections from the $\nu/\Snu$ sector
  to $\Mh$.

The behavior of  
the Higgs mass corrections as a function of the $\Bnu$ parameter is displayed
in the left plot of Fig.~\ref{fig:mhcorrectionsBnumnu}. Again,  
$\Delta m_h^{\mDRbar}$ gets negative and large for large $\Bnu$, reaching the 
maximum size at $\Bnu \simeq \mM$. For instance, for the input model parameters
in this plot, and $\Bnu=4 \times 10^{12} \gev$, $\mM= 10 ^{13} \gev$, we find
$\Delta m_h^{\mDRbar}=-21 \gev$.

%%%%%%%%%%%%%%%%%%%%%%%%%% F I G U R E %%%%%%%%%%%%%%%%%%%%%%%%%%%%%%%%%%%% 

%%%%%%%%%%%%%%%%%%%%%%%%%%%%%%%%%%%%%%%%%%%%%%%%%%%%%%%%%%%%%%%%%%%%%%%%%%%%%%%

The dependence of the mass corrections with the light Majorana neutrino mass 
is illustrated in the right panel of Fig.~\ref{fig:mhcorrectionsBnumnu}. The
size of the corrections grow with $|\mnu|$, as expected, and can be either
positive in the low region, close to $|\mnu|\sim 0.1$ eV, or negative 
in the high region, close to  $|\mnu|\sim 1$ eV.

These same interesting features of the Higgs mass corrections in terms of the two relevant
physical Majorana neutrino masses, $m_N$ and $\mnu$, are summarized in the 
contour-plot in \reffi{masscontours1}. Here we have fixed all 
the soft parameters, including $\mR$, to be at 1 TeV. 
The contour-lines for fixed  
$\Delta m_h^{\mDRbar}$ range from positive values around $0.1 \gev$ in the left
lower corner of the plot, corresponding to neutrino mass values of
$|\mnu|= 0.1-0.3$ eV and
$m_N= 3 \times 10^{13} \gev$, up to negative values around $-5 \gev$ in the right
upper corner of the plot, corresponding to, for instance, $|\mnu|=1$ eV and $m_N=10^{15} \gev$. 
It should be noticed that the contour-line with fixed $\Delta m_h^{\mDRbar}=0.09$ (drawn
with a wider black line in this plot)
coincides with the prediction for the case where just the gauge part in the
self-energies have been included. This means that 'the distance' of any other
contour-line respect to this line represents the difference in the
radiative corrections respect to the MSSM prediction.

%%%%%%%%%%%%%%%%%%%%%%%%%% F I G U R E %%%%%%%%%%%%%%%%%%%%%%%%%%%%%%%%%%%% 

%%%%%%%%%%%%%%%%%%%%%%%%%%%%%%%%%%%%%%%%%%%%%%%%%%%%%%%%%%%%%%%%%%%%%%%%%%%%%%%

We plot in Fig.~\ref{masscontours_largemR}, the contour-lines for fixed
$\Delta m_h^{\mDRbar}$ in the less conservative case where $\mR$ is
close to $\mM$. These are displayed as a function of $|\mnu|$ and the ratio
$\mR/\mM$. $\mM$ is fixed here to the reference value, $\mM=10^{14}$
GeV. For the interval studied here, we see again that the radiative corrections 
can be negative and as large as tens of GeV in the upper right corner of the
plot. For instance, $\Delta m_h^{\mDRbar}=-30 \gev$ for $\mM=10^{14}$
GeV, $|\mnu|= 0.6$ eV and 
$\mR/\mM= 0.7$.

Finally, given our previous simple analytical results of
the renormalized self-energies in the 
seesaw limit, see \refeqs{mD0}, (\ref{mD2}), 
it is interesting to derive a simple analytical expression 
for the contribution of the heavy
neutrino-sneutrino sector to the one-loop radiatively corrected Higgs mass in
the limit of large $m_M$. 
Neglecting in \refeq{eq:proppole-new} the contributions from 
$\hSi_{HH}^{\nu/\Snu}$ and $\hSi_{hH}^{\nu/\Snu}$ one finds,
\begin{eqnarray} 
\Delta  m_h^{\mDRbar} \simeq  - \frac{\hSi_{hh}^{\nu/\Snu}(M_h^2)}{2M_h} 
\label{dmhapprox}
\end{eqnarray} 
where 
$\hSi_{hh}^{\nu/\Snu}$ denotes the full corrections to the
renormalized Higgs-boson self-energy from the $\nu/\Snu$ sector and 
obtained in the \mDRbar\ scheme as described in the present work.
We have found that this yields a very good approximation to the full
  result, i.e.\ the pole obtained from \refeq{eq:proppole-new}.
In a next step in the above expression $\hSi_{hh}^{\nu/\Snu}$ has to be 
replaced
by our simplified results in the large $m_M$ limit, namely, those in 
\refeqs{mD0b} and (\ref{mD2b}), 
providing the leading ${\cal O}(m_D^0)$ and ${\cal O}(m_D^2)$
contributions. We have compared numerically this approximate  
$\De m_h^{\mDRbar}$ with our full numerical results for large $m_M$ in 
\reffi{fig:mhcorrectionsmR}, and found
very good agreement, whenever the soft SUSY masses are well below $m_M$. In
fact, the behaviour with $m_M$ of this approximate formula is indistinguishable
from the lower line in the left plot of \reffi{fig:mhcorrectionsmR}.
 
We therefore conclude that the use of the previous \refeq{dmhapprox}
with 
\begin{eqnarray} 
\hSi_{hh}^{\nu/\Snu}(M_h^2) \simeq  
\KL \hSi_{hh}^{\mDRbar}(M_h^2) \KR_{\mD^0} + \KL \hSi_{hh}^{\mDRbar}(M_h^2) \KR_{\mD^2}
\label{renselfapprox}
\end{eqnarray} 
as given in \refeqs{mD0b} and (\ref{mD2b}), 
respectively, provides an excellent approximation to the full result for 
large Majorana mass values, 
$10^{13} \gev < m_M < 10^{15} \gev$ and soft masses well
below $m_M$, $\msusy \lsim 10^4 \gev$. Furthermore, the above simple
approximation can also be 
used for estimates of the differences in the mass correction when
applied to the $\DRbar$ scheme versus the $\mDRbar$ scheme for different 
choices of the $\mu_{\DRbar}$
scale. For instance, for $m_M=10^{14} \gev$ and the other parameters set to our
reference values as defined in section \ref{sec:analysis}, we got small
differences of $|(\Delta  m_h^{\DRbar}-\Delta  m_h^{\mDRbar})/ M_h|< 1 \%$ for 
$0.1<\mu_{\DRbar} / m_M<1$.

%%%%%%%%%%%%%%%%%%%%%%%%%%%%%%%%%%%%%%%%%%%%%%%%%%%%%%%%%%%%%%%%%%%%%%%%%%%%%%
%%%%%%%%%%%%%%%%%%%%%%%%%%%%%%%%%%%%%%%%%%%%%%%%%%%%%%%%%%%%%%%%%%%%%%%%%%%%%%

\clearpage

\section{Conclusions}
In this paper we have presented the one-loop radiative corrections to the
renormalized $\cp$-even Higgs boson self-energies and to the lightest Higgs boson
mass from the one-generation 
neutrino-sneutrino sector within the context of the MSSM-seesaw. The
most interesting features in this scenario are that the neutrinos,
differently to other fermions,  
 are assumed to be
Majorana particles, and that the origin for the light neutrino mass  
is not as for the other fermions either, but it is instead generated by means 
of the seesaw mechanism with the addition of heavy right handed neutrinos 
with a large Majorana mass.

As a first useful result, we have included here the complete set of Feynman
rules in this MSSM-seesaw context that are relevant for this work, which to our
knowledge are not available in the literature. These include all
vertices for the interactions among the Higgs sector and the
neutrinos/sneutrinos and for the $Z$ gauge boson and the neutrinos/sneutrinos.
These Feynman rules have been presented
in terms of all the physical masses and mixing angles of the 
particles involved, namely, the $\cp$-even Higgs bosons $h$ and $H$, the $\cp$-odd Higgs boson
$A$, the light and heavy
Majorana neutrinos $\nu$ and $N$, their SUSY partners ${\Snu}_{\pm}$, 
${\tilde N}_{\pm}$ and the neutral gauge boson $Z$.     
 
The computation presented here is a full one-loop 
Feynman diagrammatic one and does not make use of any of the 
approximations applied in the literature. In particular, we do not use
the mass insertion approximation 
for any of the involved soft mass parameters, nor we neglect the external momentum
in the self-energies, which we have found to be relevant for the final
computation of the Higgs mass corrections. We have presented our analytical 
results in
terms of the physical neutrinos, sneutrinos, $Z$, and Higgs bosons masses. In
addition we have analyzed the role played by the heavy Majorana mass scale
$\mM$, and emphasized the differences between the Majorana and Dirac neutrino
cases.

We have fully analyzed the behavior of the neutrino/sneutrino corrections to
the renormalized $\cp$-even Higgs 
self-energies with all the involved masses and parameters: 
$\mM$, $\tb$, $M_A$, $\mL$, $\mR$, $\Anu$, $\mnu$
and $\Bnu$. Our numerical study of the size of these corrections 
has been performed over a wide interval for all
these parameters, so that our conclusions can be considered as general.
From this exhaustive study we have concluded that the
most relevant parameters are $\mM$, $\mnu$, $\mR$ and $\Bnu$. In
particular, the Majorana
mass is by far the most crucial one. In general, we have found sizeable corrections 
to the
self-energies, indeed comparable or even larger than the other relevant 
one-loop corrections, as the ones from the MSSM top-stop sector, at the 
highest explored 
values of $\mM$, $\mnu$, $\mR$ and $\Bnu$. We have explained here the 
large size of these corrections in terms of the neutrino Yukawa couplings, 
which are
typically large, $\Ynu \sim {\cal O}(1)$ in these seesaw scenarios with heavy Majorana 
neutrinos.  For comparison, we have further included the predictions in two 
renormalization schemes, the on-shell and the $\DRbar$ schemes, where we have 
found
interesting differences. These differences have been analyzed and
explained with the 
help of simple formulas that are valid in the seesaw limit where $\mM$ is much
larger than all the other mass scales involved.  

The main conclusions on the corrections to the lightest Higgs boson mass are 
summarized in the contour-plots shown in Figs.~\ref{masscontours1} and 
~\ref{masscontours_largemR}. For the most conservative scenario of 
Fig.~\ref{masscontours1}, where all the soft mass parameters are at  
the TeV scale, the corrections are positive and smaller than 0.1 GeV if 
$10^{13}\,\,{\rm GeV}<\mM< 10^{14} \gev$ (or, equivalently, 
$10^{13}\,\,{\rm GeV}<m_N<10^{14} \gev$) and $0.1\,\,{\rm eV}<|\mnu|<1$ eV. 
For larger 
$\mM$
and/or $|\mnu|$ values the corrections change to negative sign and grow in size
with these two masses up to values of around $-5 \gev$ for 
$\mM= 10^{15} \gev$ and $|\mnu|=1$~eV. For the less conservative
scenario of Fig.~\ref{masscontours_largemR}, where the soft mass associated to
the right handed neutrino sector, $\mR$ is of the order of the Majorana
mass scale, we find very large negative corrections,  at the right 
upper corner of the plot, that is for large $\mM$ and $\mR$, of 
${\cal O}(10^{14}) \gev$, and $|\mnu|$ of  ${\cal O}(1)$ eV.
For instance, they are around $-30 \gev$ , for $\mM=10^{14} \gev$, 
$\mR/\mM =0.7$ and $|\mnu|=0.6$ eV.        
In view of the anticipated experimental precisions at the LHC and
  the ILC these corrections are very large and should be taken into
  account if the experimental data indicate the existence of Majorana
  (s)neutrinos. 

In summary, we conclude that the one-loop corrections from heavy Majorana 
neutrinos to the
Higgs boson masses are important in this MSSM-seesaw scenario, and overwhelm by
many orders of magnitude the corresponding corrections in the case
of Dirac massive neutrinos. These have also been estimated here and are 
extremely tiny, smaller than $10^{-22} \gev$. 

Finally, we briefly remark on the interesting and more formal
issue of decoupling/non-decoupling effects from the heavy Majorana
neutrinos/sneutrinos sector in the low energy MSSM Higgs boson physics. It is
clear that our results in the present paper, showing large one-loop
corrections $\Delta  m_h^{\mDRbar}$ to the $h$  boson mass for large $m_M$, 
suggest that there could be indeed non-decoupling effects from the heavy
particles in the low energy MSSM Higgs bosons physics. Particularly suggesting
are the numerical results shown in Figs.~12-15 where it is clearly manifested a
growing of $\Delta  m_h^{\mDRbar}$ with $m_M$. Also our simplified analytical
results for $\Delta  m_h^{\mDRbar}$ in \refeqs{mD0b}, (\ref{mD2b}), \refeqs{dmhapprox} and 
\refeqs{renselfapprox}  suggest a non-decoupling 
effect, since the mass correction does not vanish in the asymptotic limit $m_M
\to \infty$, even for $Y_\nu$ (or $m_D$) kept fixed. However, we believe that
one should not conclude on non-decoupling effects based just on the behaviour of
the Higgs mass corrections with $m_M$. It is well known that the mass itself is
not the proper physical observable to study the decoupling/non-decoupling issue.
A more proper tool for that study would be the use of Effective Field Theory
techniques, and more concretely the computation of the one-loop effective action
by integration in the path integral of the heavy degrees of freedom. An
expansion, valid to low external momenta, $p\ll m_M$, of the derived 1PI
renormalized Green functions with Higgs bosons in the external legs would 
provide the definite answer to the issue of 
decoupling/non-decoupling of the heavy $\nu_R$, ${\tilde \nu}_R$, degrees of
freedom in the low energy Higgs boson physics. Alternatively one could  
perform one-loop predictions within
the present MSSM-seesaw model for other more proper
observables for this issue like, for instance,  cross sections involving Higgs 
particles in the
external legs,  decay rates of Higgs bosons, etc. The behaviour of these kind of
radiative corrections at asymptotically large $m_M$ could also be conclusive 
on this issue. All these proposed studies 
are extremely interesting but are far beyond the scope of the present work.               

\label{sec:concl}

%%%%%%%%%%%%%%%%%%%%%%%%%%%%%%%%%%%%%%%%%%%%%%%%%%%%%%%%%%%%%%%%%%%%%%%%%%%%%%

\subsection*{Acknowledgements}

We thank M.~Hirsch and W.~Hollik for helpful discussions.  
The work of S.H. was partially supported by CICYT (grant FPA 2007--66387) and
by the Spanish Consolider-Ingenio 2010 Program under grant MultiDark
CSD2009-00064.
The work of M.H. and A.R.-S. was partially supported by CICYT (grants
FPA2006-05423 and FPA2009-09017)
and  the Comunidad de Madrid project HEPHACOS, S2009/ESP-1473.
A.R.-S. thanks the Spanish Ministry of Science and Education for her FPU 
fellowship
Ref. AP2006-02535.
The work of S.P. was supported by a \textit{Ram{\'o}n y Cajal} contract 
from MEC (Spain) (PDRYC-2006-000930) and partially
by CICYT (grants FPA2006-2315 and FPA2009-09638) and 
the Comunidad de Arag\'on project DCYT-DGA E24/2.
The work is also supported in part by  
the European Community's Marie-Curie Research
Training Network under contract MRTN-CT-2006-035505 and also 
by the Spanish Consolider-Ingenio 2010 Programme CPAN (CSD2007-00042).

%%%%%%%%%%%%%%%%%%%%%%%%%%%%%%%%%%%%%%%%%%%%%%%%%%%%%%%%%%%%%%%%%%%%%%%%%%%%%%%
%%%%%%%%%%%%%%%%%%%%%%%%%%%%%%%%%%%%%%%%%%%%%%%%%%%%%%%%%%%%%%%%%%%%%%%%%%%%%%%
\newpage

\begin{appendix}

\section*{Appendix A: New Feynman rules} 

In this appendix we collect the Feynman rules within the MSSM-seesaw 
that are relevant for the present work. These correspond to the 
interactions between the neutrinos and
sneutrinos with the MSSM Higgs bosons and between the neutrinos and
sneutrinos with the $Z$~gauge bosons. We write all the Feynman rules here 
in the  physical basis. Here $\cw=\cos\theta_W$.\\

\begin{table}[h!]
\begin{tabular}{ll}
\parbox[c]{1em}{\includegraphics{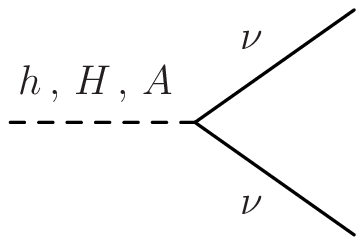}}
 &$i\frac{g}{2\MW} \mD \sin 2 \theta  \left(\frac{\Ca}{\Sbe},\frac{\Sa}{\Sbe},
-i\ga_5 \CTb\right)$\\
&\\
%&\\
\parbox[c]{1em}{\includegraphics{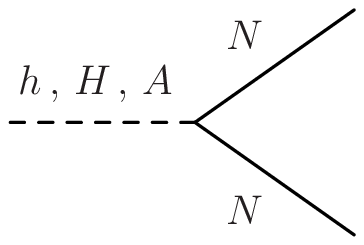}}
 & $-i\frac{g}{2\MW} \mD \sin 2 \theta  \left(\frac{\Ca}{\Sbe},\frac{\Sa}{\Sbe},
-i\ga_5 \CTb\right)$\\
&\\
%&\\
\parbox[c]{1em}{\includegraphics{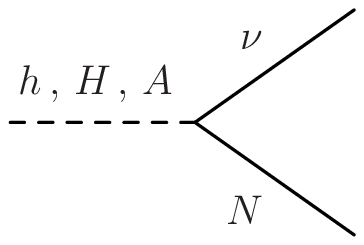}}
 & $-i\frac{g}{2\MW} \mM \sin \theta \cos\theta 
\left(\frac{\Ca}{\Sbe},\frac{\Sa}{\Sbe},
-i\ga_5 \CTb\right)$\\
&\\
%&\\
\parbox[c]{1em}{\includegraphics{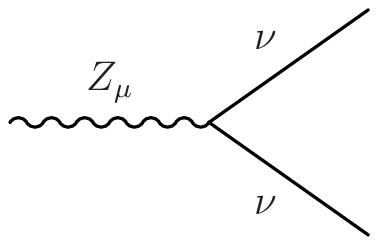}}
 & $\frac{ig}{2\cw}\cos^2 \theta \, \ga_{\mu}\ga_5$\\
&\\
%&\\
\parbox[c]{1em}{\includegraphics{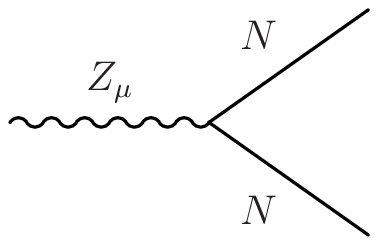}}
 & $\frac{ig}{2\cw}\sin^2 \theta \, \ga_{\mu}\ga_5$\\
&\\
{\includegraphics{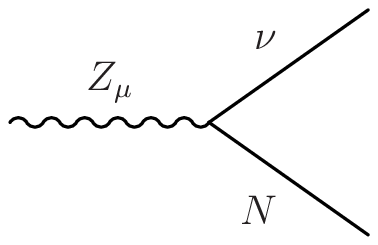}}
& 
\put(10,30){$\frac{ig}{2\cw}\sin \theta \cos \theta \, \ga_{\mu}\ga_5$}\\
&\\
\end{tabular}
\caption{Three-point couplings of two Majorana neutrinos to one MSSM Higgs boson 
and of two Majorana neutrinos to the 
$Z$ gauge boson.}
\end{table}

\begin{table}[p!]
\begin{tabular}{ll}
\parbox[c]{1em}{\includegraphics{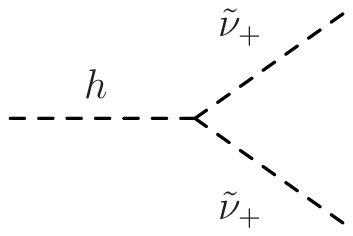}}
 & $i\frac{g}{4\cw\MW \Sbe}[-4 \cw\Ca\, \mD^2 
+2 \cw \Ca\, \mD (\Anu+ \mM + \mu \tana) \stwoTp$\vspace*{-0.6cm}\\
& $+ \frac{\MW^2}{\cw} \Sbe\,\Sab \,(1+ \ctwoTp)\,]$\\
&\\
\parbox[c]{1em}{\includegraphics{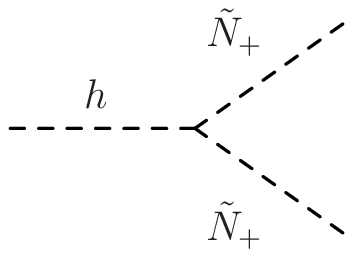}}& 
$-i\frac{g}{4\cw\MW \Sbe}[4 \cw\Ca\, \mD^2 
+2 \cw \Ca\, \mD (\Anu+ \mM + \mu \tana) \stwoTp$\vspace*{-0.6cm}\\
& $ - \frac{\MW^2}{\cw} \Sbe\,\Sab\, (1- \ctwoTp)\,]$\\
&\\
\parbox[c]{1em}{\includegraphics{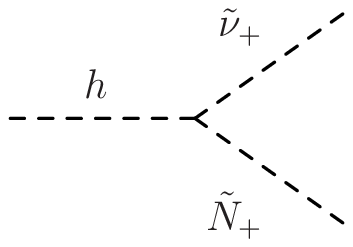}}
& $-i\frac{g}{2\cw\MW \Sbe}[\cw \Ca\, \mD (\Anu+\mM + \mu \tana) \ctwoTp
$\vspace*{-0.6cm}\\
& $-\frac{\MW^2}{\cw} \Sbe\,\Sab\, \cTp \sTp]$\\
&\\
\parbox[c]{1em}{\includegraphics{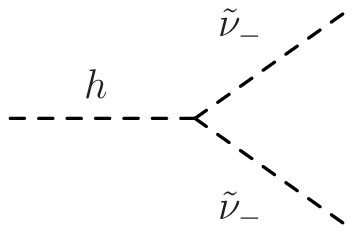}}
 & $i\frac{g}{4\cw\MW \Sbe}[-4\cw  \Ca\, \mD^2 
+2 \cw \Ca\, \mD (\Anu- \mM + \mu \tana) \stwoTm$\vspace*{-0.6cm}\\
& $ + \frac{\MW^2}{\cw} \Sbe\,\Sab\, (1+ \ctwoTm)\,]$\\
&\\
\parbox[c]{1em}{\includegraphics{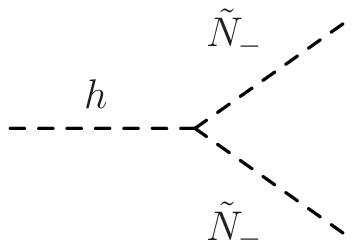}}& 
$-i\frac{g}{4\cw\MW \Sbe}[4 \cw\Ca\, \mD^2 
+2 \cw \Ca\, \mD (\Anu- \mM + \mu \tana) \stwoTm$\vspace*{-0.6cm}\\
& $ - \frac{ \MW^2}{\cw} \Sbe\,\Sab\, (1- \ctwoTm)\,]$\\
&\\
 \includegraphics{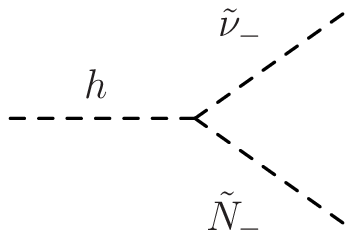}
 & \put(8,30){$-i\frac{g}{2\cw\MW \Sbe}[\cw\Ca \, \mD\, (\Anu-\mM 
+ \mu \tana) \ctwoTm$}\vspace*{-0.8cm}\\
 & $- \frac{\MW^2}{\cw} \Sbe\,\Sab\, \cTm \sTm]$\\
&\\
&\\\end{tabular}
\caption{Three-point couplings of two sneutrinos to the Higgs boson $h$. 
The corresponding couplings to the Higgs boson $H$ are obtained
from the ones here by replacing 
$\Ca \to \Sa\,, \Sa \to -\Ca\,, \Sab  \to -\Cab$. All the couplings not shown 
here vanish.}
\end{table}

\begin{table}[h!]
\begin{tabular}{ll}
\parbox[c]{1em}{\includegraphics{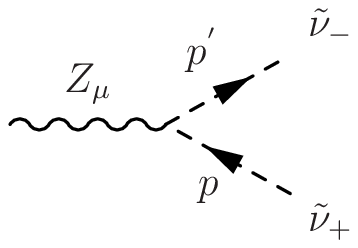}}&
$\frac{g}{2\cw} \cTp \cTm \,(p+p')_{\mu}$\\
&\\
\parbox[c]{1em}{\includegraphics{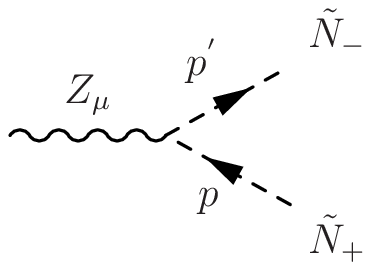}}& 
$\frac{g}{2\cw} \sTp \sTm \,(p+p')_{\mu}$\\
&\\
{\includegraphics{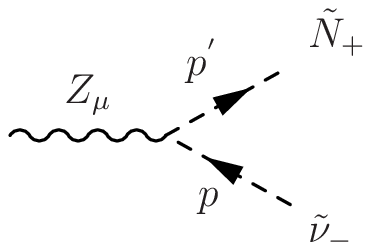}}&
\put(8,35){$\frac{g}{2\cw} \sTp \cTm \,(p+p')_{\mu}$}\\
&\\
\includegraphics{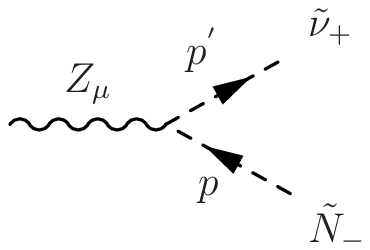}&
\put(8,35){$\frac{g}{2\cw} \cTp \sTm \,(p+p')_{\mu}$}\\
\end{tabular}
\caption{Three-point couplings of two sneutrinos to the $Z$ gauge boson. 
All the couplings not shown here
vanish.}
\end{table}

\begin{table}[p]%[h!]
\begin{tabular}{ll}
\parbox[c]{1em}{\includegraphics{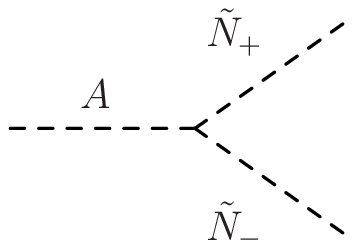}}& 
$i\frac{g}{2\MW} \CTb \,\mD [(\Anu+ \mu \tb) \STdifmp + \mM \STsummp]$\\
&\\
\parbox[c]{1em}{\includegraphics{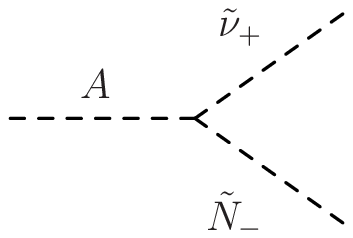}}&
$i\frac{g}{2\MW} \CTb \,\mD [-(\Anu+ \mu \tb)\CTdifmp + \mM \CTsummp]$\\
&\\
\includegraphics{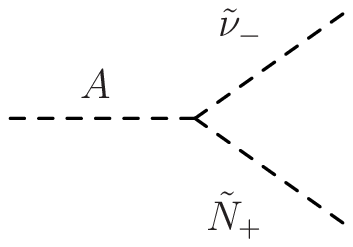}
&
\put(8,30){$i\frac{g}{2\MW} \CTb\, \mD 
           [(\Anu+ \mu \tb)\CTdifmp + \mM\CTsummp]$}\\
&\\
{\includegraphics{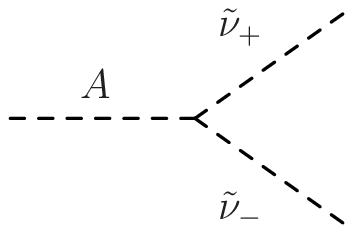}}&
\put(8,30){$i\frac{g}{2\MW} \CTb \, \mD 
           [(\Anu+ \mu \tb) \STdifmp - \mM \STsummp]$}\\	   
\end{tabular}
\caption{Three-point couplings of two sneutrinos to the Higgs boson $A$. 
All the couplings not shown here
vanish.}
\end{table}

 \begin{table}[p!]
\begin{tabular}{ll}
\parbox[c]{1em}{\includegraphics{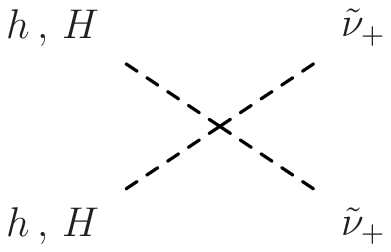}}& 
$i\frac{g^2}{8\cw^2\MW^2 \sin^2\be}[4 (-\CQa,\SQa) \cw^2{\mD^2} 
+ \CZa \MW^2 \SQb (1+\ctwoTp)]$\\
&\\
\parbox[c]{1em}{\includegraphics{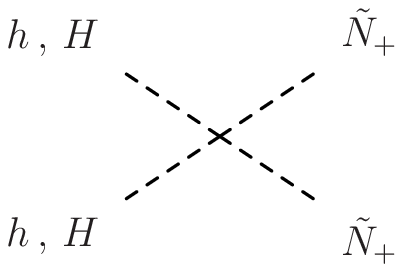}}&
$-i\frac{g^2}{8\cw^2\MW^2 \sin^2\be}[4 (\CQa,\SQa) \cw^2{\mD^2} 
(-,+) \CZa \MW^2 \SQb (1-\ctwoTp)]$\\
&\\
\parbox[c]{1em}{\includegraphics{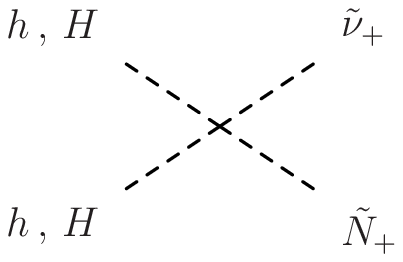}}&
$(+,-) \,i\frac{g^2}{4 \cw^2}\CZa\, \cTp\, \sTp$\\
&\\
\parbox[c]{1em}{\includegraphics{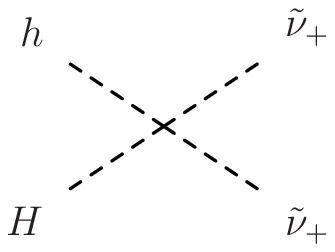}}&
{$i\frac{g^2}{8\cw^2\MW^2 \sin^2\be}\SZa\, 
[-2\cw^2{\mD^2} + \MW^2 \SQb\,(1+\ctwoTp)]$}\\
&\\
{\includegraphics{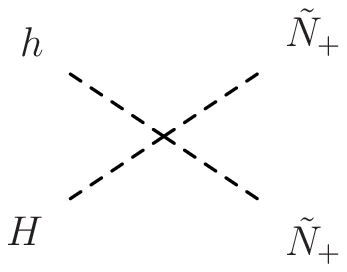}}&
\put(5,40){$-i\frac{g^2}{8\cw^2\MW^2 \sin^2\be}\,\SZa\, 
           [2\cw^2{\mD^2} - \MW^2 \SQb\, (1-\ctwoTp)]$}\\
&\\
\parbox[c]{1em}{\includegraphics{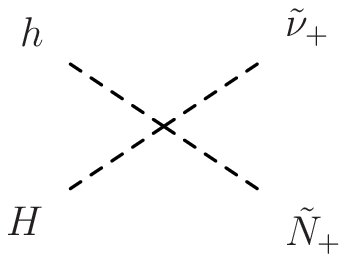}}&
{$i\frac{g^2}{4 \cw^2}\SZa \cTp \sTp$}\\
&\\
\end{tabular}
\caption{Four-point couplings of two sneutrinos to two $\cp$-even Higgs bosons. 
 The corresponding couplings for $\tilde{\nu}_-$ and $\tilde{N}_-$ can be
 obtained from these by replacing $\theta_+ \rightarrow \theta_-$.
 All the couplings not shown here vanish}
\end{table}

\begin{table}[p]%[t!]
\begin{tabular}{ll}
\parbox[c]{1em}{\includegraphics{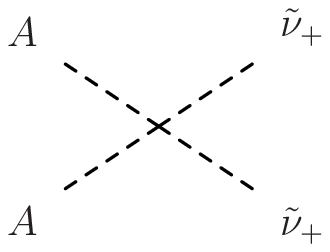}} &
$i\frac{g^2}{8\cw^2\MW^2 \sin^2\be}
[-4 \CQb \,\cw^2\,{\mD^2} + \CZb\, \MW^2\, \SQb\,(1+\ctwoTp)]$\\
&\\
\parbox[c]{1em}{\includegraphics{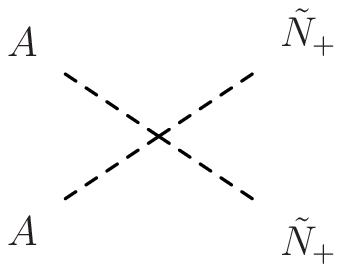}} &
$-i\frac{g^2}{8\cw^2\MW^2 \sin^2\be}
[4 \CQb \,\cw^2\,{\mD^2} - \CZb \, \MW^2\, \SQb\,(1-\ctwoTp)]$\\
&\\
\parbox[c]{1em}{\includegraphics{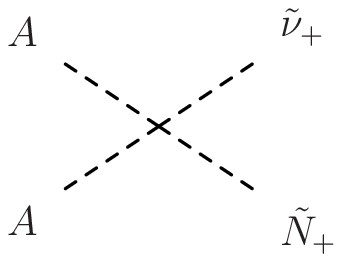}} &
{$i\frac{g^2}{4 \cw^2}\CZb\, \cTp\, \sTp$}\\
&\\
\parbox[c]{1em}{\includegraphics{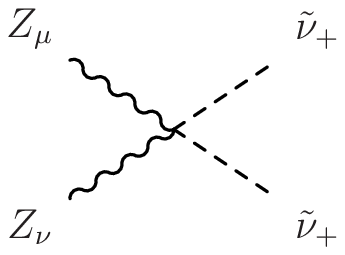}} & 
{$i\frac{g^2}{2\cw} \cQTp g_{\mu\nu}$}\\
&\\
\parbox[c]{1em}{\includegraphics{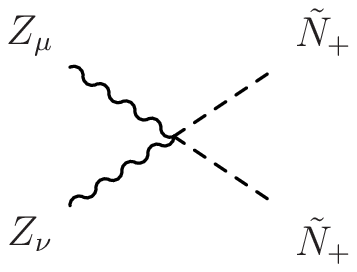}} & 
{$i\frac{g^2}{2\cw} \sQTp g_{\mu\nu}$}\\
&\\
\includegraphics{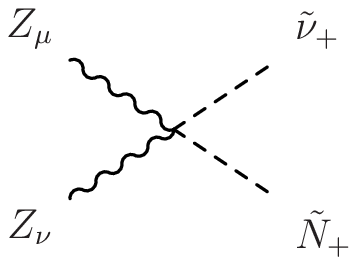} & 
\put(0,40){$i\frac{g^2}{2\cw} \cTp \sTp g_{\mu\nu}$}\\
\end{tabular}
\caption{Four-point couplings of two sneutrinos to two $\cp$-odd Higgs bosons 
and of two sneutrinos to two $Z$ gauge bosons. 
The corresponding couplings for $\tilde{\nu}_-$ and $\tilde{N}_-$ can be
 obtained from these by replacing $\theta_+ \rightarrow \theta_-$.
All the couplings not shown here
vanish.}
\end{table}

\cleardoublepage

%%%%%%%%%%%%%%%%%%%%%%%%%%%%%%%%%%%%%%%%%%%%%%%%%%%%%%%%%%%%%%%%%%%%%%%%%%%%%%%
%%%%%%%%%%%%%%%%%%%%%%%%%%%%%%%%%%%%%%%%%%%%%%%%%%%%%%%%%%%%%%%%%%%%%
\section*{Appendix B: Majorana case. One-loop neutrino/sneutrino corrections
  to the unrenormalized self-energies and tadpoles}

In this Appendix we collect all the analytical results for the neutrino and 
sneutrino  one-loop corrections to the Higgs boson tadpoles and 
unrenormalized self-energies, and to the $Z$~self-energies, within the
MSSM-seesaw.   
The contributions from neutrinos ($\nu$) and sneutrinos (${\Snu}$) 
are presented separately for clearness. Here $\cw=\cos\theta_W$.

\BEA
{T^{\nu}_{h}} &=&
\frac{g}{16 \cw \MZ \pi^2}\frac{\Ca \sin2\theta}{\Sbe} \mD 
(\mnu A_0[\mnu^2] - m_N A_0[m_N^2])\\
&&\non\\
{T^{\Snu}_{h}}  &=&
-\frac{g}{64 \cw \MZ \pi^2}\frac{1}{\Sbe}
(A_0[m_{\Snu_+}^2](\MZ^2 \cQTp \Sbe \Sab\non\\ 
&& +\mD \mu \Sa \stwoTp + 
 \mD \Ca (-2 \mD + (\Anu + \mM) \stwoTp))\non\\
&+& A_0[m_{\Snu_-}^2](\MZ^2 \cQTm \Sbe \Sab\non\\
&&+ \mD \mu \Sa \stwoTm - 
 \mD \Ca (2 \mD - (\Anu - \mM) \stwoTm))\non\\
&-& A_0[m_{\tilde N_+}^2](-\MZ^2 \Sbe \Sab  \sQTp\non\\
&& +2\mD \Ca (\mD + \frac{1}{2}(\Anu + \mM) \stwoTp) + \mD \mu \Sa \stwoTp)\non\\
&-& A_0[m_{\tilde N_-}^2](-\MZ^2 \Sbe \Sab  \sQTm\non\\
&& + 
   2\mD \Ca (\mD + \frac{1}{2}(\Anu - \mM) \stwoTm) + \mD \mu \Sa \stwoTm))\\
&&\non\\
{\se{hh}^{\nu}(p^2)}&=&-\frac{g^2}{64 \cw^2 \MZ^2 \pi^2}
\frac{\cos^2 \alpha \sin^2 2\theta}{\sin^2 \beta} 
\left[2 \mD^2 A_0[\mnu^2] + 
(2 \mD^2  + \mM^2) A_0[m_N^2]\right.\non\\
& +& 4 \mD^2 \mnu^2 B_0[p^2, \mnu^2, \mnu^2]
+\mM^2 (\mnu^2+\mnu m_N) B_0[p^2, \mnu^2, m_N^2]\non\\
& +& 4 \mD^2 m_N^2 B_0[p^2, m_N^2, m_N^2]\non\\
&+&\left.p^2 (2 \mD^2 B_1[p^2, \mnu^2, \mnu^2]+ \mM^2 B_1[p^2,  \mnu^2, m_N^2] + 
 2 \mD^2 B_1[p^2, m_N^2, m_N^2])\right]\\
&&\non\\
{\se{hh}^{\Snu}(p^2)} &=&\frac{g^2}{512 \cw^2 \MZ^2 \pi^2 \SQb}
[-4 A_0[m_{\Snu_+}^2] (-2 \mD^2 \CQa + \MZ^2 \SQb \CZa \cQTp) \non\\
&-&4 A_0[m_{\tilde N_+}^2] (-2 \mD^2 \CQa + \MZ^2 \SQb \CZa \sQTp)\non\\
&-&4 A_0[m_{\Snu_-}^2] (-2 \mD^2 \CQa + \MZ^2 \SQb \CZa \cQTm)\non\\
&-&4 A_0[m_{\tilde N_-}^2] (-2 \mD^2 \CQa + \MZ^2 \SQb \CZa \sQTm)]\non\\
&+& 2 B_0[p^2, m_{\tilde N_+}^2, m_{\Snu_+}^2]
(4 \mD^2 \cQtwoTp \CQa \,(\Anu +\mM+\mu \tan \alpha)^2\non\\
&& +\MZ^2 \Sbe \Sab
(\MZ^2 \Sbe \Sab \sQtwoTp\non\\
&&-2 \mD \Ca(\Anu + \mM +\mu \tan \alpha)\sin 4 \theta_{+})\non\\
&+& 2 B_0[p^2, m_{\tilde N_-}^2, m_{\Snu_-}^2]
(4 \mD^2 \cQtwoTm \CQa \,(\Anu -\mM+\mu \tan \alpha)^2\non\\
&& +\MZ^2 \Sbe \Sab
(\MZ^2 \Sbe \Sab \sQtwoTm\non\\
&&-2 \mD \Ca(\Anu - \mM +\mu \tan \alpha)\sin 4 \theta_{-})\non\\
&+& 4 B_0[p^2, m_{\tilde N_+}^2, m_{\tilde N_+}^2]
(\mD \Ca (2\mD +\stwoTp (\Anu + \mM + \mu \tan \alpha))\non\\
&&- \MZ^2 \Sbe \Sab \sQTp)^2\non\\
&+& 4 B_0[p^2, m_{\tilde N_-}^2, m_{\tilde N_-}^2]
(\mD \Ca (2\mD +\stwoTm (\Anu - \mM + \mu \tan \alpha))\non\\
&&- \MZ^2 \Sbe \Sab \sQTm)^2\non\\
&+& 4 B_0[p^2, m_{\Snu_+}^2, m_{\Snu_+}^2]
(\mD \Ca (-2\mD +\stwoTp (\Anu + \mM + \mu \tan \alpha))\non\\
&&- \MZ^2 \Sbe \Sab \cQTp)^2\non\\ 
&+& 4 B_0[p^2, m_{\Snu_-}^2, m_{\Snu_-}^2]
(\mD \Ca (-2\mD +\stwoTm (\Anu - \mM + \mu \tan \alpha))\non\\
&&- \MZ^2 \Sbe \Sab \cQTm)^2]
\EEA

The corresponding results for the tadpole $T_{H}$, and the unrenormalized 
self-energy $\se{HH}$ are obtained from the above formulas by replacing $\Ca \to \Sa\,, \Sa \to -\Ca\,, \Sab  \to -\Cab$. 

\BEA
{\se{hH}^{\nu}(p^2)} &=&-\frac{g^2}{128 \cw^2 \MZ^2 \pi^2}
\frac{\SZa \sin^2 2\theta}{\sin^2 \beta} 
\left[2 \mD^2 A_0[\mnu^2] + 
(2 \mD^2  + \mM^2) A_0[m_N^2]\right.\non\\
& +& 4 \mD^2 \mnu^2 B_0[p^2, \mnu^2, \mnu^2]
+\mM^2 (\mnu^2+\mnu m_N) B_0[p^2, \mnu^2, m_N^2]\non\\
& +& 4 \mD^2 m_N^2 B_0[p^2, m_N^2, m_N^2]]\non\\
&+&\left.p^2 (2 \mD^2 B_1[p^2, \mnu^2, \mnu^2]+ \mM^2 B_1[p^2,  \mnu^2, m_N^2] + 
 2 \mD^2 B_1[p^2, m_N^2, m_N^2])\right]\\
&&\non\\
{\se{hH}^{\Snu}(p^2)} &=&
\frac{g^2}{512 \cw^2 \MZ^2 \pi^2 \SQb}
[4 A_0[m_{\Snu_+}^2] \SZa (\mD^2- \MZ^2 \SQb \cQTp)\non\\
&+& 4 A_0[m_{\tilde N_+}^2] \SZa (\mD^2- \MZ^2 \SQb \sQTp)\non\\
&+&4 A_0[m_{\Snu_-}^2] \SZa (\mD^2 - \MZ^2 \SQb \cQTm)\non\\
&+&4 A_0[m_{\tilde N_-}^2] \SZa (\mD^2 - \MZ^2 \SQb \sQTm)\non\\
&+& 2 B_0[p^2, m_{\tilde N_+}^2, m_{\Snu_+}^2] \times \non\\
&\quad& (2 \mD^2 \cQtwoTp (-2 (\Anu + \mM) \mu \CZa + 
((\Anu + \mM)^2 - \mu^2) \SZa)\non\\
&& +\MZ^2 \Sbe (-\MZ^2 \Sbe \Sab \Cab \sQtwoTp\non\\
&&+ \mD ((\Anu + \mM) \cos(2\alpha+\beta) +\mu \sin(2\alpha+\beta))
\sin 4 \theta_{+}))\non\\
&+& 2 B_0[p^2, m_{\tilde N_-}^2, m_{\Snu_-}^2] \times \non\\
&\quad& (2 \mD^2 \cQtwoTm (-2 (\Anu - \mM) \mu \CZa + 
((\Anu - \mM)^2 - \mu^2) \SZa)\non\\
&& +\MZ^2 \Sbe (-\MZ^2 \Sbe \Sab \Cab \sQtwoTm\non\\
&&+ \mD ((\Anu - \mM) \cos(2\alpha+\beta) +\mu \sin(2\alpha+\beta))
\sin 4 \theta_{-}))\non\\
&+& 2 B_0[p^2, m_{\tilde N_+}^2, m_{\tilde N_+}^2]
(\mD^2 
(-2  \mu \CZa \stwoTp (2 \mD + (\Anu + \mM) \stwoTp)\non\\ 
&& + \SZa (4 \mD^2 + 4 \mD (\Anu + \mM) \stwoTp + 
((\Anu + \mM)^2 - \mu^2) \sQtwoTp))\non\\
&&+ \MZ^2 \mD \Sbe \sQTp (2 \mu \sin(2\alpha+\beta) \stwoTp\non\\
&&+2 
(2 \mD + (\Anu + \mM) \stwoTp) \cos(2\alpha+\beta))\non\\
&&- \MZ^4 \SQb \sin^4\theta_{+} \sin 2(\alpha+\beta))\non\\
&+& 2 B_0[p^2, m_{\tilde N_-}^2, m_{\tilde N_-}^2]
(-\mD^2 
(2 \mu \CZa \stwoTm (2 \mD + (\Anu - \mM) \stwoTm)\non\\
&& - \SZa (4 \mD^2 + 4 \mD (\Anu - \mM) \stwoTm + 
((\Anu - \mM)^2 - \mu^2) \sQtwoTm))\non\\
&&+ \MZ^2 \mD \Sbe \sQTm (2\mu \sin(2\alpha+\beta) \stwoTm\non\\
&&+2 
(2 \mD + (\Anu - \mM) \stwoTm) \cos(2\alpha+\beta))\non\\
&&- \MZ^4 \SQb \sin^4\theta_{-} \sin 2(\alpha+\beta))\non\\
&+& 2 B_0[p^2, m_{\Snu_+}^2, m_{\Snu_+}^2]
(-\mD^2 
(2  \mu \CZa \stwoTp (-2 \mD + (\Anu + \mM) \stwoTp)\non\\
&& - \SZa (4 \mD^2 - 4 \mD (\Anu + \mM) \stwoTp + 
((\Anu + \mM)^2 - \mu^2) \sQtwoTp))\non\\
&&+ \MZ^2 \mD \Sbe \cQTp (-2 \mu \sin(2\alpha+\beta) \stwoTp\non\\
&&+2 
(2 \mD - (\Anu + \mM) \stwoTp) \cos(2\alpha+\beta))\non\\
&&- \MZ^4 \SQb \cos^4\theta_{+} \sin 2(\alpha+\beta))\non\\
&+& 2 B_0[p^2, m_{\Snu_-}^2, m_{\Snu_-}^2]
(-\mD^2 
(2  \mu \CZa \stwoTm (-2 \mD + (\Anu - \mM) \stwoTm)\non\\
&& - \SZa (4 \mD^2 - 4 \mD (\Anu - \mM) \stwoTm + 
((\Anu - \mM)^2 - \mu^2) \sQtwoTm))\non\\
&&+ \MZ^2 \mD \Sbe \cQTm (-2 \mu \sin(2\alpha+\beta) \stwoTm\non\\
&&+2 
(2 \mD - (\Anu - \mM) \stwoTm) \cos(2\alpha+\beta))\non\\
&&- \MZ^4 \SQb \cos^4\theta_{-} \sin 2(\alpha+\beta))]
\EEA
\BEA
{\se{AA}^{\nu}(\MA^2)} &=&-\frac{g^2}{64 \cw^2 \MZ^2 \pi^2}
\frac{\cos^2 \beta \sin^2 2\theta}{\sin^2 \beta} 
[2 \mD^2 A_0[\mnu^2] + 
(2 \mD^2  + \mM^2) A_0[m_N^2]\\
& +& \mM^2 (\mnu^2- \mnu m_N) B_0[\MA^2, \mnu^2, m_N^2]\non\\
&+&\MA^2 (2\mD^2 (B_1[\MA^2,  \mnu^2,  \mnu^2] + 
B_1[\MA^2, m_N^2, m_N^2]) +\mM^2 B_1[\MA^2, \mnu^2, m_N^2])]\non\\[1em]
{\se{AA}^{\Snu}(\MA^2)} &=&\frac{g^2}{256 \cw^2 \MZ^2 \pi^2}
\frac{1}{\SQb}[
 A_0[m_{\Snu_+}^2] (4 \mD^2 \CQb -2 \MZ^2 \cos 2\beta \SQb \cQTp)\non\\
&+&A_0[m_{\tilde N_+}^2] (4 \mD^2 \CQb - 2 \MZ^2 \cos 2\beta
 \SQb \sQTp)\non\\
&+&A_0[m_{\Snu_-}^2] (4 \mD^2 \CQb -2 \MZ^2 \cos 2\beta \SQb \cQTm)\non\\
&+& A_0[m_{\tilde N_-}^2] (4 \mD^2 \CQb - 2 \MZ^2 \cos 2\beta
 \SQb \sQTm)\non\\
&+& 4 \mD^2 \left[B_0[\MA^2, m_{\Snu_+}^2, m_{\Snu_-}^2] (\mu \Sbe \STdifmp
\right.\non\\
&& +\Cb (\Anu \STdifmp - \mM \STsummp))^2\non\\
&+& B_0[\MA^2, m_{\tilde N_+}^2, m_{\tilde N_-}^2] 
(\mu \Sbe \STdifmp\non\\
&&+ \Cb (\Anu \STdifmp + \mM \STsummp))^2\non\\
&+& B_0[\MA^2, m_{\tilde N_-}^2, m_{\Snu_+}^2] 
(\mu \Sbe \CTdifmp\non\\
&& +\Cb (\Anu \CTdifmp - \mM \CTsummp))^2\non\\
&+& B_0[\MA^2, m_{\tilde N_+}^2, m_{\Snu_-}^2]
(\mu \Sbe \CTdifmp\non\\
&&\left.+\Cb (\Anu \CTdifmp + \mM \CTsummp))^2\right]]
\EEA
\BEA
{\se{ZZ}^{\nu}(\MZ^2)} &=&-\frac{g^2}{32 \cw^2\pi^2}
\left[\cos^4\theta A_0[\mnu^2] +\frac{1}{2} 
(3+\cos 2\theta)\sin^2\theta A_0[m_N^2]\right.\non\\
&+&2 \cos^4\theta \left(\mnu^2 B_0[\MZ^2,  \mnu^2,  \mnu^2]
- B_{00}[\MZ^2,  \mnu^2,  \mnu^2]
+\frac{\MZ^2}{2} B_1[\MZ^2,  \mnu^2,  \mnu^2]\right)\non\\
&+&2 \sin^4\theta \left(m_N^2 B_0[\MZ^2,  m_N^2,  m_N^2]
- B_{00}[\MZ^2,  m_N^2,  m_N^2]
+\frac{\MZ^2}{2} B_1[\MZ^2,  m_N^2,  m_N^2]\right)\non\\
&+&\frac{1}{2}\sin^2 2\theta \left(\mnu(\mnu+m_N) B_0[\MZ^2,  \mnu^2,  m_N^2]
- 2 B_{00}[\MZ^2, \mnu^2,  m_N^2]\right.\non\\
&&\left.\left.
+\MZ^2 B_1[\MZ^2, \mnu^2,  m_N^2]\right)\right]\\
&&\non\\
{\se{ZZ}^{\Snu}(\MZ^2)} &=&\frac{g^2}{64 \cw^2\pi^2}
\Big[ A_0[m_{\Snu_-}^2]\cQTm + A_0[m_{\Snu_+}^2] \cQTp \non \\
&+& A_0[m_{\tilde N_-}^2] \sQTm + A_0[m_{\tilde N_+}^2] \sQTp \non \\
&-& 4 (B_{00}[\MZ^2, m_{\Snu_+}^2, m_{\Snu_-}^2] \cQTm \cQTp
+ B_{00}[\MZ^2, m_{\tilde N_-}^2, m_{\Snu_+}^2] \cQTp \sQTm\non\\
&+& B_{00}[\MZ^2, m_{\tilde N_+}^2, m_{\Snu_-}^2] \cQTm \sQTp
+ B_{00}[\MZ^2, m_{\tilde N_+}^2, m_{\tilde N_-}^2] \sQTm \sQTp)
\Big]
\EEA

The definitions of the loop functions $A_0$, $B_0$, $B_1$ and $B_{00}$ appearing 
in this and the next appendices can be found, for instance, in
~\citere{loopfunctions} (where $B_{00}=B_{22}$). 

%%%%%%%%%%%%%%%%%%%%%%%%%%%%%%%%%%%%%%%%%%%%%%%%%%%%%%%%%%%%%%%%%%%%%%%%%%%%%
%%%%%%%%%%%%%%%%%%%%%%%%%%%%%%%%%%%%%%%%%%%%%%%%%%%%%%%%%%%%%%%%%%%%%%%%%%%%%

\section*{Appendix C: Dirac case.
One-loop contributions from neutrinos and sneutrinos to the 
renormalized $h$ Higgs boson self-energy}

We present here the result for the one-loop corrections 
from neutrinos ($\nu$) and sneutrinos (${\Snu}$) to the renormalized 
$hh$ self-energy in the case of Dirac neutrinos, obtained in 
the \drbar\ scheme. Here $\cw=\cos\theta_W$.
\BEA
\ser{hh}^{\nu}(p^2)_{\mathrm{Dirac}} &=&
\frac{g^2}{32 \cw^2 \MZ^2 \pi^2}\Big\{
 A_0[\mD^2]\, (\sin^2(\al +\be)
  \MZ^2 \non\\
&& +\frac{1}{\Sbe}(\sin(2\alpha-3\beta) + 
3 \sin(2\alpha- \beta) -2 \Sbe))\mD^2 \non\\
&+&  \sin^2(\al +\be)
\MZ^2 (\mD^2 B_0[\MZ^2, \mD^2, \mD^2]\non\\
&&-2 B_{00}[\MZ^2, \mD^2, \mD^2] +\MZ^2 B_1[\MZ^2, \mD^2, \mD^2])\non\\
&-&2 \frac{\cos^2 \alpha}{\sin^2 \beta} (2\mD^4 B_0[p^2, \mD^2, \mD^2]+
p^2 B_1[p^2, \mD^2, \mD^2])\non\\
&+& 2 \MA^2 \mD^2 \frac{\cos^2(\al -\be)
\cos^2\beta}{\sin^2 \beta}
B_1[\MA^2, \mD^2, \mD^2]\Big\}\\[1em]
\ser{hh}^{\Snu}(p^2)_{\mathrm{Dirac}} &=&
-\frac{g^2}{256 \cw^2 \MZ^2 \pi^2} \Bigg\{ A_0[m_{\tilde{\nu}_1}^2]
\Big[ 8\MZ^2 \sin^2(\al +\be)
\cos^2 {\tilde \theta} 
+2\mD \frac{\Samb \sin 2 {\tilde \theta}}{\Sbe} \times \non\\
&& \qquad
 (\mu(3\Sa - \sin(\alpha-2\beta))+\Anu
(3\Ca + \cos(\alpha-2\beta)))\Big] \non\\
&+&A_0[m_{\tilde{\nu}_2}^2]
\Big[8\MZ^2 \sin^2(\al +\be) 
\sin^2 {\tilde \theta}
-2\mD \frac{\Samb \sin 2 {\tilde \theta}}{\Sbe} \times \non\\
&& \qquad (\mu(3\Sa - \sin(\alpha-2\beta))+\Anu
(3\Ca + \cos(\alpha-2\beta)))\Big] \non\\
&-&\frac{1}{16}\frac{1}{\sin^2\beta}
B_0[p^2, m_{\tilde{\nu}_1}^2, m_{\tilde{\nu}_1}^2]
 \Big[ 2 (8 \mD^2 - \MZ^2) \Ca \non\\
&& \qquad +2 \MZ^2 (\cos(\alpha+ 2 \beta) - 2 \cos 2{\tilde \theta} \Sbe \Sab)
\non \\
&& \qquad - 8 \mD \sin 2{\tilde \theta} \Ca(\Anu + \mu \tan \alpha)\Big]^2\non\\
%&& \htr{Are we sure about the $[]^2$ ??} \non \\
&-&\frac{1}{16}\frac{1}{\sin^2\beta}
B_0[p^2, m_{\tilde{\nu}_2}^2, m_{\tilde{\nu}_2}^2]
 \Big[ 2 (8 \mD^2 - \MZ^2) \Ca \non\\
&&\qquad +2 \MZ^2 (\cos(\alpha+ 2 \beta) + 2 \cos 2{\tilde \theta} \Sbe \Sab)
\non \\
&& \qquad 
+ 8 \mD \sin 2{\tilde \theta} \Ca(\Anu + \mu \tan \alpha)\Big]^2\non\\
&-&\frac{1}{8}\frac{1}{\sin^2\beta}
B_0[p^2, m_{\tilde{\nu}_2}^2, m_{\tilde{\nu}_1}^2]
\Big[ -4 \MZ^2 \sin 2{\tilde \theta} \Sbe \Sab \non\\
&&\qquad +8 \mD \cos 2{\tilde \theta} \Ca (\Anu + \mu \tan\alpha)\Big]^2\non\\
&+& 8 \mD^2 \cos^2(\al-\be)
\cot^2 \beta \,
B_0[\MA^2, m_{\tilde{\nu}_2}^2, m_{\tilde{\nu}_1}^2]\,
 (\Anu + \mu \tan \beta)^2\non\\
&-& 8 \MZ^2 \sin^2(\al+\be)
\left(
2\cos^4 {\tilde \theta} 
B_{00}[\MZ^2, m_{\tilde{\nu}_1}^2, m_{\tilde{\nu}_1}^2]+
2\sin^4 {\tilde \theta} 
B_{00}[\MZ^2, m_{\tilde{\nu}_2}^2, m_{\tilde{\nu}_2}^2]\right.\non\\
&&\left.+
\sin^2 2{\tilde \theta} 
B_{00}[\MZ^2, m_{\tilde{\nu}_2}^2, m_{\tilde{\nu}_1}^2]\right)\Bigg\}
\EEA
\end{appendix}

%%%%%%%%%%%
%%%%%%%%%%%
%%%%%%%%%%%

%%%%%%%%%%%%%%%%%%%%%%%%%%%%%%%%%%%%%%%%%%%%%%%%%%%%%%%%%%%%%%%%%%%%%%%%%%%%%%%
%%%%%%%%%%%%%%%%%%%%%%%%%%%%%%%%%%%%%%%%%%%%%%%%%%%%%%%%%%%%%%%%%%%%%%%%%%%%%%%

%\newpage
%\pagebreak
%\clearpage

\end{document}